\documentclass[12pt]{article} 
\usepackage{graphicx,color} 
\usepackage{times,latexsym,amssymb,overcite}
\usepackage{phifthen}

\graphicspath{{./}{efqcgr/}{graphics/}{cliff/finaldata/}}
\definecolor{purple}{rgb}{0.625,0.125,0.9375}
\definecolor{grey}{rgb}{0.8,0.8,0.8}

\usepackage{hyperref}
\ifx\hypersetup\undefined\relax\else\hypersetup{colorlinks=true,urlcolor=blue,linkcolor=purple,pageanchor=false}\fi % _hyperref_
\usepackage{geometry}
\newcommand{\ifnature}[2]{\ifthenelse{\isundefined{\fornature}}{#2}{#1}}
\ifnature{\linespread{2}}{\linespread{1.1}}

\newcommand{\ignore}[1]{}
\ignore{
rm -R /tmp/efqc; mkdir /tmp/efqc
cp `texfls efqc.log | perl -e '$a=<>; $a =~ s:/\S*(revtex4|natbib)\S*(\s|$)::g; print "$a\n"'` /tmp/efqc
find /tmp/efqc -type f -name '*.pdf' -exec pdftops -eps {} \; -exec rm {} \;
find /tmp/efqc -type f -name '*.jpg' -exec perl -e '$f=shift; $g=f$; $g=~s/.jpg/.eps/; exec("convert $f $g");' {} \; -exec rm {} \;
perl -i -n -e 's/^%(.*)% _dvitex_/$1/; print;' /tmp/efqc/efqc.tex

cp /usr/share/texmf/tex/latex/base/ifthen.sty /tmp/efqc/phifthen.sty
perl -i -n -e 's/ProvidesPackage\{ifthen\}/ProvidesPackage\{phifthen\}/; print;' /tmp/efqc/phifthen.sty
perl -i -n -e 's/usepackage\{ifthen\}/usepackage\{phifthen\}/; print;' /tmp/efqc/efqc.tex

(cd /tmp/efqc; tar czvf efqc.tar.gz *)
}

\newcommand{\Sy}{\mathbb{S}}
\newcommand{\sysfnt}{\mathsf}

\newcommand{\ket}[1]{{|}{#1}{\rangle}}

\newcommand{\kets}[2]{{|}{#1}{\rangle}_{{}_{\!\!\scriptstyle{\sysfnt{#2}}}}}

\newcommand{\slb}[2]{{{#1}^{({\sysfnt{#2}})}}}

\newcommand{\expdata}[2]{#1} 
\newcommand{\cA}{{\cal A}}
\newcommand{\cC}{{\cal C}}
\newcommand{\cE}{{\cal E}}

\newcommand{\cG}{{\cal G}}

\newcommand{\one}{I}
\newcommand{\dg}{{}^{\circ}}
\newcommand{\serrb}[1]{{\pm}#1}
\newcommand{\aerrb}[2]{\raisebox{1pt}{$\pm$\scalebox{.8}{\mbox{${}^{#1}_{#2}$}}}}

\newcommand{\nputbox}[3]{\put(#1){\makebox(0,0)[#2]{#3}}}
\newcommand{\nputgr}[4]{\put(#1){\makebox(0,0)[#2]{\includegraphics[#3]{#4}}}}
\newcounter{herefignum}
\newenvironment{herefig}{\begin{center}\refstepcounter{herefignum}}{\end{center}}
\newcommand{\herefigcap}[1]{\\\begin{minipage}{\textwidth}{FIGURE.~\theherefignum: #1}\end{minipage}}

\newcounter{heretabnum}
\newenvironment{heretab}{\begin{center}\refstepcounter{heretabnum}}{\end{center}}
\newcommand{\heretabcap}[1]{{TABLE.~\theheretabnum: #1}}

\newcommand{\phead}[1]{\par\noindent{\bf #1}}

\newcommand{\minCSS}{\cG_{\textnormal{\tiny min}}}

\begin{document}

\title{
Quantum Computing with Very Noisy Devices
}

\author{E. Knill \\[-6pt]
\normalsize \emph{
   Mathematical and Computational Sciences Division,}\\[-8pt]
\normalsize
   \emph{National Institute of Standards and Technology, Boulder, CO 80305}
}

\date{}
\maketitle

% Summary. Up to 150 words, no references.
\textbf{ In theory, quantum computers can efficiently simulate quantum
physics, factor large numbers and estimate integrals, thus solving
otherwise intractable computational problems.  In practice, quantum
computers must operate with noisy devices called ``gates'' that tend
to destroy the fragile quantum states needed for computation.  The
goal of fault-tolerant quantum computing is to compute accurately even
when gates have a high probability of error each time they are used.
Here we give evidence that accurate quantum computing is possible with
error probabilities above $3\,\%$ per gate, which is significantly
higher than what was previously thought possible.  However, the
resources required for computing at such high error probabilities are
excessive. Fortunately, they decrease rapidly with decreasing error
probabilities.  If we had quantum resources comparable to the
considerable resources available in today's digital computers, we
could implement non-trivial quantum computations at error
probabilities as high as $1\,\%$ per gate.  }

% Up to 5 pages of text. 
% Without figures etc, this is about 5*1300 words.
% Guideline: 3000 words of text
% Begin with about 500 words of referenced background.
%%%
% 10/21/04: 5762 words outside of figures and captions.
% 10/31/04: 5789 words
% 11/02/04: 5704 words (detex, w/o \ignore's)

\ignore{
Remaining terminological issues:
* Two architectures: Postselected and error-correcting $C_4/C_6$,
clarify terminology and use it?
* ``postselected threshold''?
}

Research in quantum computing is motivated by the great increase in
computational power offered by quantum
computers~\cite{shor:qc1995a,feynman:qc1982a,abrams:qc1999b}. There is
a large and still growing number of experimental efforts whose
ultimate goal is to demonstrate scalable quantum computing.  Scalable
quantum computing requires that arbitrarily large computations can be
efficiently implemented with little error in the output.
Criteria that need to be satisfied by devices used for scalable
quantum computing have been specified by
DiVincenzo~\cite{divincenzo:qc2000a}.  One of the criteria is that the
level of noise affecting the physical gates is sufficiently low. The
type of noise affecting the gates in a given implementation is called
the ``error model''. A scheme for scalable quantum computing in the
presence of noise is called a ``fault-tolerant architecture''. In view of
the low-noise criterion, studies of scalable quantum computing involve
constructing fault-tolerant architectures and providing answers to
questions such as the following: Q1: Is scalable quantum computing
possible for error model $\cE$?  Q2: Can fault-tolerant
architecture $\cA$ be used for scalable quantum computing with error
model $\cE$?  Q3: What resources are required to implement quantum
computation $\cC$ using fault-tolerant architecture $\cA$ with
error model $\cE$?

To obtain broadly applicable results, fault-tolerant architectures are
constructed for generic error models. Here, the error model is
parametrized by an error probability per gate (or simply error per
gate, EPG), where the errors are unbiased and independent.  The
fundamental theorem of scalable quantum computing is the threshold
theorem and answers question Q1 as follows: If the EPG is smaller than
a threshold, then scalable quantum computing is
possible~\cite{preskill:qc1998a,kitaev:qc1997a,aharonov:qc1996a,knill:qc1998a}.
Thresholds depend on additional assumptions on the error
model and device capabilities.  Estimated thresholds vary from
below $10^{-6}$
[\citeonline{preskill:qc1998a,kitaev:qc1997a,aharonov:qc1996a,knill:qc1998a}]
to $3\times 10^{-3}$ [\citeonline{steane:qc2002a}], with $10^{-4}$
[\citeonline{gottesman:qc1997b}] often quoted as the EPG to be
achieved in experimental quantum computing.

Many experimental proposals for quantum computing claim to achieve
EPGs below $10^{-4}$ in theory. However, in the few cases where
experiments with two quantum bits (qubits) have been performed, the
EPGs currently achieved are much higher, $3\times 10^{-2}$ or more in
ion traps~\cite{leibfried:qc2003a,roos:qc2004a} and liquid-state
NMR~\cite{knill:qc1999a,childs:qc2001b}.  The first goal of our work
is to give evidence that scalable quantum computing is possible at
EPGs above $3\times 10^{-2}$. While this is encouraging, the
fault-tolerant architecture that achieves this is extremely
impractical because of large resource requirements.  To reduce the
resource requirements, lower EPGs are required.  The second goal of
our work is to give a fault-tolerant architecture (called the
``$C_4/C_6$ architecture'') well suited to EPGs between $10^{-4}$ and
$10^{-2}$ and to determine its resource requirements, which we compare to
the state of the art in scalable quantum computing as exemplified by
the work of Steane~\cite{steane:qc2002a}.

Fault-tolerant architectures realize low-error qubits and
gates by encoding them with error-correcting codes. A standard
technique for amplifying error reduction is concatenation.  Suppose we
have a scheme that, starting with qubits and gates at one EPG, produces
encoded qubits and gates that have a lower EPG.  Provided the error model for
encoded gates is sufficiently well behaved, we can then apply the same
scheme to the encoded qubits and gates to obtain a next level of
encoded qubits and gates with much lower EPGs. Thus, a concatenated
fault-tolerant architecture involves a hierarchy of repeatedly encoded
qubits and gates.  The hierarchy is described in terms of levels of
encoding, with the physical qubits and gates being at level $0$.  The
top level is used for implementing quantum computations and its
qubits, gates, EPGs, etc.~are referred to as being ``logical''.
Typically, the EPGs decrease superexponentially with number of levels,
provided that the physical EPG is below the threshold for the
architecture in question.

The $C_4/C_6$ architecture differs from previous ones in five
significant ways. First, we use the simplest possible error-detecting
codes, thus avoiding the complexity of even the smallest
error-correcting codes. Error correction is added naturally by
concatenation. Second, error correction is performed in one step and
combined with logical gates by means of error-correcting
teleportation.  This minimizes the number of gates contributing to
errors before they are corrected. Third, the fault-tolerant
architecture is based on a minimal set of operations with only one
unitary gate, the controlled-NOT.  Although this set does not suffice
for universal quantum computing, it is possible to bootstrap other
gates.  Fourth, verification of the needed ancillary states (logical
Bell states) largely avoids the traditional syndrome-based schemes.
Instead, we use hierarchical teleportations. Fifth, the highest
thresholds are obtained by introducing the model of postselected
computing with its own thresholds, which may be higher than those for
standard quantum computing. Our fault tolerant implementation of
postselected computing has the property that it can be used to prepare
states sufficient for (standard) scalable quantum computing.

\phead{Basics.} For an introduction to quantum information,
computing and error correction, see~[\citeonline{nielsen:qc2001a}].
The unit of quantum information is the qubit whose states are
superpositions $\alpha\ket{0}+\beta\ket{1}$.  Qubits are acted on by
the Pauli operators $X=\sigma_x$ (bit flip), $Z=\sigma_z$ (sign flip)
and $Y=\sigma_y=i\sigma_x\sigma_z$.  The identity operator is
$I$. One-qubit gates include preparation of $\ket{0}$ and
$\ket{+}=(\ket{0}+\ket{1})/\sqrt{2}$, $Z$-measurement (distinguishing
between $\ket{0}$ and $\ket{1}$), $X$-measurement (distinguishing
between $\ket{+}$ and $\ket{-}=(\ket{0}-\ket{1})/\sqrt{2}$), and the
Hadamard gate (HAD,
$\alpha\ket{0}+\beta\ket{1}\mapsto\alpha\ket{+}+\beta\ket{-}$).  We
use one unitary two-qubit gate, the controlled-NOT (CNOT), which maps
$\ket{00}\mapsto \ket{00}$, $\ket{01}\mapsto\ket{01}$,
$\ket{10}\mapsto\ket{11}$, and $\ket{11}\mapsto\ket{10}$.  This set of
gates is a subset of the so-called Clifford gates, which are
insufficient for universal quantum computing~\cite{gottesman:qc1997b}.
Our minimal gate set $\minCSS$ consists of $\ket{0}$ and $\ket{+}$
preparation, $Z$ and $X$ measurement and CNOT. Universality may be
achieved with the addition of other one-qubit preparations or
measurements, as explained below. The physical gates mentioned are
treated as being implemented in one ``step''; the actual
implementation may be more complex.

The $C_4/C_6$ architecture is based on two error-detecting stabilizer
codes, $C_4$, a four-qubit code, and $C_6$, a three qubit-pair code,
both encoding a qubit pair. A stabilizer code is a common eigenspace
of a set of commuting products of Pauli operators (the ``check
operators'').  Such products are denoted by strings of $X$, $Y$, $Z$
and $I$. For example, $XIZ$ is a Pauli product for three qubits with
$X$ acting on the first and $Z$ on the last.  The shortest
error-detecting code $C_4$ for qubits encodes two qubits in four and
has check operators $XXXX$ and $ZZZZ$.  The encoded qubits (labeled
$L$ and $S$) are defined by encoded operators $X_L=XXII$, $Z_L=ZIZI$,
$X_S=IXIX$ and $Z_S=IIZZ$.  We use this code as a first level of
encoding and call the encoded qubits ``level $1$'' qubits. Level $1$
qubits come in pairs, each encoded in a ``block'' of four physical
qubits. The second code $C_6$ is constructed as a code on three qubit
pairs able to detect any error acting on one pair. It encodes a qubit
pair and has check operators $XIIXXX$, $XXXIIX$, $ZIIZZZ$ and $ZZZIIZ$
acting on three consecutive qubit pairs. A choice for encoded
operators for $C_6$ is $X_L = IXXIII, Z_L = IIZZIZ, X_S = XIXXII, Z_S
= IIIZZI$. This code is used for the second and higher levels of
encoding. For example, the second level is obtained by using three
level one pairs to obtain a level two pair. A level $l$ qubit pair
requires a block of $4\times 3^{l-1}$ physical qubits. The
block structure is depicted in Fig.~\ref{fig:c4/c6_struct}.

The concatenation of error-detecting codes allows for a flexible use
of error detection and correction.  Given a joint eigenstate of the
check operators, its list of eigenvalues is called the ``syndrome''.
The level $l$ encoding has check operators that can be derived from
the check and encoded operators of $C_4$ and $C_6$. Ideally, the state
of a level $l$ qubit exists in the subspace with syndrome $\mathbf{0}$
(all eigenvalues are $+1$).  In the presence of errors this is
typically not the case, so the state is defined only with respect to a
current ``Pauli frame'' and an implicit recovery scheme. The Pauli
frame is defined by a Pauli product that restores the error-free state
of the block to the syndrome $\mathbf{0}$ subspace.  The implicit
recovery scheme determines the Pauli products needed to coherently map
states with other syndromes to the syndrome of the error-free state.
Defining the level $l$ state in this way makes it possible to avoid
explicitly applying Pauli products for correction or teleportation
compensation~\cite{steane:qc2002a}. Error-detection and correction are
based on measurements that retroactively determine the syndrome of the
state (the current syndrome has already been affected by further
errors).  An error is detected when the syndrome differs from that
expected according to the Pauli frame. In ``postselected'' quantum
computing, the state is then rejected and the computation restarted.
In standard quantum computing, the syndrome information must be used
to update the Pauli frame. With the $C_4/C_6$ architecture, it is
possible to do so at level 2 and above by the following method: First
check the level 1 $C_4$ syndromes of each block of four qubits. For
each block where an error is detected, mark the encoded level 1 qubit
pair as having an error. Proceed to level 2 and check the (encoded)
$C_6$ syndrome for each block of three level 1 pairs. If exactly one
of the level 1 pairs has an error, use the $C_6$ syndrome to correct
it. This works because error-detecting codes can always correct an
error at a known location. If not, mark the encoded level 2 pair as
having an error unless none of the three level 1 pairs have an error
and the $C_6$ syndrome is as expected according to the Pauli frame.
Continue in this fashion through all higher levels.  For optimizing
state preparation, we can replace the error-correction step by
error-detection at the top few levels depending on context as
explained below.

\phead{Error model and assumptions.}  All error models can be
described by inserting errors (which act as quantum operations) after
gates or before measurements.  We could model correlations between the
errors by extending the errors' quantum operations to a common
external environment.  However, here we assume that errors are
independent.  We further assume that a gate's errors consist of
applications of Pauli products with probabilities determined by the
gate.  Ideally, we would obtain a threshold that does not depend on
the details of the probability distributions of Pauli products.  This
is too difficult with available techniques, so a \emph{depolarizing}
model is assumed for each gate: $\ket{0}$ ($\ket{+}$) state
preparation erroneously produces $\ket{1}$ ($\ket{-}$) with
probability $e_p$.  A binary (e.g. $Z$ or $X$) measurement results in
the wrong outcome with probability $e_m$. CNOT is followed by one of
the $15$ possible non-identity Pauli products, each with probability
$e_c/15$. HAD is modified by one of the Pauli operators, each with
probability $e_h/3$.  We further simplify by setting $e_c=\gamma$,
$e_m=e_p= 4\gamma/15$, $e_h=4\gamma/5$. This choice is justified as
follows: $4\gamma/5$ is the one-qubit marginal probability of error
for the CNOT, and it is reasonable to expect that one-qubit gates have
error below this.  In fact, one-qubit gates have much lower error than
CNOTs in experimental systems such as ion-traps and liquid-state NMR.
As for preparation errors, if they are much larger than $4\gamma/15$,
then it is possible to purify prepared states using a CNOT. For
example, prepare $\ket{0}$ twice, apply a CNOT from the first to the
second and $Z$ measure the second. Try again if the measurement
outcome indicates $\ket{1}$, otherwise use the first state.  The
probability of error is given by $4\gamma/15 +
O(\gamma^2)$, assuming that CNOT error is as above and measurement and
preparation errors are proportional to $\gamma$.  This also works for
$\ket{+}$ preparation.  \ignore{ The only order 1 (in $\gamma$) way to
accept the wrong state involves having CNOT errors of the form
$(X|Y)(X|Y)$, which gives four possibilities.  } To improve $Z$
measurement, it is necessary to introduce an ancilla in $\ket{0}$,
apply a CNOT from the qubit to be measured to the ancilla, and measure
both qubits, accepting the answer only if the measurements agree. The
error probability conditional on acceptance is again
$4\gamma/15+O(\gamma^2)$. Detected error is much more readily managed
than undetected error~\cite{knill:qc2003b}. In our architecture,
the primary role of state preparation implies that the
conditional error is typically more relevant. To improve measurement without
possibility of rejection requires an additional
ancilla~\cite{divincenzo:qc2000a} and CNOT with majority decoding of
the three measurement outcomes. However, the error probability is now
$4\gamma/5 + O(\gamma^2)$.  \ignore{ With ancillas numbered $2$ and
$3$, use CNOT($1\rightarrow 2)$ followed by CNOT($1\rightarrow 3)$.
If the state starts out as $\ket{0}$, The following outputs of weight
$>1$ have probability $O(\gamma)$: $101$ ($8\gamma/15$), $111$
($4\gamma/15$).  }

The error model used here is idealized and does not match the error
behavior of physical qubits and gates. There are three notable
differences. First, real errors include coherent rotations, but any
error can still be expressed as a linear combination of Pauli
products. The syndrome measurements serve to ``collapse'' the linear
combinations, so that these errors can be managed.  The main problem
with such errors is that consecutive errors can add coherently rather
than probabilistically, resulting in more rapid error propagation. In
principle, this problem can be eliminated by frequently applying known
but random Pauli products, thus modulating the Pauli frame and
reducing the likelihood of coherent addition.  This also has the
beneficial effect of decoupling~\cite{kern:qc2004a} weakly interacting
environments. The second difference is that real errors on gates
nearby in time and space have correlations. These correlations are
expected to decay rapidly with distance, and their effect can be
alleviated by known coding techniques such as block
interleaving~\cite{steane:qc2002a}. The third difference is that in
many cases, qubits are defined by a subspace of a quantum system from
which amplitude can leak.  An example of this problem is photon loss
in optical quantum computing. Leakage errors, particularly if
undetected, can be problematic. One advantage of error-correcting
teleportation is that leakage is automatically controlled at each
step.

The error model does not specify ``memory'' error or the amount of
time used by a measurement~\cite{steane:qc2002a}.  We assume that
gates other than measurements take the same amount of time; that is,
the error parameter should represent the total error including any
delays for faster gates to equalize gate times.  For the $C_4/C_6$
architecture, memory is an issue only when waiting for measurement
outcomes that determine whether prepared states are good, or that are
needed after teleportation, particularly when implementing
non-Clifford gates~\cite{gottesman:qc1999a,zhou:qc2000b}. If the
architecture is used for postselected computing, we can compute
``optimistically'', anticipating but not waiting for measurement
outcomes. The output of the computation is accepted only if all
measurement outcomes are as anticipated.  Consider standard quantum
computing with maximum parallelism.  In teleportation, Bell
measurements determine correction gates that need to be
applied. If the correction gates are simple Pauli products, they can
be absorbed into the Pauli frame and do not need to be known
immediately.  For teleportations used to implement non-Clifford gates,
a non-Pauli compensation may be required. In this case, we must wait for
the measurement outcomes.  To avoid accumulating memory errors, we can
maintain the logical state by repeatedly applying error-correcting
teleportations with delays whose memory error is equivalent to that of
physical one-qubit gates. The logical errors for these steps are
comparable to logical one-qubit gates, which are small by design. The
measurement outcomes for the teleportations determine only Pauli-frame
updates and need not be known immediately.  In state preparation,
using additional resources, we may continue computing optimistically
without waiting for measurement outcomes until the state is ready to
be used for a logical gate. At this point it is necessary to wait for
measurement outcomes to make sure the prepared state has no detected
uncorrectable error. This adds at most the memory error incurred
during a measurement time to each qubit. To account for this we assume
that gate errors are set high enough to include this memory error.

Two additional assumptions are used in analyzing the $C_4/C_6$
architecture: The first is that there is no error and no speed
constraint on classical computations required to interpret measurement
outcomes and control future gates.  The second is that two-qubit gates
can be applied to any pair of qubits without delay or additional
error. This assumption is unrealistic, but the effect on
the threshold is due primarily to relatively short-range CNOTs acting
within the ancillas needed for maintaining one or two blocks. This may
be accounted for by use of a higher effective
EPG~\cite{steane:qc2002a,svore:qc2004a}.

\phead{Clifford gates for $C_4$ and $C_6$.}  The codes $C_4$, $C_6$
and their concatenations have the property that encoded CNOTs, HADs
and measurements that act in parallel on both encoded qubits in a pair
can be implemented ``transversally'' with physical qubit relabeling.
For example, to apply an encoded CNOT between two encoded qubit pairs,
it suffices to apply physical CNOTs transversally, that is between
corresponding physical qubits in the encoding blocks.  HAD requires in
addition permuting the physical qubits in a block, which can be done
by relabeling without physical manipulations or error, see the
supplementary information (S.I. Sect.~\ref{sect:networks}).  The
transversal implementation ensures that errors from the physical gates
apply independently to each physical qubit in a block so that they can
be managed by error detection or correction.  We use two methods for
encoded state preparation. The first yields low-error encoded
$\ket{0}$ and $\ket{+}$ states and follows from the Bell
state-preparation scheme needed for error correction
(S.I. Sect.~\ref{sect:networks}).  The second uses teleportation to
``inject'' physical states into encoded qubits. The resulting encoded
state has error but can be purified.

\phead{Error-correcting teleportation.}  To correct (more accurately,
to keep track of) errors, we use error-correcting teleportation, which
generalizes gate teleportation~\cite{gottesman:qc1999a}.  It involves
preparing two blocks, each encoding a logical qubit pair so that the
first pair is uniformly entangled with the second. Two such blocks
form a ``logical Bell pair'', and its logical state is the ``logical
Bell state''. Suppose that a logical Bell pair's error is as if each
physical qubit were subject to independent error of order $\gamma$. A
block used for logical computation can then be error-corrected by
applying Bell measurements transversally between corresponding qubits
in the computational block and the first block of the logical Bell
pair. This is the first step of conventional quantum
teleportation~\cite{bennett:qc1993b} and results in the transfer of
the logical state to the second block of the logical Bell pair, up to
a known change in the Pauli frame. The Bell measurement outcomes
reveal the syndromes of the products of identical check operators on
the two blocks.  Provided that the combined errors from the two
measured blocks are within the limits of what the codes used can
handle, they can be determined to update the Pauli frame
(S.I. Sect.~\ref{sect:teleportation}). Compared to the syndrome
extraction methods of Steane~\cite{steane:qc2002a}, error-correcting
teleportation involves only one step instead of at least two but
requires preparing more complex states.

\phead{Logical Bell state preparation.}  State preparation networks
are detailed in S.I Sect.~\ref{sect:networks}. It is necessary to
prepare logical Bell states so that any errors introduced are similar
to independent physical one-qubit errors. We prepare such states by
constructing encoded Bell states at each level, using them as a
resource for constructing Bell states at the next level.  An encoded
Bell state can be obtained by preparing and verifying encoded
$\ket{++}\ket{00}$ in two encoded qubit pairs and applying an encoded
CNOT from the first to the second.  The encoded CNOT is applied
transversally but can introduce correlations between the first and
second block.  To limit these correlations to the current level, the
subblocks are teleported using lower-level Bell states with error
detection or correction depending on context.  The remaining
correlations do not appear to significantly affect logical errors but
can be reduced by purification~\cite{duer:qc2003a} or by entanglement
swapping~\cite{zukowski:qc1993a} with two encoded Bell states.  A key
observation for preparing encoded $\ket{00}$ and $\ket{++}$ is that
for both $C_4$ and $C_6$, they are close to cat states such as
$(\ket{0\ldots 0}+\ket{1\ldots 1})/\sqrt{2}$. In the case of $C_4$,
encoded $\ket{00}$ and $\ket{++}$ are cat states on four qubits. For
$C_6$, they are parallel three-qubit cat states on three qubit pairs
modified by internal CNOTs on two of the pairs.  To prepare verified
cat states we use a minimal variant of the methods of
Shor~\cite{shor:qc1996a} starting with Bell states.  For the
concatenations used here, the internal CNOTs can be implemented by
relabeling the physical qubits.

\phead{Low error logical $\minCSS$ gates.}  The first step in
establishing fault tolerance of the $C_4/C_6$ architecture is to
implement logical $\minCSS$ gates with low EPGs. For the purpose of
establishing high thresholds, we first consider postselected $\minCSS$
computing.  Postselected computing is like standard quantum computing
except that when a gate is applied, the gate may fail. If it fails,
this is known. The probability of success must be non-zero. There may
be gate errors conditional on success, but fault-tolerant postselected
computing requires that such errors are small. Purely postselected
computing has little computational power, but we can use it to prepare
states needed to enable scalable quantum computing.  A fault-tolerant
architecture for postselected computing can be implemented by use of
the $C_4/C_6$ architecture without error correction, aborting the
computation whenever an error is detected.  We have used two methods
to determine threshold values for $\gamma$ below which fault-tolerant
postselected $\minCSS$ computing is possible.  The first involves a
computer-assisted heuristic analysis of the conditional errors in
prepared encoded Bell pairs. The analysis is described for a $C_4$
architecture in~[\citeonline{knill:qc2004b}].  It requires that
encoded Bell pairs are purified to ensure that errors are
approximately independent between each Bell pair's two blocks.  We
obtained exact conditional errors for level $1$ encoded Bell pairs and
then heuristically bounded them from above with an error model that is
independent between the two blocks.  This independence implies that the
error model for gates at the next level also satisfies strict
independence, so the process can be repeated at each level to bound
the conditional logical errors.  With this analysis, thresholds of
above $\gamma=0.03$ were obtained.  The second method involves direct
simulation of the error behavior of postselected encoded CNOTs with
error-detecting teleportation at up to two levels of encoding and
physical EPGs of $.01 \leq \gamma \leq .0375$. The simulation method
is outlined in S.I. Sect.~\ref{sect:simulation}. The resulting
conditional logical errors are shown in Fig.~\ref{fig:condEPG} and
suggest a threshold of above $\gamma = 0.06$ by extrapolation.  At
$\gamma=0.03$, the logical preparation and measurement errors were
found to be consistent with being below the threshold.

Scalable $\minCSS$ computing with the $C_4/C_6$ architecture requires
lower EPGs and the use of error correction to increase the probability
of success to near $1$. To optimize the resource requirements needed
to achieve a given logical EPG, the last level at which error
correction is used is $dl$ levels below the relevant top level, where
$dl$ depends on context and $\gamma$. At higher levels, errors are
only detected.  For simplicity and to enable extrapolation by
modeling, we examined a fixed strategy with $dl=1$ in all
state-preparation contexts and $dl=0$ (maximum error correction) in
the context of logical computation.  The relevant top level in a state
preparation context is the level of a block measurement or
error-correcting teleportation of a subblock, not the logical level of
the state that is eventually prepared.  Each logical gate now has a
probability of detected but uncorrectable error, and a probability of
logical error conditional on not having detected an
error. Fig.~\ref{fig:ecEPG} shows both error probabilities up to level
$4$ for a logical CNOT with error-correcting teleportation and EPGs
$\gamma \leq 0.01$. The data indicate that the threshold for this
architecture is above $0.01$. The logical preparation and measurement
errors were found to be comparatively low.

In designing and analyzing fault-tolerant architectures, particularly
those based on concatenation, care must be taken to ensure that
logical errors do not have correlations that lead to larger than
expected errors when gates are composed. Such effects can be missed
when inferring thresholds from analysis or simulation of just one
level of concatenation.  An additional complication is that the
$C_4/C_6$ architecture's level $l+1$ gates are not implemented solely
in terms of level $l$ gates. We therefore simulated the architecture
at the highest levels possible. To verify that logical errors are
sufficiently uncorrelated, we simulated sequential teleportation and
checked the incremental error behavior of each step as shown in
Fig.~\ref{fig:ecChain}.

\phead{Universal computation.}  To complete the $\minCSS$ gate set so
that we can implement arbitrary quantum computations, it suffices to
add HAD and preparation of the state
$\ket{\pi/8}=\cos(\pi/8)\ket{0}+\sin(\pi/8)\ket{1}$~\cite{knill:qc1997a,knill:qc2004a}.
We treat the qubits in a logical qubit pair identically and ignore one
of them for the purpose of computation. See
S.I. Sect.~\ref{sect:universal} for how to take full advantage of both
qubits.  The logical HAD is implemented similarly to the logical CNOT
and uses one error-correcting teleportation.  Its logical errors are
are less than those of the logical CNOT.  To prepare logical
$\ket{\pi/8}$ in both qubits of a logical qubit pair, we obtain a
logical Bell pair, decode the first block of the Bell pair into two
physical qubits and make measurements to project the physical qubits'
states onto $\ket{\pi/8}$ or the orthogonal state. If an orthogonal
state is obtained, we adjust the Pauli frame by $Y$'s accordingly.
Because of the entanglement between the physical qubits and the
logical ones, this prepares the desired logical state, albeit with
error. This procedure is called ``state injection''.  To decode the
first block of the Bell pair, we first decode the $C_4$ subblocks and
continue by decoding six-qubit subblocks of $C_6$.  Syndrome
information is obtained in each step and can be used for error
detection or correction.  The error in decoding is expected to be
dominated by the last decoding steps. Consequently, the error in the
injected state should be bounded as the number of levels increase,
which we verified by simulation to the extent possible.  To remove
errors from the injected states, logical purification can be
used~\cite{bravyi:qc2004a,knill:qc2004a} and is effective if
the error of the injected state is less than
$0.141$~\cite{bravyi:qc2004a}. The purification method can be
implemented fault tolerantly to ensure that the purified logical
$\ket{\pi/8}$ states have errors similar to those of logical CNOTs
(S.I. Sect.~\ref{sect:resources}).  To simplify the implementation of
quantum computations, other states can be prepared similarly.

Consider the threshold for postselected universal quantum computing.
The logical HAD and injection errors at $\gamma=0.03$ and level $2$
are shown in Fig.~\ref{fig:condEPG}.  The injection error
is well below the maximum allowed and is not expected to increase
substantially for higher levels.  The injection error
should scale approximately linearly with EPG, so the extrapolated
threshold of $\gamma \geq 0.06$ may apply to universal postselected
quantum computing.

The injection and purification method for preparing states needed to
complete the gate set works with the error-correcting $C_4/C_6$
architecture. Consider state injection at $\gamma=0.01$.  The context
for injection is state preparation, which determines the combination
of error-correction and detection as discussed above.  The conditional
logical error after state injection was determined to be
$\expdata{8.6\aerrb{0.6}{0.5}\times
10^{-3}}{cliff/finaldata/injeec_l3_be.tex[0.01,:]}$ at level 3 and
$\expdata{1.1\aerrb{0.1}{0.1}\times
10^{-2}}{cliff/finaldata/injeec_l4_be.tex[0.01,:]}$ at level 4,
comparable to $\gamma$ and sufficiently low for $\ket{\pi/8}$
purification. As a result, the $C_4/C_6$ architecture enables scalable
quantum computing at EPGs above $0.01$.  To obtain higher thresholds,
we use fault-tolerant postselected computing to prepare states
in a code that can handle higher EPGs than $C_4/C_6$ concatenated
codes can.  The states are chosen so that we can implement a universal
set of gates by error-correcting teleportation.  Suppose that
arbitrarily low logical EPGs are achievable with the $C_4/C_6$
architecture for universal postselected computing. To compute
scalably, we choose a sufficiently high level $l$ for the $C_4/C_6$
architecture and a very good error-correcting quantum code $C_e$.  The
first step is to prepare the desired $C_e$-encoded states using level
$l$ encoded qubits, in essence concatenating $C_e$ with level $l$ of
the $C_4/C_6$ architecture. The second step is to decode each block of
the $C_4/C_6$ architecture to physical qubits to obtain unconcatenated
$C_e$-logical states.  Once these states are successfully prepared,
they can be used to implement each logical gate by error-correcting
teleportation.  Simulations show that the postselected decoding
introduces an error $\lesssim\gamma$ for each decoded qubit
(Fig.~\ref{fig:condEPG}).  There is no postselection in
error-correcting teleportation with $C_e$, and it is sensitive to
decoding error in two blocks ($\approx 2\gamma$) as well as the error
of the CNOT ($\approx\gamma$) and the two physical measurements
($\approx 8\gamma/15$) required for the Bell measurement. Hence, the
effective error per qubit that needs to be corrected is $\approx
3.53\gamma$.  The maximum error probability per qubit correctable by
known codes $C_e$ is $\approx
0.19$~[\citeonline{divincenzo:qc1998c}]. Provided that $3.53\gamma
\lesssim 0.19$ and $\gamma$ is below the postselected threshold for
the $C_4/C_6$ architecture, the error in the state preparation before
decoding together with the logical error in error-correcting
teleportation can be made smaller than $10^{-3}$
(S.I. Sect.~\ref{sect:resources}). The $C_e$ architecture can
therefore be concatenated with the error-correcting $C_4/C_6$
architecture to arbitrarily reduce the logical EPG.  In view of the
postselected threshold indicated by Fig.~\ref{fig:condEPG}, scalable
quantum computing is possible at $\gamma = 0.03$ and perhaps up to
$\gamma \approx 0.05$.  Although the postselection overheads are
extreme, this method is theoretically efficient.

\phead{Resources} The resource requirements for the error-correcting
$C_4/C_6$ architecture can be mapped out as a function of $\gamma$ for
different sizes of computations. Since we do not have analytical
expressions for the resources for logical Bell state preparation or
for the logical errors as a function of $\gamma$ and, with our current
capabilities, we are not able to determine them in enough detail by
simulation, we use naive models to approximate the needed
expressions. The resources required are related to the number of
physical CNOTs used, which dominates the number of state state
preparations and measurements. HADs are used only for universality at
the logical level. The number of physical CNOTs used in a logical Bell
state preparation is modeled by functions of the form
$C/(1-\gamma)^k$, which would be correct on average if the
state-preparation network had $C$ gates of which $k$ failed
independently with probability $\gamma$, and the network were repeatedly
applied until none of the $k$ gates fail. $C$ and $k$ depend on the
level of concatenation.  The logical error probabilities are modeled
at level $l\geq 1$ by $p_d(l) = d(l)\gamma^{f(l+1)}$ (detected error)
and $p_c(l) = c(l)\gamma^{f(l+2)}$ (conditional logical error), where
$f(0)=0,f(1) = 1, f(l+1) = f(l)+f(l-1)$ is the Fibonacci
sequence. These expressions are asymptotically correct as
$\gamma\rightarrow 0$. We verified that they model the desired values
well and determined the constants at the lower levels by simulation
(S.I. Sect.~\ref{sect:resources}). At high levels, the constants were
estimated by extrapolating their level-dependent behavior.  Using
these expressions, we determined the level of concatenation that
requires the fewest resources to implement a computation of a given
size.  The resulting resource graph is shown in
Fig.~\ref{fig:resource_graph}.

Since interesting quantum computations use many non-Clifford gates, it
is necessary to estimate the average resources required for preparing
states such as the $\ket{\pi/8}$ state.  One instance of this state
suffices for implementing a $45\dg$ $Y$ rotation. Two are required for
a phase-variant of a Toffoli gate.  Consider $\gamma=0.01$.  At level
$4$ of the $C_4/C_6$ architecture, one purification stage requires
$\approx\expdata{370}{$\ket{\pi/8}$ state preparation overhead
determined in S.I.}$ logical CNOTs (S.I. Sect.~\ref{sect:resources}).
It is likely that this overhead can be significantly improved, but it
must be accounted for when using the graphs of
Fig.~\ref{fig:resource_graph} as discussed in the caption.  It is
possible to implement a computation with $100$ logical qubits and up
to $1000$ $\ket{\pi/8}$-preparations using $\expdata{1.23\times
10^{14}}{see S.I.}$ physical CNOTs
(S.I. Sect.~\ref{sect:resources}). This takes into account the
probability that the computation fails with a detected error.  The
conditional probability of obtaining an incorrect output is
$\approx\expdata{0.02}{see S.I.}$. Such a computation is non-trivial
in the sense that its output is not efficiently predictable
using known classical algorithms. The resource requirements are large
but would be reasonable in the context of classical computing: Central
processing units have $10^8$ or more transistors operating at rates
faster than $10^9$ bit operations per second~\cite{intel.com:qc2004a}.

We compare our resource requirements to those of Steane's architecture
based on an example at $\gamma=10^{-4}$ detailed
in~[\citeonline{steane:qc2002a}].  Steane's architecture is based on
non-concatenated block codes, which are expected to be more efficient
at such low EPGs~\cite{preskill:qc1998a}.  Steane's example has an
effective logical error per qubit of $\approx 7\times 10^{-12}$ using
$\approx 420$ physical CNOTs per qubit per gate.  Our architecture
achieves detected errors of $\expdata{5.5\times 10^{-9}}{see S.I.}$
(level $3$) or $\expdata{6\times 10^{-14}}{see S.I.}$ (level $4$)
using respectively $\approx\expdata{2100}{see S.I.}$ or
$\expdata{2.6\times 10^4}{see S.I.}$ physical CNOTs per qubit
(S.I. Sect.~\ref{sect:resources}). The conditional logical errors are
much smaller.  The $C_4/C_6$ architecture's resource requirements are
still within two orders of magnitude of Steane's at $\gamma=10^{-4}$.
The $C_4/C_6$ architecture has the advantage of simplicity,
of yielding more reliable answers conditional on having no detected
errors, and of operating at higher EPGs.

\phead{Discussion.}  
How high must EPGs be so that it is not possible to scalably quantum
compute?  It is known that if unbiased one-qubit EPGs exceed $.5$,
then we can simulate the effect of gates
classically~\cite{bennett:qc1999c,harrow:qc2003a}.  Furthermore, if
one-qubit EPGs exceed $0.25$, then we cannot realize a quantum
computation ``faithfully'', that is by encoding the computation's
qubits with quantum codes~\cite{bennett:qc1996a,bruss:qc1998a}.  This
is because the quantum channel capacity vanishes at a depolarizing
error probability above $0.25$.  Faithful techniques are likely to
require at least three sequential gates before an error can be
eliminated (in our case these are preparation gates whose errors
remain in the logical Bell pairs, a CNOT and a measurement for
teleportation). Thus one would not expect to obtain thresholds above
$\sim 0.09$ using faithful methods. This is not far from the
extrapolated $0.05$ evidenced by our work.  Note that the thresholds
obtained here are similar to those for quantum
communication~\cite{briegel:qc1999a}.

An important use of studies of fault-tolerant architectures is to
provide guidelines for EPGs that should be achieved to meet the
low-error criterion for scalability.  Such guidelines should depend on
the details of the relevant error models and constraints on
two-qubit gates. Nevertheless, the value of $\gamma=10^{-4}$ has often
been cited as the EPG to be achieved.  With architectures such as
Steane's~\cite{steane:qc2002a,reichardt:qc2004a} and the one
introduced here, resource requirements at $\gamma=10^{-3}$ are now
comparable to what they were for
$\gamma=10^{-4}$~[\citeonline{steane:qc1997a}] at the time this value
was starting to be cited.

\ignore{ E.g. Steane~\cite{steane:qc1997a} has overheads of order
$10^5$ for error rates below $3*10^{-5}$.  More specifically:
$O(10^9)$ Toffoli gates, Steane took that as an algorithm requiring
$2{\times}10^{12}$ cnots ($KQ = 4 10^{12}$ on our terms.)  Preskills
three-fold concatenated $[[7,1,3]]$ code at $\gamma=10^{-6}$ would
require $10^17$ gates (multiply by two if these are mostly cnots, as I
think they are).  The highest error-tolerating code seems in this
paper appears to be a $[[87,1,15]]$ code which at
$\gamma=1.9{\times}10^{-5}$ requires $3.9\times 10^16$ gates. This
gives overheads at these error rates of perhaps a few times $10^{4}$
on our terms.  The question is, what would they be at
$\gamma=10^{-4}$?  The $[[7,1,3]]$ architecture would require at least
one more level of concatenation, one would guess another order of
magnitude or two in overhead (assuming it was even doable at the
time).  Ok, I should be able to get the behavior from Steane's
equations (9) and (11). I think he must have used a threshold of
$10^{-4}$ for the $[[7,1,3]]$ architecture, though I don't see it
stated explicitly. Taking the more optimistic value of $10^{-3}$ in
Eqs. (10) and (11), I get: $N/K \approx 26$, $L\approx 2$, $T/Q\approx
7*480*2150/\eta$, where $\eta$ is the ``average number of
computational steps per correction of the whole computer'', which he
``safely'' chooses as $\eta=w$.  This is an optimization, do more
computational steps (i.e.  logical gates) before correcting. $w$ is
the average weight in a row of the classical parity check matrix for a
CSS code.  $w$ is said to be $d+1$, where $d$ is the minimum distance.
So $9$ in my example. This makes $T/Q\approx 8*10^{5}$, and this is
optimistic, assuming a speculative threshold for this architecture
that may not have applied to the known ways of implementing it at the
time. To look at the other code, he writes $m=(n-1)/2$, $n$ is the
length of the code, $m$ occurs in the formulas.  I am looking at
eq. (9) for the $[[87,1,15]]$ code.  $T/Q \approx
(86*16+435)*2*r*2150/16 \approx 3.9*10^6$, where is the dependence on
$\gamma$? He assumes that on average, $r=(d+1)/2$, it is the number of
syndromes needed.  The dependence on $\gamma$ only shows up in the
logical error probability, which presumably is not good enough at a
$5$ times higher $\gamma$ than given in the tables (probably decreases
by at least $5^7$). It looks like somewhere between $10^{-7}$ and
$10^{-8}$ from Fig. 4 (per correction, per block).  For comparison, my
resources at $\gamma=10^{-3}$ are $5.3\times 10^{5}$ using level $5$
for $KQ = 10^10$ and $3\times 10^6$ with level $5$ at $KQ = 10^12$.  }

Several open problems arise from the work presented here. Can
the high thresholds evidenced by our simulations be mathematically
proven? Are thresholds for postselected computing strictly
higher than thresholds for scalable standard quantum computing?
Recent work by Reichardt~\cite{reichardt:qc2004a} shows that Steane's
architecture can be made more efficient by the judicious use of error
detection, improving Steane's threshold estimates to around $10^{-2}$.
How do the available fault-tolerant architectures compare for EPGs
between $10^{-3}$ and $10^{-2}$?  It would be helpful to significantly
improve the resource requirements of fault-tolerant architectures,
particularly at high EPGs.

% up to 50 references
\bibliographystyle{nature}
\bibliography{journalDefs,qc}

This work is a contribution of NIST, an agency of the
U.S. government, and is not subject to U.S. copyright.  Correspondence
and requests for materials should be sent to E. Knill
(knill@boulder.nist.gov).

\pagebreak

% figures

\begin{herefig}
\label{fig:c4/c6_struct}
\begin{picture}(-1.4,3.5)(0,-3.3)
\nputgr{0,0}{t}{scale=.5}{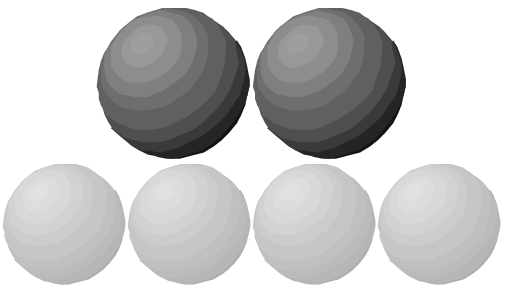}
\nputbox{0,-1}{b}{$\overbrace{\rule{4.2in}{0pt}}$}
\nputgr{0,-1}{t}{scale=.5}{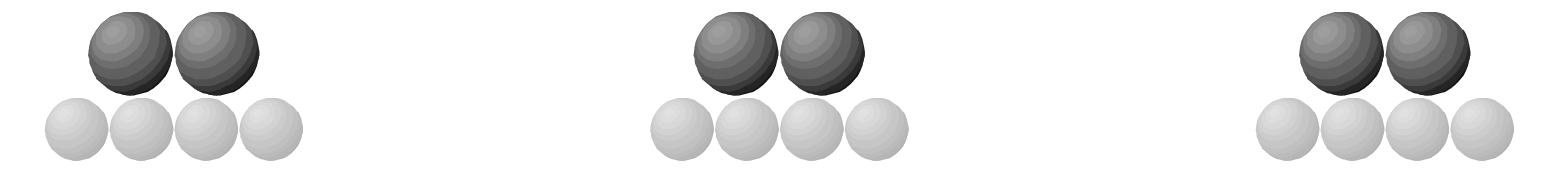}
\nputbox{-1.75,-1.8}{b}{$\overbrace{\rule{1.5in}{0pt}}$}
\nputbox{0,-1.8}{b}{$\overbrace{\rule{1.5in}{0pt}}$}
\nputbox{1.75,-1.8}{b}{$\overbrace{\rule{1.5in}{0pt}}$}
\nputgr{0,-1.8}{t}{scale=.5}{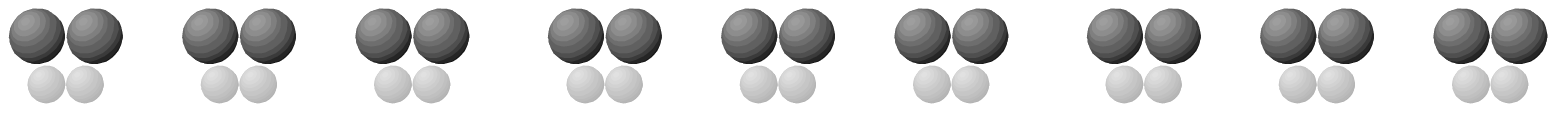}

\nputbox{-2.3,-2.53}{b}{$\overbrace{\rule{.5in}{0pt}}$}
\nputbox{-1.725,-2.53}{b}{$\overbrace{\rule{.5in}{0pt}}$}
\nputbox{-1.15,-2.53}{b}{$\overbrace{\rule{.5in}{0pt}}$}
\nputbox{-.575,-2.53}{b}{$\overbrace{\rule{.5in}{0pt}}$}
\nputbox{0,-2.53}{b}{$\overbrace{\rule{.5in}{0pt}}$}
\nputbox{.575,-2.53}{b}{$\overbrace{\rule{.5in}{0pt}}$}
\nputbox{1.15,-2.53}{b}{$\overbrace{\rule{.5in}{0pt}}$}
\nputbox{1.725,-2.53}{b}{$\overbrace{\rule{.5in}{0pt}}$}
\nputbox{2.3,-2.53}{b}{$\overbrace{\rule{.5in}{0pt}}$}
\nputgr{0,-2.5}{t}{scale=.5}{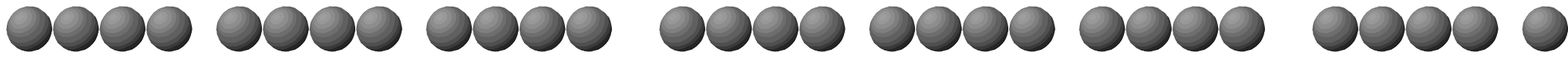}

\nputbox{-4,-2.635}{l}{  $\left.\begin{array}{@{}l@{}}
\textsf{Physical qubits:}
    \end{array}\right\}$}
\nputbox{-2.58,-2.635}{rb}{\rule{.14in}{1pt}}

\nputbox{-4,-2.2}{l}{   $\left.\begin{array}{@{}l@{}}
\textsf{\small Level $1$} \\[-3pt] \textsf{\small syndrome bits:}
    \end{array}\right\}$}
\nputbox{-2.5,-2.1}{rt}{\rotatebox{12}{\rule{.33in}{1pt}}}
\nputbox{-4,-1.8}{l}{
   $\left.\begin{array}{@{}l@{}}
\textsf{\small Level $1$ encoded} \\[-3pt] \textsf{\small qubit pairs:}
    \end{array}\right\}$}
\nputbox{-2.53,-1.89}{rb}{\rotatebox{-25}{\rule{.2in}{1pt}}}

\nputbox{-4,-1.3}{l}{   $\left.\begin{array}{@{}l@{}}
\textsf{\small Level $2$} \\[-3pt] \textsf{\small syndrome bits:}
    \end{array}\right\}$}
\nputbox{-2.18,-1.35}{rb}{\rotatebox{-4}{\rule{.65in}{1pt}}}
\nputbox{-4,-.9}{l}{
   $\left.\begin{array}{@{}l@{}}
\textsf{\small Level $2$ encoded} \\[-3pt] \textsf{\small qubit pairs:}
    \end{array}\right\}$}
\nputbox{-2.0,-1.15}{rb}{\rotatebox{-19}{\rule{.76in}{1pt}}}

\nputbox{-2,-.6}{l}{   $\left.\begin{array}{@{}l@{}}
\textsf{\small Level $3$} \\[-3pt] \textsf{\small syndrome bits:}
    \end{array}\right\}$}
\nputbox{-.5,-.52}{rt}{\rotatebox{8}{\rule{.32in}{1pt}}}
\nputbox{-2,-.2}{l}{
   $\left.\begin{array}{@{}l@{}}
\textsf{\small Level $3$ encoded} \\[-3pt] \textsf{\small qubit pairs:}
    \end{array}\right\}$}
\nputbox{-.35,-.22}{rb}{\rotatebox{-5}{\rule{.34in}{1pt}}}

\end{picture}
\ifnature{\pagebreak}{} \herefigcap{Block structure of $C_4/C_6$
concatenated codes.  The bottom line shows $9$ blocks of four physical
qubits. Each block encodes a level $1$ qubit pair with $C_4$.  The
encoded qubit pairs are shown in the line above.  Formally, each such
pair is associated with two syndrome bits, shown below the encoded
pair in a lighter shade, which are accessible by syndrome measurements
or decoding for the purpose of error detection and correction. The
next level groups three level $1$ qubit pairs into a block, encoding a
level $2$ qubit pair with $C_6$ that is associated with $4$ syndrome
bits.  A level $2$ block consists of a total of $12$ physical qubits.
Three level $2$ qubit pairs are used to form a level $3$ qubit pair,
again with $C_6$ and associated with $4$ syndrome bits.  The total
number of physical qubits in a level $3$ block is $36$.  }
\end{herefig}

\pagebreak

\begin{herefig}
\label{fig:condEPG}
\begin{picture}(0,3.7)(0,-3.5)
\nputbox{.17,-1.1}{tl}{\color{grey}\rule{2pt}{.25in}}

\nputgr{0,0}{t}{height=3in}{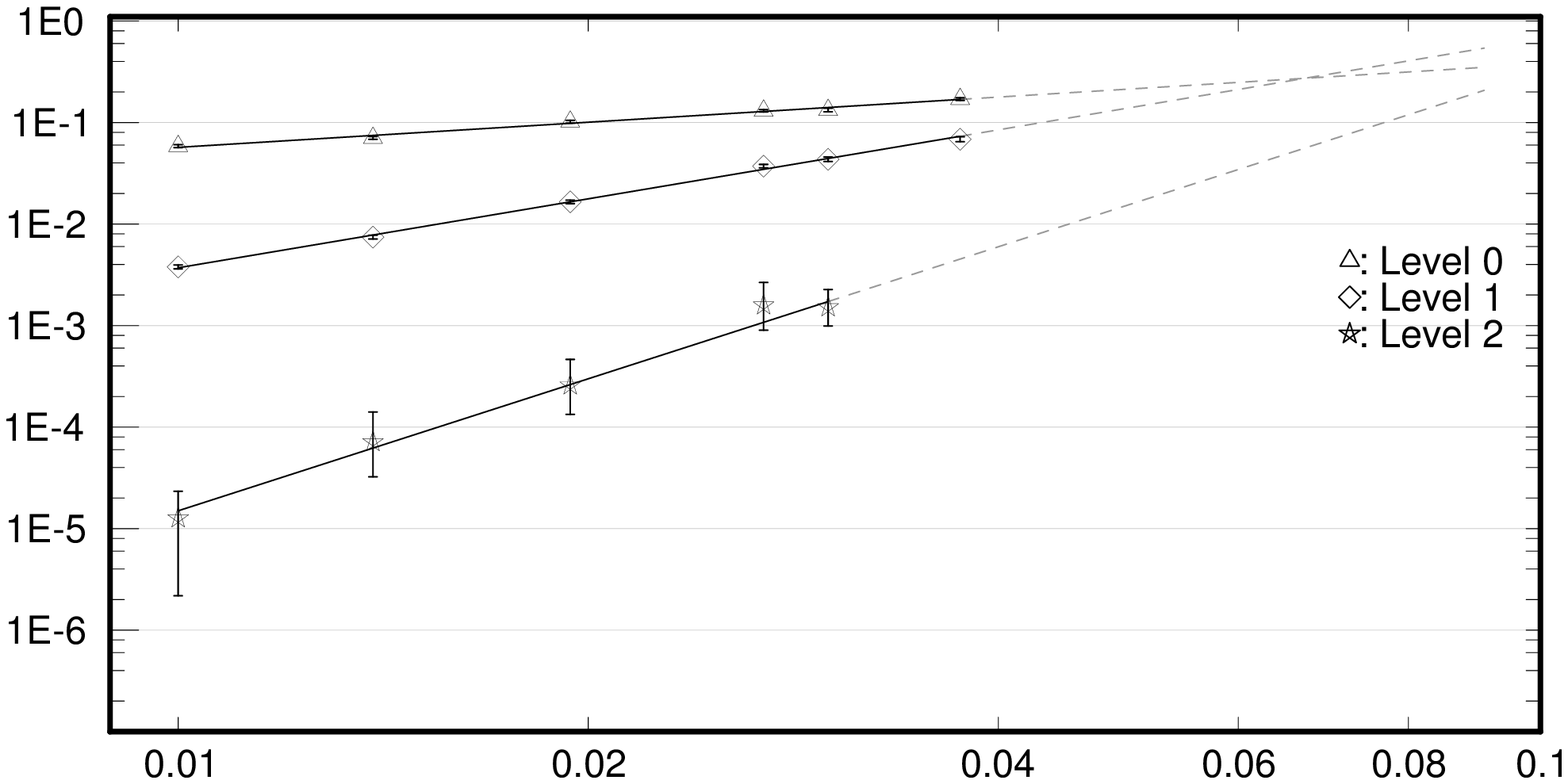}
\nputbox{-3,0}{tr}{\rotatebox{90}{\textsf{\small Logical CNOT conditional error probability}}}
\nputbox{3,-3.0}{tr}{\textsf{\small Physical CNOT error probability $\gamma$.}}

\nputbox{-.47,-1.35}{tl}{\setlength{\fboxrule}{2pt}\setlength{\fboxsep}{2pt}%
    \fcolorbox{grey}{white}{%
      \textsf{\small
      \begin{tabular}{lll@{}}
        \multicolumn{3}{@{}l@{}}{Level $2$ errors at $\gamma=0.03$.}\\[-3pt]
        &CNOT : & $\expdata{1.5\aerrb{0.7}{0.5}\times 10^{-3}}{cliff/finaldata/2teledet_l2_be.dat[0.03,:]} $\\[-2pt]
        &preparation : & $\expdata{2.1\aerrb{0.3}{0.2}\times 10^{-4}}{cliff/finaldata/prepedet_l2_be.dat[0.03,:]} $\\[-2pt]
        &measurement : & $\expdata{4.1\aerrb{1.4}{1.1}\times 10^{-5}}{cliff/finaldata/measedet_l2_be.dat[0.03,:]} $\\[-2pt]
        &HAD : & $\expdata{1.8\aerrb{0.4}{0.3}\times 10^{-3}}{cliff/finaldata/1teledet_l2_be.dat[0.03,:]}$\\[-2pt]
        &decoding : & $\expdata{2.2\aerrb{0.2}{0.2}\times 10^{-2}}{cliff/finaldata/decedet_l2_be.dat[0.03,:]}$\\[-2pt]
        &injection : & $\expdata{3.1\aerrb{0.2}{0.2}\times 10^{-2}}{cliff/finaldata/injedet_l2_be.dat[0.03,:]}$\\[-2pt]
      \end{tabular}
      }
    }%
  }
\end{picture}
\ifnature{\pagebreak}{} \herefigcap{Conditional logical errors with
postselection.  The plot shows logical CNOT errors conditional on not
detecting any errors as a function of EPG parameter $\gamma$ at levels
$0$, $1$ and $2$. The logical CNOT is implemented with transversal
physical CNOTs and two error-detecting teleportations, where the
output state is accepted only if no errors are detected in the
teleportations. The data show the incremental error attributable to
the logical CNOT in the context of a longer computation, as explained
in S.I. Sect.~\ref{sect:simulation}.  The error bars are $68\,\%$
confidence intervals.  The solid lines are obtained by least-squares
interpolation followed by gradient-descent likelihood maximization.
Extrapolations are shown with dashed lines and suggest that logical
EPG improvements with increasing levels are possible above
$\expdata{\gamma=0.06}{Determined by conservative visual inspection of
the graph.}$.  The error in the slope
($\expdata{4.32}{cliff/finaldata/2teledet_l2_be.ipol[1,2]}$) of the
level $2$ line is estimated as
$\expdata{0.52}{sqrt(cliff/finaldata/2teledet_l2_be.ipol[1,4])}$ by
resampling. The smallest number of undetectable errors at level $2$ is
$4$, which should be the slope as $\gamma$ goes to $0$. At high
$\gamma$, the curves are expected to level
off~\cite{steane:qc2002a}. Other operations' errors for $\gamma=0.03$
and level $2$ are shown in the inset table. Ratios between the
preparation or measurement and CNOT errors are smaller than those
assumed for the physical error model. The logical HAD error is
expected to be between $0.5$ and $0.8$ of the logical CNOT error,
which could not be confirmed because of the large error bars.  The
decoding error is the incremental error introduced by decoding a block
into two physical qubits.  The injection error is the error in a
logical state that we prepare by decoding one block of a logical Bell
pair and measuring the decoded qubits.  The measurement
error per qubit is assumed to be the same as that of $X$- and
$Z$-measurements. Decoding and injection errors were found to decrease
from level 1 (decoding error $\expdata{4.4\aerrb{0.4}{0.4}\times
10^{-2}}{cliff/finaldata/decedet_l1_be.tex[0.03,:]}$, injection error
$\expdata{5.5\aerrb{0.5}{0.4}\times
10^{-2}}{cliff/finaldata/injedet_l1_be.tex[0.03,:]}$ ) to level 2.}
\end{herefig}

\pagebreak

\begin{herefig}
\label{fig:ecEPG}\label{fig:decEPG}
\begin{picture}(0,4.2)(0,-4)
\nputgr{-1.8,0}{t}{height=3in}{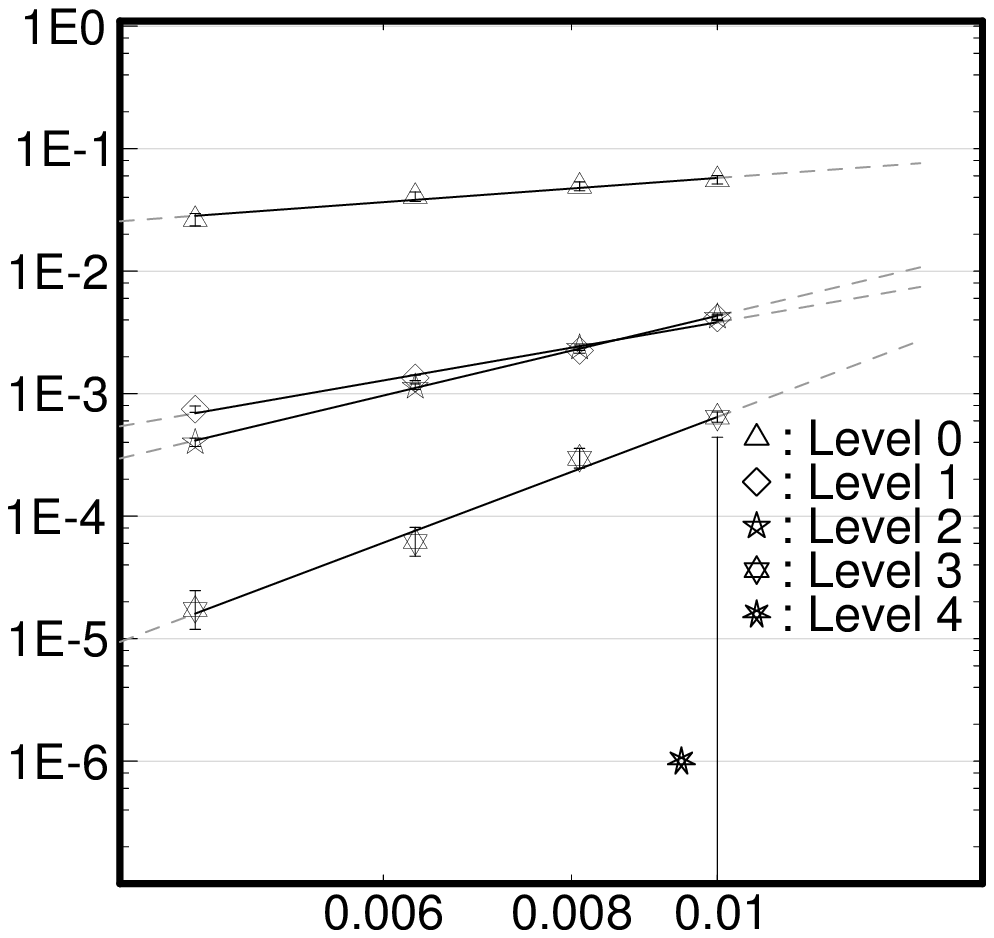}
\nputbox{-3.35,0}{tr}{\rotatebox{90}{\textsf{\small Logical CNOT conditional error probability}}}
\nputbox{-2.8,-.2}{c}{\Large\textbf{a.}}
\nputbox{-1.22,-2.5}{tl}{\rotatebox{9}{\scalebox{1.8}{$\downarrow$}{}}}

\nputbox{.25,0}{tr}{\rotatebox{90}{\textsf{\small Logical CNOT detected error probability}}}
\nputgr{1.8,0}{t}{height=3in}{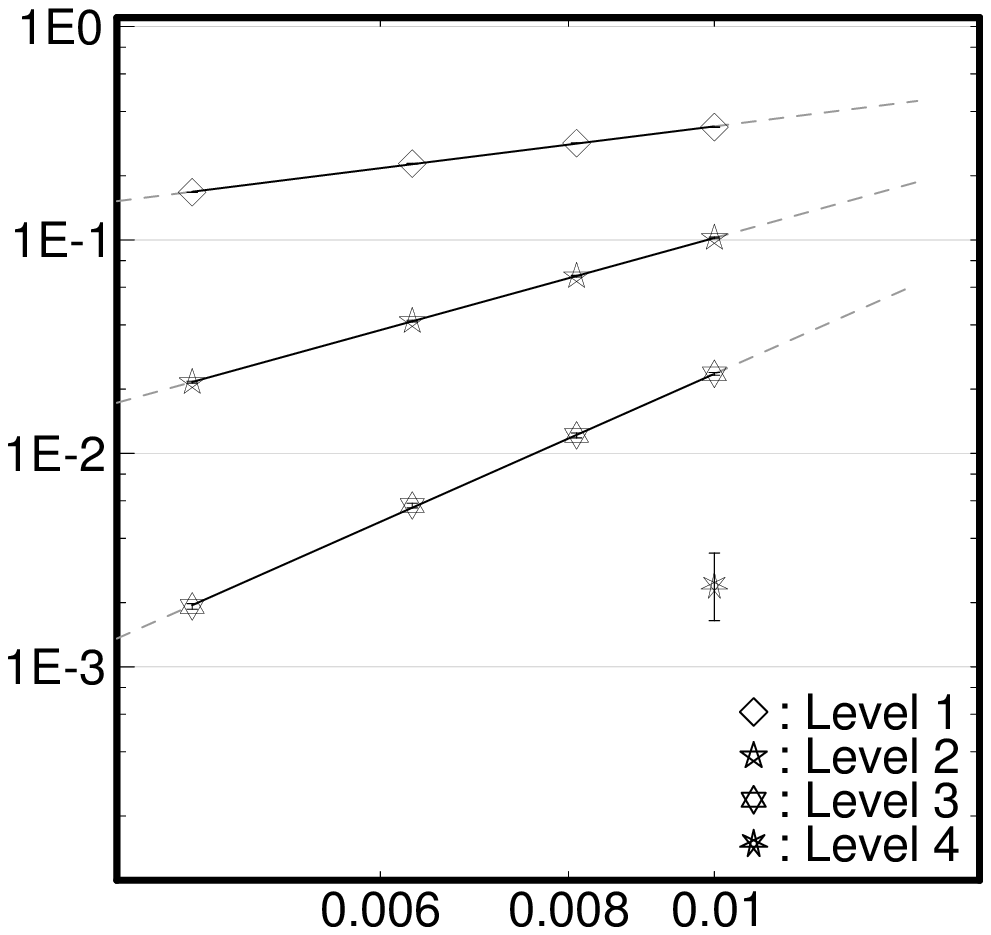}
\nputbox{3.4,-3.0}{tr}{\textsf{\small Physical CNOT error probability $\gamma$.}}
\nputbox{.8,-.2}{c}{\Large\textbf{b.}}
\end{picture}
\ifnature{\pagebreak}{} \herefigcap{Conditional and detected logical
errors with error correction.  The plot shows incremental detected and
conditional logical errors for a logical CNOT as a function of EPG
parameter $\gamma$ up to level $4$. Error bars and lines are as
described in the caption of Fig.~\ref{fig:condEPG}. 
The data is obtained as described in
S.I. Sect.~\ref{sect:simulation}. The combination of error correction
and detection is as required for the error-correcting
$C_4/C_6$ architecture.  Plot \textbf{a.} shows the logical CNOT's
error conditional on not detecting an uncorrectable error.  Plot
\textbf{b.}  shows the probability of detecting an uncorrectable
error.  At $\gamma=0.01$, the detected errors are are
$\expdata{2.4\aerrb{0.0}{0.0}\times
10^{-2}}{cliff/finaldata/2teleec_l3_de.tex[0.01,:]}$ (level 3) and
$\expdata{2.4\aerrb{1.0}{0.7}\times
10^{-3}}{cliff/finaldata/2teleec_l4_de.tex[0.01,:]}$ (level 4).  The
conditional errors are $\expdata{6.4\aerrb{0.6}{0.6}\times
10^{-4}}{cliff/finaldata/2teleec_l3_be.tex[0.01,:]}$ (level 3) and
$\expdata{0.0\aerrb{4.4}{0.0}\times
10^{-4}}{cliff/finaldata/2teleec_l4_be.tex[0.01,:]}$ (level 4).  For
comparison, the preparation errors at levels 3,4 were found to be
$\expdata{2.1\aerrb{0.3}{0.3}\times10^{-4}}{cliff/finaldata/prepec_l3_de.tex[0.01,:]}$,
$\expdata{0.0\aerrb{1.0}{0.0}\times10^{-4}}{cliff/finaldata/prepec_l4_de.tex[0.01,:]}$
(detected error) and
$\expdata{3.3\aerrb{7.5}{2.7}\times10^{-6}}{cliff/finaldata/prepec_l3_be.tex[0.01,:]}$,
$\expdata{0.0\aerrb{1.0}{0.0}\times10^{-4}}{cliff/finaldata/prepec_l4_be.tex[0.01,:]}$
(conditional error).  The measurement errors are
$\expdata{4.7\aerrb{0.4}{0.4}\times10^{-4}}{cliff/finaldata/measec_l3_de.tex[0.01,:]}$,
$\expdata{5.6\aerrb{12.8}{4.6}\times10^{-5}}{cliff/finaldata/measec_l4_de.tex[0.01,:]}$
(detected error) and
$\expdata{3.3\aerrb{7.4}{2.7}\times10^{-6}}{cliff/finaldata/measec_l3_be.tex[0.01,:]}$,
$\expdata{0.0\aerrb{1.0}{0.0}\times10^{-4}}{cliff/finaldata/measec_l4_be.tex[0.01,:]}$
(conditional error).  Finally, the HAD errors at level 3 are
$\expdata{1.3\aerrb{0.0}{0.0}\times
10^{-2}}{cliff/finaldata/1teleec_l3_de.tex[0.01,:]}$ (detected error)
and $\expdata{3.5\aerrb{0.6}{0.5}\times
10^{-4}}{cliff/finaldata/1teleec_l3_be.tex[0.01,:]}$ (conditional
error).  }
\end{herefig}

\pagebreak

\begin{herefig}
\label{fig:ecChain}
\begin{picture}(0,4.2)(0,-4)
\nputgr{-1.8,0}{t}{height=3in}{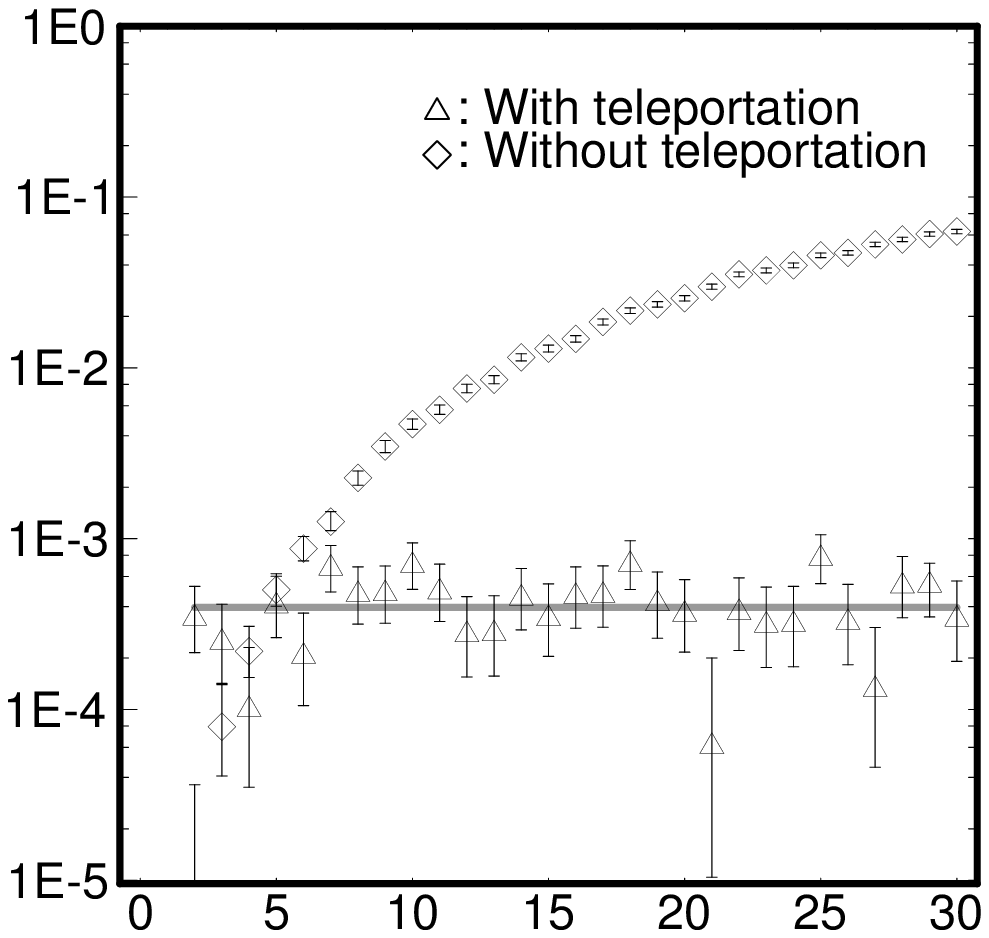}
\nputbox{-3.35,0}{tr}{\rotatebox{90}{\textsf{\small Logical conditional error probability}}}

\nputbox{.25,0}{tr}{\rotatebox{90}{\textsf{\small Logical detected error probability}}}
\nputbox{-2.8,-.2}{c}{\Large\textbf{a.}}

\nputgr{1.8,0}{t}{height=3in}{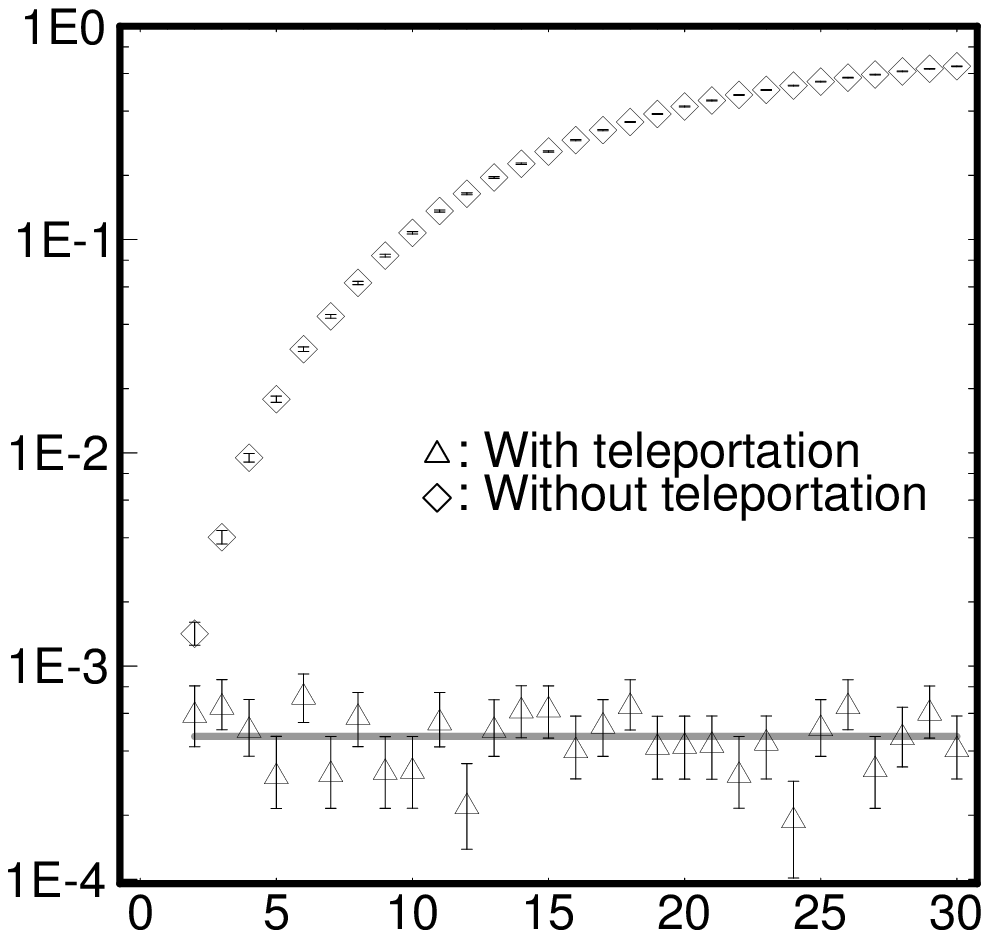}
\nputbox{3.4,-3.0}{tr}{\textsf{\small Number of steps.}}
\nputbox{.8,-.2}{c}{\Large\textbf{b.}}
\end{picture}
\ifnature{\pagebreak}{} \herefigcap{Error-compounding behavior with
and without error-correcting teleportation.  Incremental conditional
(\textbf{a.}) and detected (\textbf{b.})  error probabilities are
shown for each step of a sequence of 30 steps of applying the
one-qubit error associated with HAD to each physical qubit and
teleporting or not teleporting the logical qubit pair's block.  Error
bars are $68\,\%$ confidence intervals.  Level $3$ of the
error-correcting architecture is used.  The first step is omitted
since it is biased by the error-free reference-state preparation as
discussed in S.I. Sect.~\ref{sect:simulation}. The horizontal gray
lines show the average incremental error if teleportation is
used. Note that for the first four steps, the incremental conditional
error is smaller if no teleportation is used.  This may be
exploited when optimizing networks, provided one takes account of the
resulting spreading of otherwise localized error
events~\cite{zalka:qc1996a,steane:qc2002a}. }
\end{herefig}

\pagebreak

\begin{herefig}
\label{fig:resource_graph}
\begin{picture}(0,3.7)(0,-3.5)

\nputgr{0,0}{t}{height=3in}{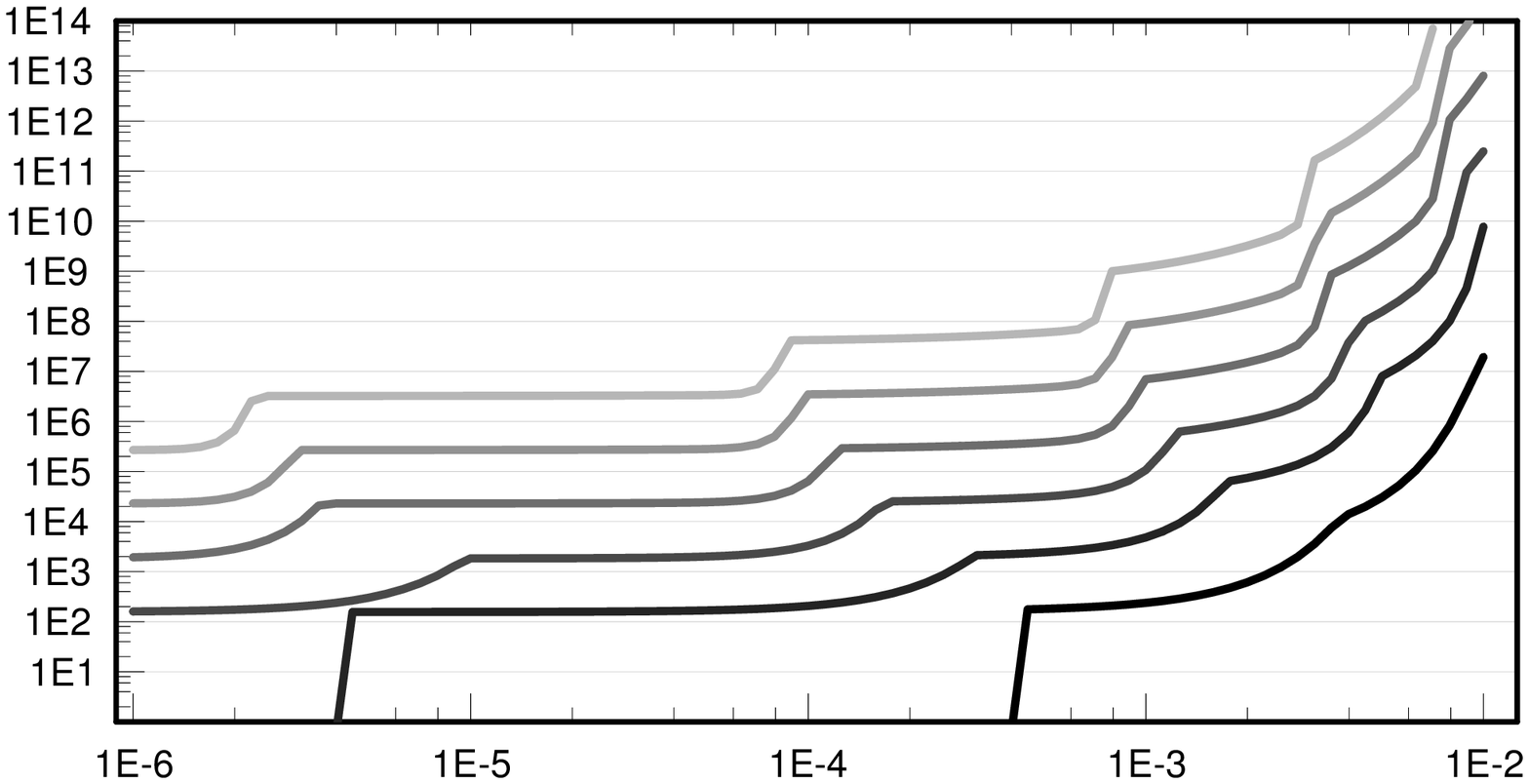}
\expdata{}{Output generated by matlab code in cliff/finaldata, see cliff/finaldata/catalogue.txt. Requires file prepared in octave code below}
\nputbox{-3,0}{tr}{\rotatebox{90}{\textsf{\small
  Resources per qubit and gate}}}
\nputbox{3,-3.0}{tr}{\textsf{\small Physical CNOT error probability $\gamma$.}}
\nputbox{-2.2,-1.6}{c}{\textsf{\small $10^{34}$}}
\nputbox{-2,-1.8}{c}{\textsf{\small $10^{21}$}}
\nputbox{-1.9,-1.98}{c}{\textsf{\small $10^{13}$}}
\nputbox{-1.4,-2.14}{c}{\textsf{\small $10^{8}$}}
\nputbox{-1.73,-2.45}{c}{\textsf{\small $10^{5}$}}
\nputbox{.86,-2.4}{c}{\textsf{\small $10^{3}$}}

% Mark transition to level 4.
% See cliff/finaldata/optcnts_opt_k*.dat
\nputbox{2.82,-1.3}{c}{\circle{.17}}
\nputbox{2.73,-1.3}{c}{\textsf{\small 4}}
\nputbox{1.87,-1.75}{c}{\circle{.17}}
\nputbox{1.78,-1.75}{c}{\textsf{\small 4}}
\nputbox{.54,-1.85}{c}{\circle{.17}}
\nputbox{.45,-1.85}{c}{\textsf{\small 4}}
\nputbox{-1.61,-1.85}{c}{\circle{.17}}
\nputbox{-1.7,-1.85}{c}{\textsf{\small 4}}
\nputbox{-1.69,-1.63}{c}{\circle{.17}}
\nputbox{-1.78,-1.63}{c}{\textsf{\small 5}}
\nputbox{-1.86,-1.43}{c}{\circle{.17}}
\nputbox{-1.95,-1.43}{c}{\textsf{\small 6}}

\end{picture}
\ifnature{\pagebreak}{} \herefigcap{ Resources per qubit and gate for
different computation sizes.  The size $KQ$ of a computation is the
product of the number of gates (including ``memory'' gates) and the
average number of qubits per gate. Each curve is labeled by the
computation size and shows the number $\textsl{pcnot}$ of physical
CNOTs required per qubit and gate to implement a computation of size
$KQ$ with the $C_4/C_6$ architecture. The curves are based on naive
models of resource usage in state preparation and of the logical
errors (S.I. Sect.~\ref{sect:resources}).  The circled numbers
are at the point above which the indicated or a higher level must be
used. The curves are most reliable for levels $<4$.
To obtain the total computational resources, multiply
$\textsl{pcnot}\times KQ$ by twice the average number of logical CNOTs
needed for implementing a gate of the computation.  It is assumed that
these logical CNOTs are involved in state preparation required for
universality but do not contribute to the error
(S.I. Sect.~\ref{sect:resources}).  The ``scale-up'' (number of
physical qubits per logical qubit) depends on parallelism and level
$l$ of concatenation.  With maximum parallelism, the scale-up is of
the same order as $\textsl{pcnot}$. For a completely sequential
algorithm such as could be used if there is no memory error, this can
be reduced to $3^{l-1}2$. With some memory error and logical gate
parallelism, $\approx (1+2*(l-1))3^{l-1}2$ is more realistic
(S.I. Sect.~\ref{sect:resources}).  The steps in the curve arise from
increasing the number of levels.  The first step is to level $2$, and
each subsequent step increments the level by $1$.  The steps are
smoothed because we can exploit error-detection to avoid using the
next level. Improvements of only one to two orders of magnitude are
obtained by reducing $\gamma$ from $0.001$ to $0.0001$, compared to at
least five orders by reducing $\gamma$ from $0.01$ to $0.001$.}
\end{herefig}

\pagebreak

\appendix

\begin{center}
\huge\textbf{ Supplementary Information}
\end{center}

\section{Explanation of Error-Correcting Teleportation}
\label{sect:tec}\label{sect:teleportation}

For the basic theory of stabilizer codes, see
[\citeonline{nielsen:qc2001a}]. Let $Q$ be the $l\times 2n$ binary
check matrix with entries defining a stabilizer code on $n$ qubits for
encoding $k=n-l$ qubits with good error-detecting or -correcting
properties.  The check matrix is obtained from an independent set of
check operators $P_1, \ldots, P_l$ by placing a binary representation
of $P_k$ into row $k$. The binary representation of $P$ is obtained by
replacing the Pauli operator symbols according to $I\mapsto 00$,
$X\mapsto 10$, $Z\mapsto 01$ and $Y\mapsto 11$.  For example, the
check operator $XIY$ is represented by the row vector $[100011]$.
Commas are omitted in binary vectors, and square brackets are used to
distinguish them from binary strings.  The syndrome of a joint
eigenstate of the check operators is denoted by a binary column
vector, with $0$ ($1$) in the $k$'th position denoting a $1$ ($-1$,
respectively) eigenvalue of $P_k$.  The projection operator onto the
eigenspace with syndrome $\mathbf{x}$ is denoted by
$\Pi(Q,\mathbf{x})$. If a Pauli product $P$ with binary representation
$\mathbf{g}$ is applied to a state with syndrome $\mathbf{x}$, the new
syndrome $\mathbf{x}'$ is given by $\mathbf{x}' =
\mathbf{x}+Q\Sy\mathbf{g}^T$.  Arithmetic with binary vectors and
matrices is modulo 2 and $\Sy$ is the $2n\times 2n$ block-diagonal
matrix with blocks
$\left(\begin{array}{cc}0&1\\1&0\end{array}\right)$.

Consider an $n$ qubit ``input'' block carrying $l$ qubits encoded in
the stabilizer code for $Q$, where the block has been affected by
errors. An effective way of detecting or correcting errors is to
teleport each of the $n$ qubits of the input
block using two blocks of $n$ qubits that form an
``encoded Bell pair''. That is, both blocks have syndrome $\mathbf{0}$
with respect to $Q$ and corresponding qubits encoded in the two blocks
are in the state $(\ket{00}+\ket{11})/\sqrt{2}$.  The state of the two
blocks is defined by the following preparation procedure: Start with
$n$ pairs of qubits in the standard Bell state
$(\ket{00}+\ket{11})/\sqrt{2}$.  The two blocks are formed from the
first and second members of each pair, respectively. Use a
$Q$-syndrome measurement on the $n$ second members of each pair to
project them into one of the joint eigenspaces of $Q$. Finally, apply
identical Pauli matrices to both members of pairs in such a way as to
reset the syndromes to $\mathbf{0}$.  To teleport, apply the usual
protocol to corresponding qubits in the three blocks. In the absence
of errors, this copies the encoded input state to the second block of
the encoded Bell pair.  We show that errors are revealed by parities
of the teleportation measurement outcomes.

The standard quantum teleportation protocol begins with an arbitrary
state $\kets{\psi}{1}$ in qubit $\sysfnt{1}$ and the Bell state
$(\kets{00}{23}+\kets{11}{23})/\sqrt{2}$ in qubits
$\sysfnt{2},\sysfnt{3}$. The global initial state can be viewed as
$\ket{\psi}$ encoded in the stabilizer code generated by $IXX$ and
$IZZ$, whose check matrix has rows $\mathbf{b}_1=[001010]$ and
$\mathbf{b}_2=[000101]$.
Let $\slb{B}{23}$ be the check matrix whose rows are the $\mathbf{b}_i$.
The stabilizer consists of the system-labeled Pauli products
$\one,\slb{\sigma_x}{2}\slb{\sigma_x}{3},\slb{\sigma_y}{2}\slb{\sigma_y}{3}$
and $\slb{\sigma_z}{2}\slb{\sigma_z}{3}$.  To teleport, one makes a
Bell-basis measurement on the first two qubits.  This is equivalent to
making a $\slb{B}{12}$-syndrome measurement, where $\slb{B}{12}$ has
as rows $[101000]=[XXI]$ and $[010100]=[ZZI]$.  This is identical to
$\slb{B}{23}$ with qubits $\sysfnt{2},\sysfnt{3}$ exchanged for qubits
$\sysfnt{1},\sysfnt{2}$. Depending on the syndrome $\mathbf{e}$ that
results from the measurement, one applies correcting Pauli matrices to
qubit $\sysfnt{3}$ to restore $\ket{\psi}$ in qubit $\sysfnt{3}$.

Consider the teleportation of $n$ qubits in a block as described
above. The protocol is such that the $2n$ binary measurement outcomes
linearly (with respect to computation modulo $2$) determine the Pauli
product correction to be applied to the second block of the encoded
Bell pair. Let $\mathbf{g}$ be the binary representation of the Pauli
product correction.  The syndrome of the input block constrains
$\mathbf{g}$ as shown in Fig.~\ref{fig:teleec}.  The principle is as
explained in~[\citeonline{gottesman:qc1999a}] for unitary gates, but
generalized to measurements.  In this case, a stabilizer projection on
the destination qubits before teleportation is equivalent to a
projection after teleportation, where the syndrome associated with the
projection is modified by the correction Pauli product used at the end
of teleportation.  The expression $Q\Sy \mathbf{g}^T$ must match the
syndrome of the input block. Consequently, the syndrome of the input
block can be deduced from $\mathbf{g}$, a function of the
teleportation Bell measurement.  Errors can be detected or corrected
accordingly.

It is necessary to consider the effects of errors in the prepared
encoded Bell pair. Errors on the second block propagate forward and
must be handled by future teleportations.  Because of the Bell
measurement, errors on the first block have an effect equivalent
to the same errors on the input block.  Thus, using the inferred
syndrome for detection or correction of errors deals with errors in both
blocks, as long as their combination is within the capabilities of the
code.

Error correction or detection by teleportation handles leakage errors
in the same way as other errors. If a qubit ``leaked'', the outcome of
its Bell measurement becomes undetermined.  The Bell measurement can
be filled in arbitrarily, because for the purpose of interpreting the
syndrome, the effect is the same as if a Pauli error occurred
depending on how the measurement result is filled in.

Note that as usual, none of the Pauli corrections actually have to be
implemented explicitly. One can just update the Pauli frame as needed.

\pagebreak
\begin{herefig}
\label{fig:teleec}
\begin{picture}(0,5.8)(0,-5.9)
\nputgr{-3,-1}{l}{height=2.4in}{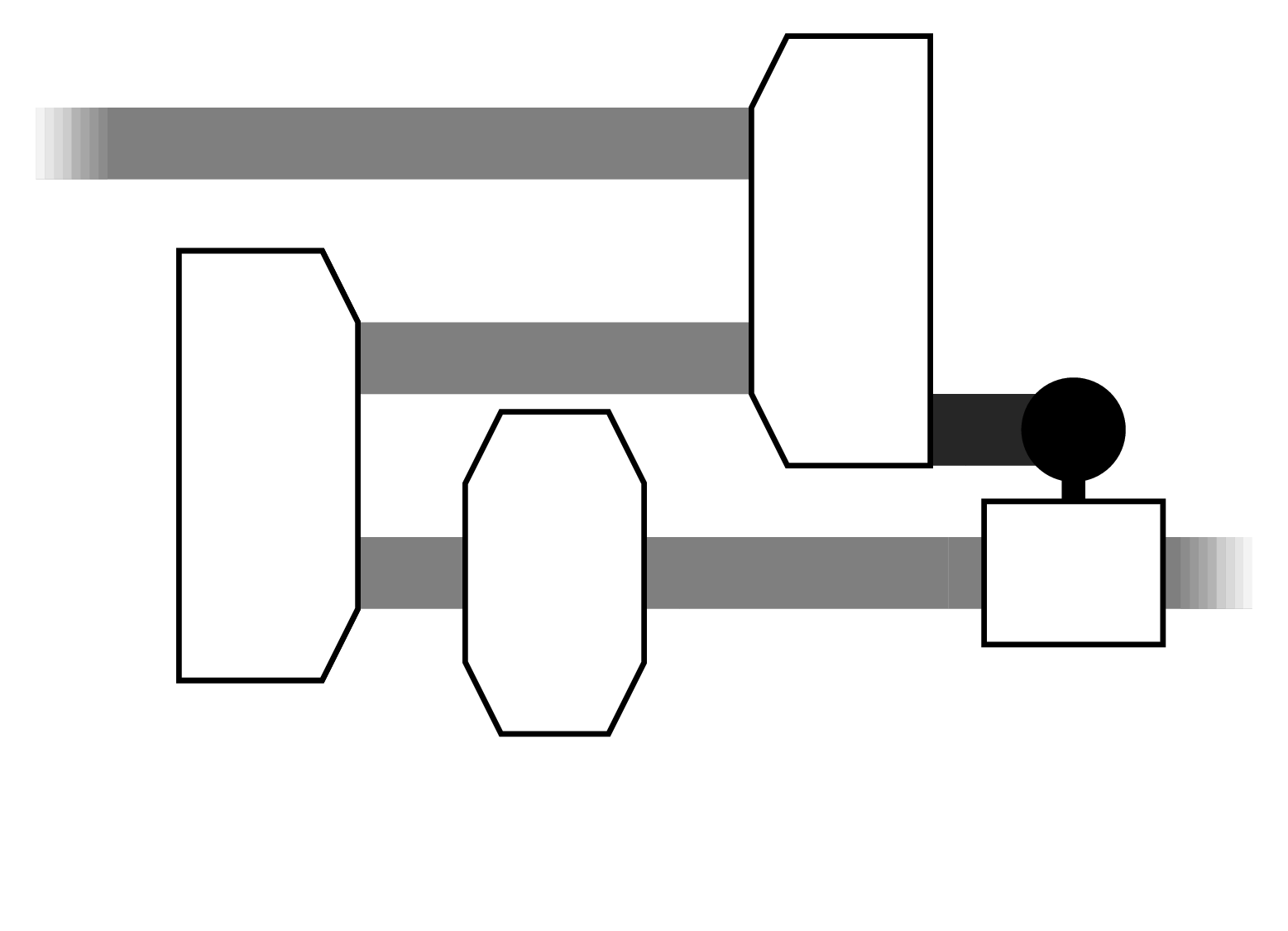} \nputbox{-3.3,.1}{lt}{$n$
input qubits} \nputbox{0.1,-.97}{lt}{$n$ output qubits}
\nputbox{-2.3,-1.04}{c}{\rotatebox{-90}{Bell}}
\nputbox{-1.58,-1.31}{c}{\small\rotatebox{-90}{$\Pi(Q,0)$}}
\nputbox{-.82,-.5}{c}{\rotatebox{-90}{Bell}}
\nputbox{-.21,-1.27}{c}{$P(\mathbf{g})$}

\nputbox{-1.7,-3}{r}{\scalebox{2.5}{$\Leftrightarrow$}}
\nputgr{-1.5,-3}{l}{height=2.4in}{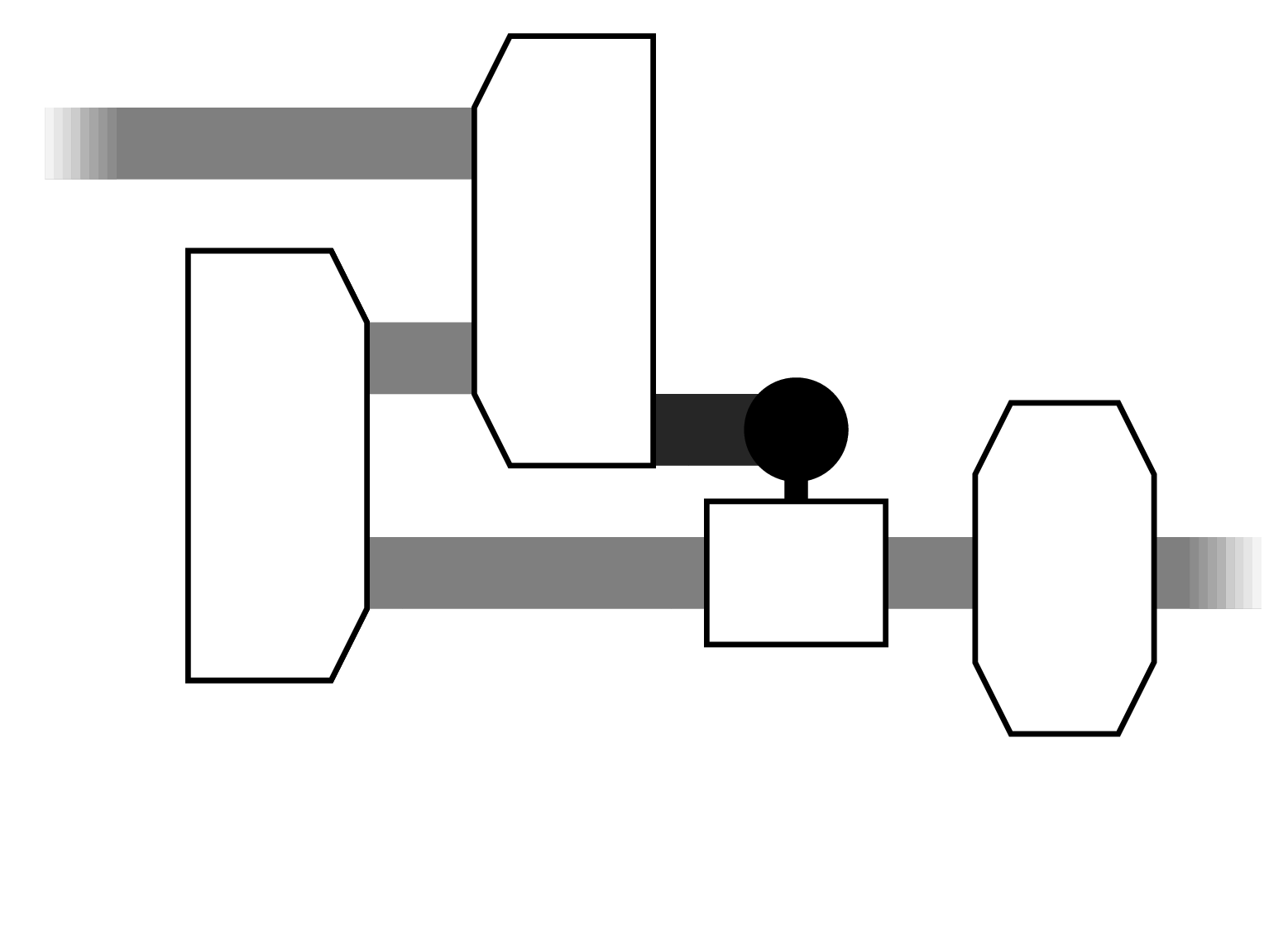}
\nputbox{-.8,-3}{c}{\rotatebox{-90}{Bell}}
\nputbox{-.05,-2.45}{c}{\rotatebox{-90}{Bell}}
\nputbox{.55,-3.26}{c}{$P(\mathbf{g})$}
\nputbox{1.25,-3.26}{c}{\small\rotatebox{-90}{$\Pi(Q,Q\Sy\mathbf{g}^T)$}}

\nputbox{-.2,-5}{r}{\scalebox{2.5}{$\Leftrightarrow$}}
\nputgr{0,-5}{l}{height=2.4in}{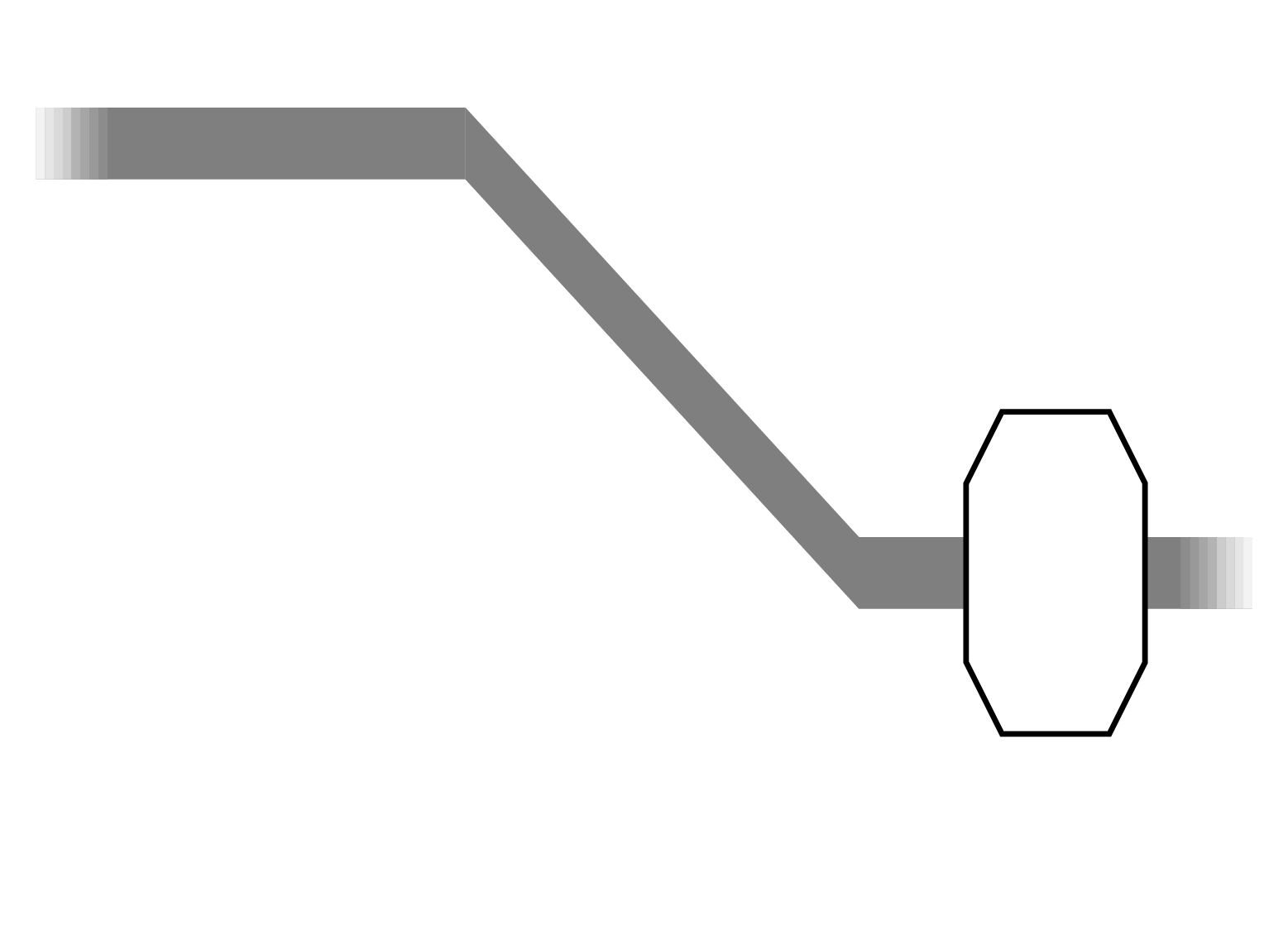}
\nputbox{2.75,-5.28}{c}{\small\rotatebox{-90}{$\Pi(Q,Q\Sy\mathbf{g}^T)$}}
\end{picture}
\nopagebreak
\herefigcap{Teleporting with an encoded entangled state is equivalent
to a syndrome measurement.  The gray lines are the time lines of
blocks of $n$ qubits. The boxes denote various operations.  The
Bell-state preparation on corresponding pairs of qubits in two blocks
is depicted with a box angled to the right and labeled ``Bell''.  The
state used for teleportation in the top diagram is obtained after
Bell-state preparation by projecting one of the blocks with $\Pi(Q,0)$
(the actual preparation procedure is different but has the same
output). Projection operators are shown with boxes angled both ways
with the operator written in the box.  Bell measurement of
corresponding pairs of qubits in two blocks is depicted with a box
angled to the left and labeled ``Bell''. A Bell measurement on qubits
$1$ and $2$ is implemented by applying a CNOT from qubit $1$ to $2$,
performing an $X$-measurement on qubit $1$ and a $Z$-measurement on
qubit $2$.  The top diagram is the actual network implemented. The
other two are logically equivalent. The Bell measurement outcome
$\mathbf{g}$ is correlated with the effective projection in the bottom
diagram. If the input state has a particular syndrome, then only
$\mathbf{g}$ for which the projection is onto the subspace with this
syndrome have non-zero probababilities.}
\end{herefig}
\pagebreak

\section{Networks for $C_4$ and $C_6$ State Preparation and Gates}
\label{sect:networks}

\begin{herefig}
\label{fig:elements}
\begin{picture}(0,4.4)(0,-4.2)
  \nputgr{-2,-.3}{r}{scale=.7}{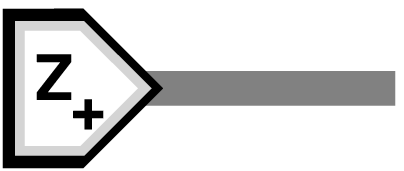}
    \nputbox{-1.9,-.3}{l}{
      \begin{tabular}{@{}l}
      Encoded $\ket{0}$\\[-4pt] preparation.
      \end{tabular}
    }
  \nputgr{1.8,-.3}{r}{scale=.7}{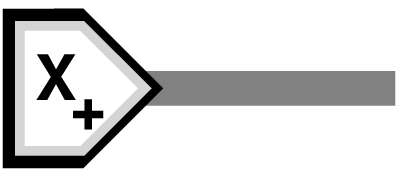}
    \nputbox{1.9,-.3}{l}{
      \begin{tabular}{@{}l}
      Encoded $\ket{+}$\\[-4pt] preparation.
      \end{tabular}
    }

  \nputgr{-2,-.9}{r}{scale=.7}{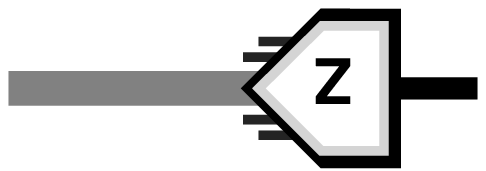}
    \nputbox{-1.9,-.9}{l}{
      \begin{tabular}{@{}l}
        Transversal $Z$\\[-4pt] measurement.
      \end{tabular}
    }
  \nputgr{1.8,-.9}{r}{scale=.7}{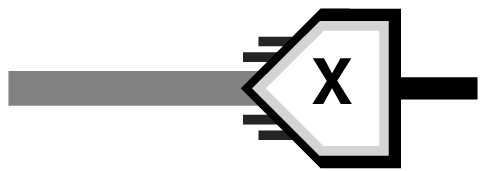}
    \nputbox{1.9,-.9}{l}{
      \begin{tabular}{@{}l}
        Transversal $X$\\[-4pt] measurement.
      \end{tabular}
    }

  \nputgr{-2,-1.8}{r}{scale=.7}{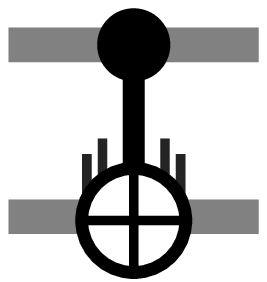}
    \nputbox{-1.9,-1.8}{l}{
      \begin{tabular}{@{}l}
        Transversal CNOT.
      \end{tabular}
    }
   
  \nputgr{1.8,-1.53}{r}{scale=.7}{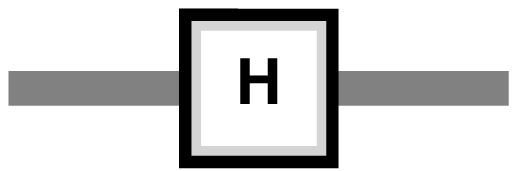}
    \nputbox{1.9,-1.53}{l}{
      \begin{tabular}{@{}l}
        Encoded HAD.
      \end{tabular}
    }
  \nputgr{1.8,-2.07}{r}{scale=.7}{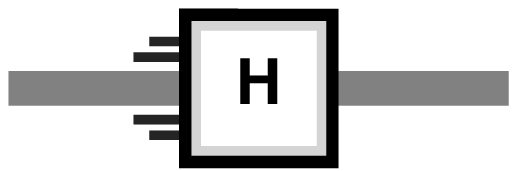}
    \nputbox{1.9,-2.07}{l}{
      \begin{tabular}{@{}l}
        Transversal HAD.
      \end{tabular}
    }

  \nputgr{-2,-2.8}{r}{scale=.7}{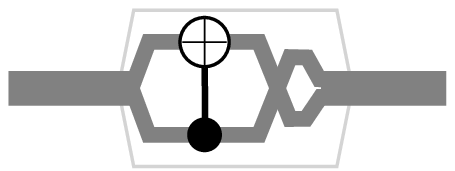}
    \nputbox{-1.9,-2.8}{l}{
      \begin{tabular}{@{}l}
        $*u$ on a block\\[-4pt]
        encoding two qubits
      \end{tabular}
    }

  \nputgr{1.8,-2.8}{r}{scale=.7}{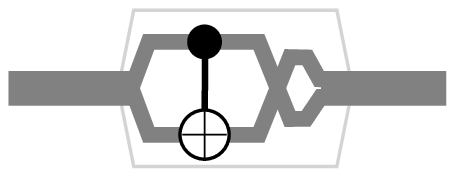}
    \nputbox{1.9,-2.8}{l}{
      \begin{tabular}{@{}l}
        $*u^2$ on a block\\[-4pt]
        encoding two qubits
      \end{tabular}
    }

  \nputgr{-2,-3.6}{r}{scale=.7}{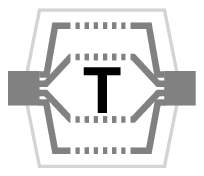}
    \nputbox{-1.9,-3.6}{l}{
      \begin{tabular}{@{}l}
         $C_4$ subblock teleportation,\\[-4pt]
         involves four lower-\\[-4pt]
         level blocks.
      \end{tabular}
    }
  \nputgr{1.8,-3.6}{r}{scale=.7}{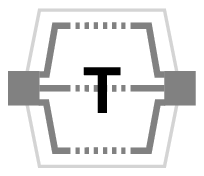}
    \nputbox{1.9,-3.6}{l}{
      \begin{tabular}{@{}l}
         $C_6$ subblock teleportation,\\[-4pt]
         involves three lower\\[-4pt]
         level blocks
      \end{tabular}
    }
 
\end{picture}
\nopagebreak
\herefigcap{Network elements.  The elements shown represent networks
  acting on blocks of qubits. Blocks (shown by thick gray lines) may
  consist of only one physical qubit, so the elements can also
  represent physical gates.  Elements with ``fringes'' are transversal
  gates: The indicated gate is applied to each physical qubit, or to
  corresponding physical qubits in the input blocks. The measurements
  have classical output indicated with a black line. Because they are
  transversal, the output contains as many bits as there are physical
  qubits.  Because the codes used here are CSS codes, the check
  operators and the encoded Pauli operators contain only one type
  of non-identity Pauli operator. The output bits therefore contain
  both error-check information and the encoded-measurement
  answers. The $*u$ and $*u^2$ elements are defined as shown for
  physical qubit pairs. The notation comes from a polynomial
  construction of $C_6$ as a code on three quaternary qudits using the
  four-element field $GF(4)$. The symbol $u$ denotes a third root of unity
  over $GF(2)$. The gates transform Pauli operators by
  multiplication with $u$ or $u^2$ in a $GF(4)$ labeling of these operators.}
\end{herefig}
\pagebreak

\begin{herefig}
\label{fig:C_4_reductions:_preps}
\begin{picture}(0,6.4)(0,-6.4)
  \nputgr{-1.3,-1.6}{r}{scale=.7}{blockprepZ}
  \nputbox{-.9,-1.6}{c}{\scalebox{2.5}{$\Leftrightarrow$}}
  \nputgr{-.5,-1.6}{l}{scale=.5}{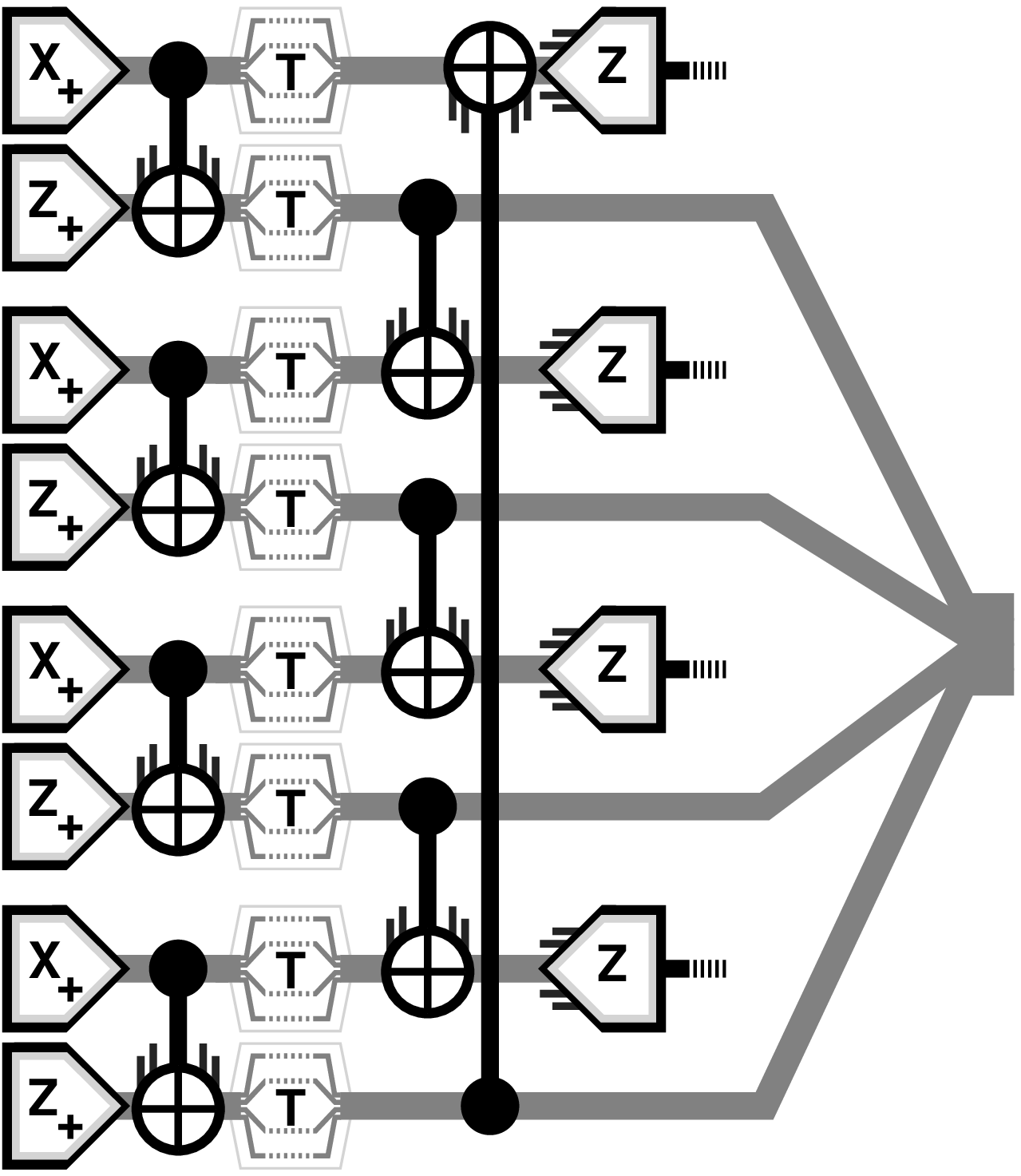}
  \nputgr{-1.3,-4.8}{r}{scale=.7}{blockprepX}
  \nputbox{-.9,-4.8}{c}{\scalebox{2.5}{$\Leftrightarrow$}}
  \nputgr{-.5,-4.8}{l}{scale=.5}{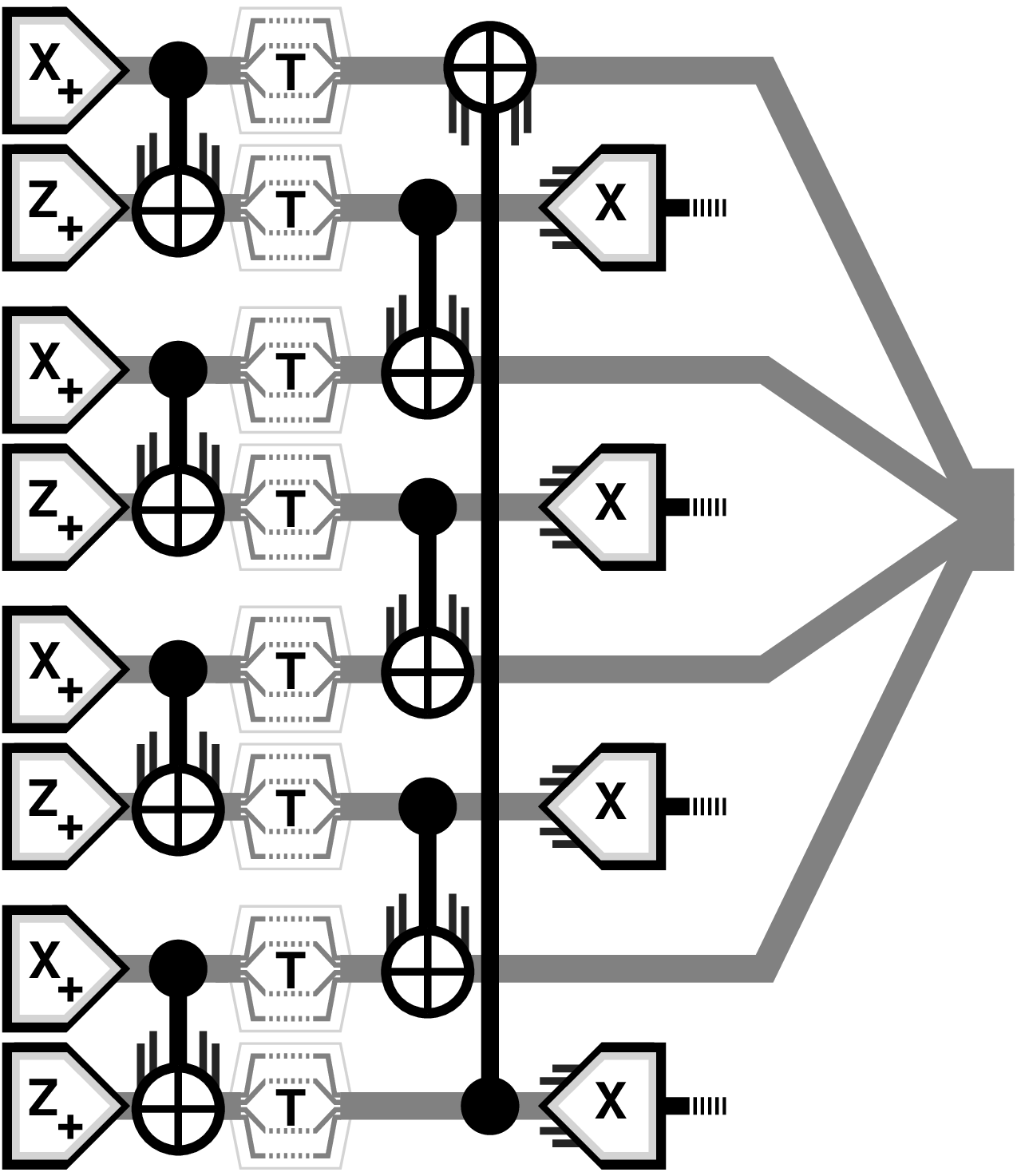}
\end{picture}
\pagebreak
\herefigcap{Encoded state preparations in terms of lower-level
elements for $C_4$. The lower-level blocks (``subblocks'') can be
either physical or encoded single qubits and are represented by the
merging lines in the networks on the right. In the $C_4/C_6$
architecture, $C_4$ is used only at level $1$, so the subblocks are
always physical qubits. In this case, the output block contains an
encoded qubit pair with each qubit in the pair in the state indicated
by the preparation gate on the left.  The physical states prepared are
four-qubit cat states ($(\ket{0000}+\ket{1111})/\sqrt{2}$ in the case
of the top network). If no error occurred, the four measurements in
each network on the right have total parity $0$.  For any single error
in the state preparation network, if this error results in an error in
the output state that is not equivalent to a single physical qubit
error, then the parity is $1$, so this event can be detected.  Thus,
if the total measurement parity is $1$, the output state is rejected.
This ensures that errors occurring with linear probability in the EPGs
introduce no undetectable errors.  Note that the networks on the
right begin with Bell-state preparations. The teleportation steps are
not implemented on physical qubits but are included for generality.
The encoded $Z$- and $X$-preparations shown assume that the next step
is a transversal CNOT followed by subblock teleportations. Otherwise
it may be necessary to teleport subblocks immediately to avoid error
propagation.  }
\end{herefig}
\pagebreak

\begin{herefig}
\label{fig:C_6_reductions:_preps}
\begin{picture}(0,5.8)(0,-5.5)
  \nputgr{-1.3,-1.3}{r}{scale=.7}{blockprepZ}
  \nputbox{-.9,-1.3}{c}{\scalebox{2.5}{$\Leftrightarrow$}}
  \nputgr{-.5,-1.3}{l}{scale=.5}{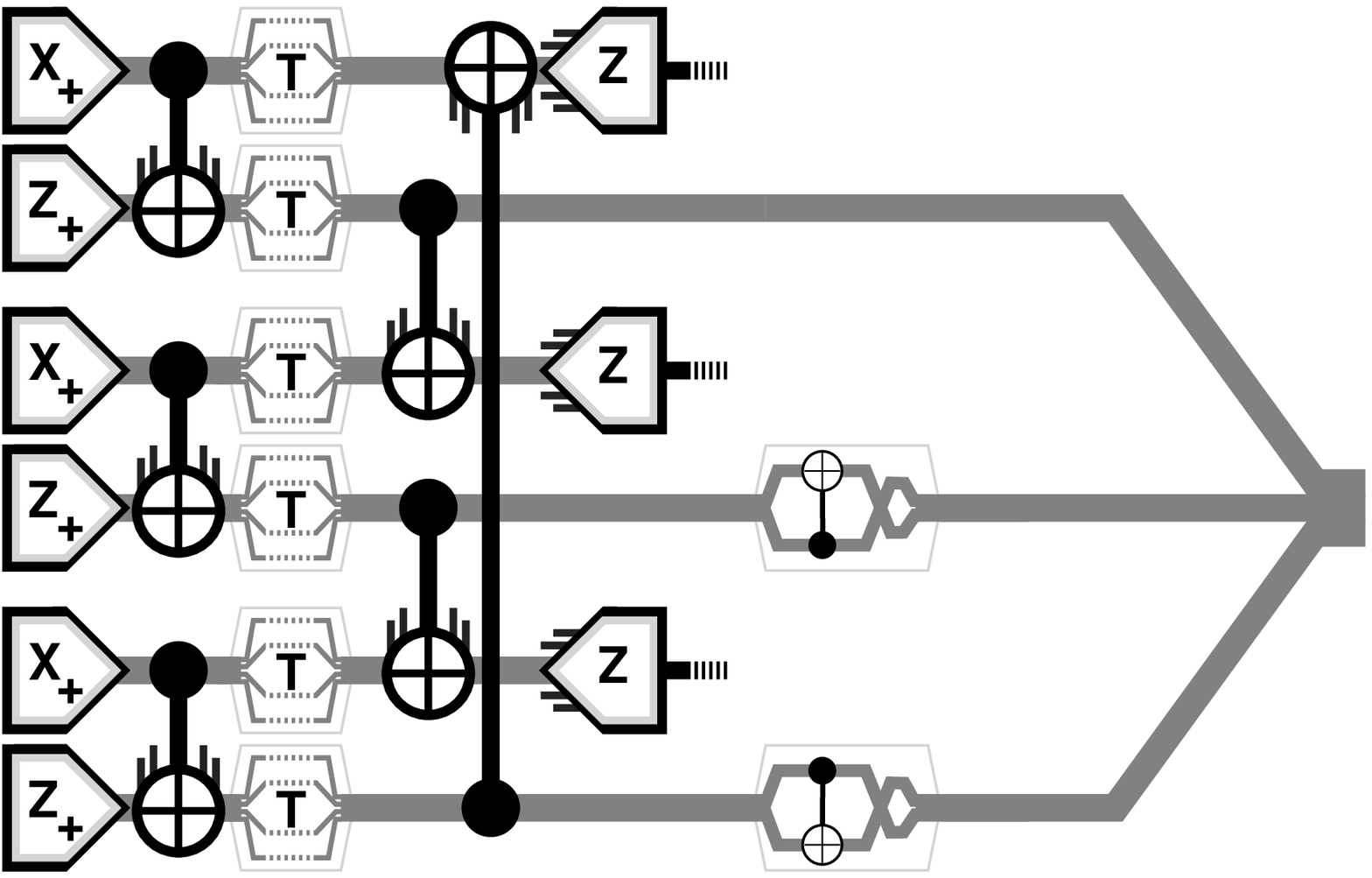}

  \nputgr{-1.3,-4.0}{r}{scale=.7}{blockprepX}
  \nputbox{-.9,-4.0}{c}{\scalebox{2.5}{$\Leftrightarrow$}}
  \nputgr{-.5,-4.0}{l}{scale=.5}{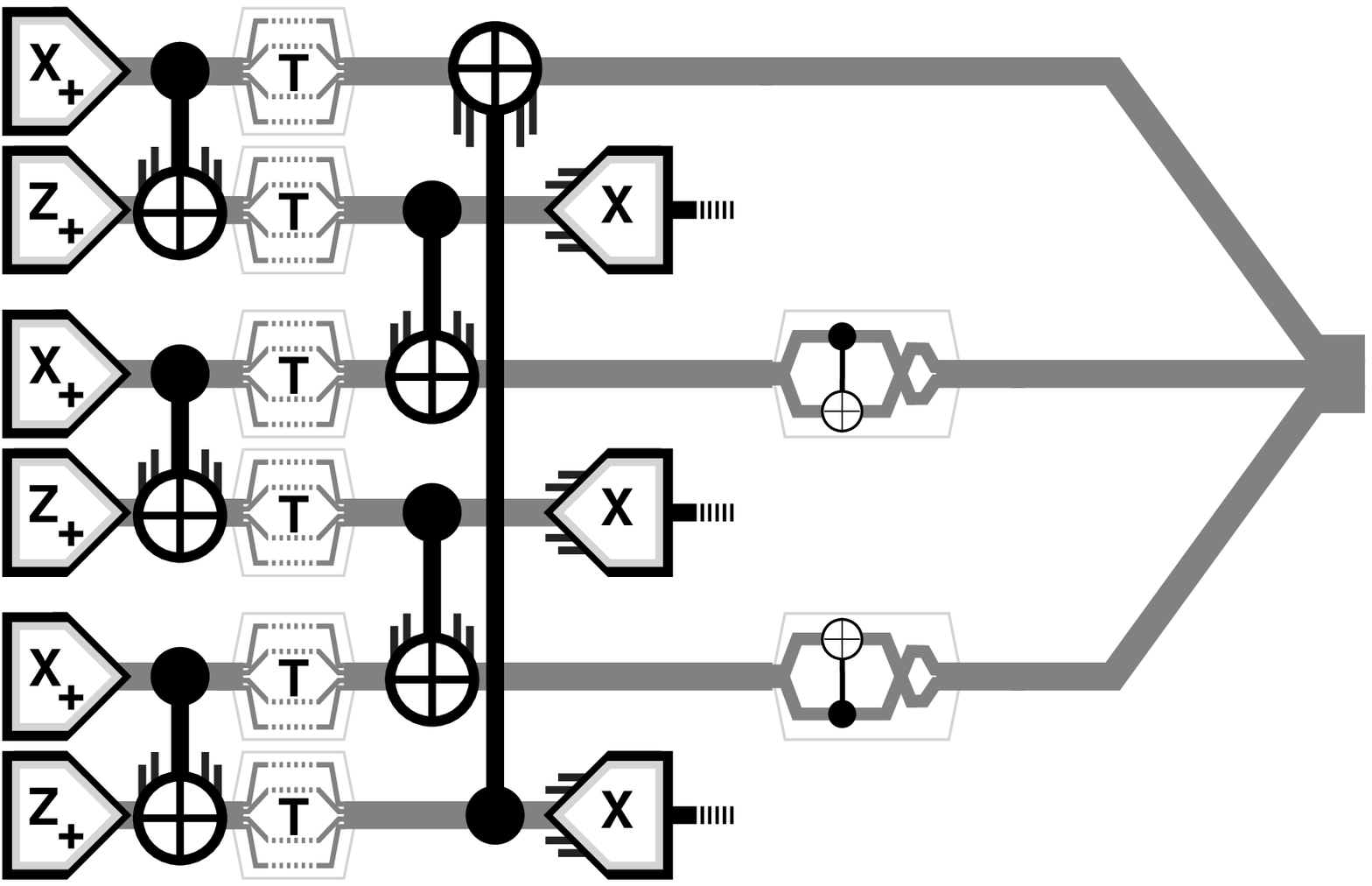}
\end{picture}
\nopagebreak
\herefigcap{Encoded state preparations in terms of lower-level elements
for $C_6$. The lower-level blocks (``subblocks'') contain encoded
qubit pairs.  The beginning of the network prepares two parallel
three-qubit cat states $(\ket{000}+\ket{111})/\sqrt{2}$ in the top
network) on corresponding members of the encoded qubit pairs.  The
encoded measurements in the cat-state preparation satisfy the parity
constraint described in the caption of
Fig.~\ref{fig:C_4_reductions:_preps} for each of the three
corresponding qubits in the encoded qubit pairs. Because the
measurements are implemented transversally, they also provide
lower-level syndrome information that can be used for error detection
or correction.  Again, the networks begin with Bell pair preparations
and the teleportations are only implemented on encoded qubit pairs.
The last elements rotate the parallel cat states into $C_6$, so that
the encoded qubit pair has both qubits in the desired state.  Because
the first level encoding uses $C_4$, they can be implemented as simple
permutations, which can be accomplished by logical relabeling without
delay or error, see Fig.~\ref{fig:C_6_reductions:_mults}.  As in
Fig.~\ref{fig:C_4_reductions:_preps}, the encoded $Z$- and
$X$-preparations shown assume that the next step is a transversal CNOT
followed by subblock teleportations. Otherwise it may be necessary to
teleport subblocks immediately to avoid error propagation.  }
\end{herefig}
\pagebreak

\begin{herefig}
\label{fig:C_6_reductions:_mults}
\begin{picture}(0,3.2)(0,-3)
  \nputgr{-1.3,-1.2}{r}{scale=.7}{blockmu}
  \nputbox{-.9,-1.2}{c}{\scalebox{2.5}{$\Leftrightarrow$}}
  \nputgr{-.5,-1.2}{l}{scale=.5}{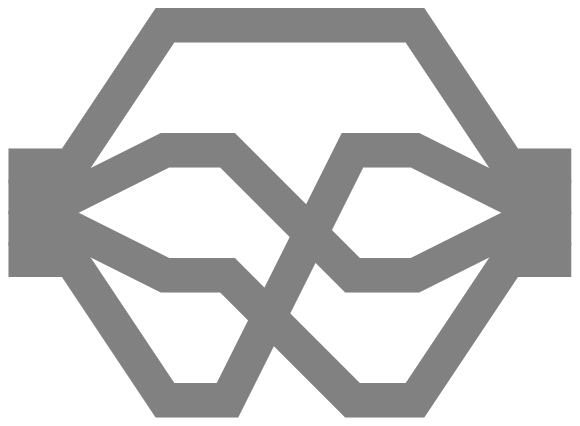}
  \nputgr{-1.3,-2.5}{r}{scale=.7}{blockmusq}
  \nputbox{-.9,-2.5}{c}{\scalebox{2.5}{$\Leftrightarrow$}}
  \nputgr{-.5,-2.5}{l}{scale=.5}{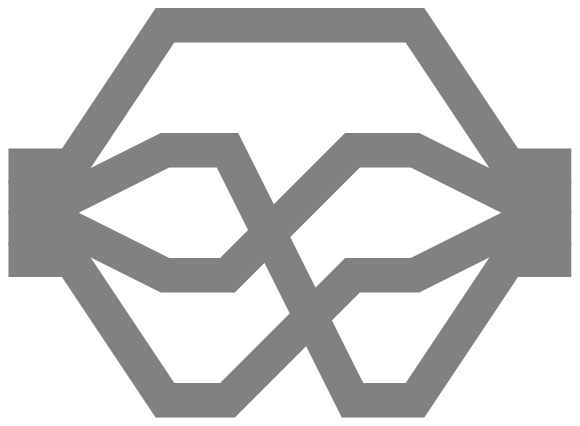}
\end{picture}
\\
\begin{picture}(0,3.8)(0,-3.6)
  \nputgr{-1.3,-1.2}{r}{scale=.7}{blockmu}
  \nputbox{-.9,-1.2}{c}{\scalebox{2.5}{$\Leftrightarrow$}}
  \nputgr{-.5,-1.2}{l}{scale=.5}{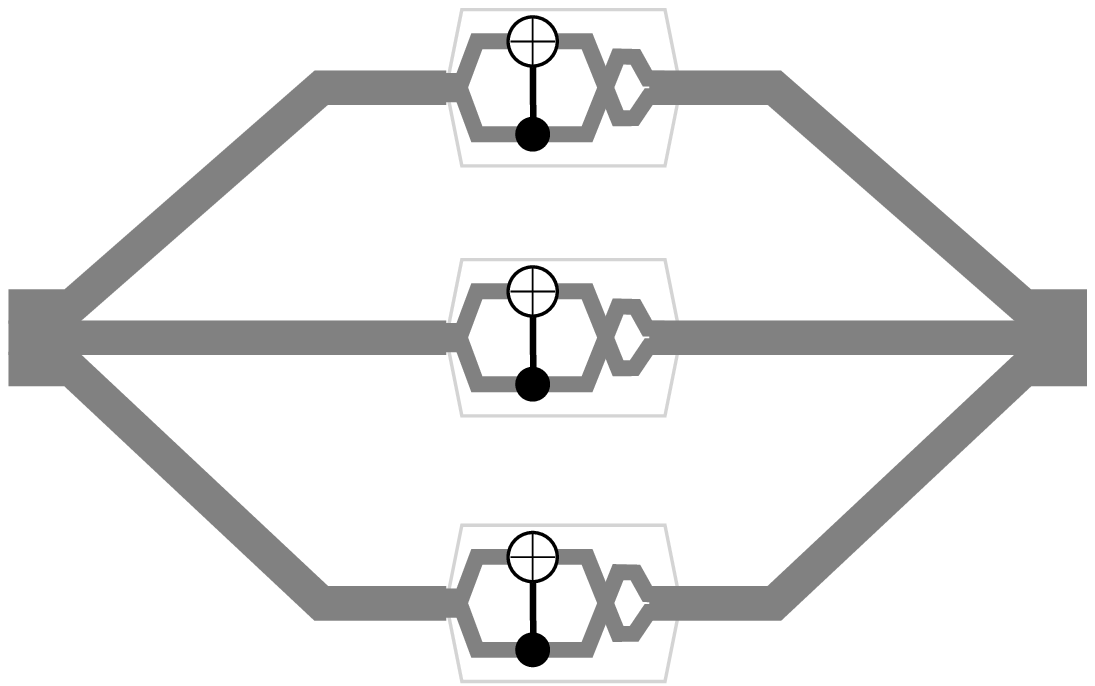}
  \nputgr{-1.3,-2.8}{r}{scale=.7}{blockmusq}
  \nputbox{-.9,-2.8}{c}{\scalebox{2.5}{$\Leftrightarrow$}}
  \nputgr{-.5,-2.8}{l}{scale=.5}{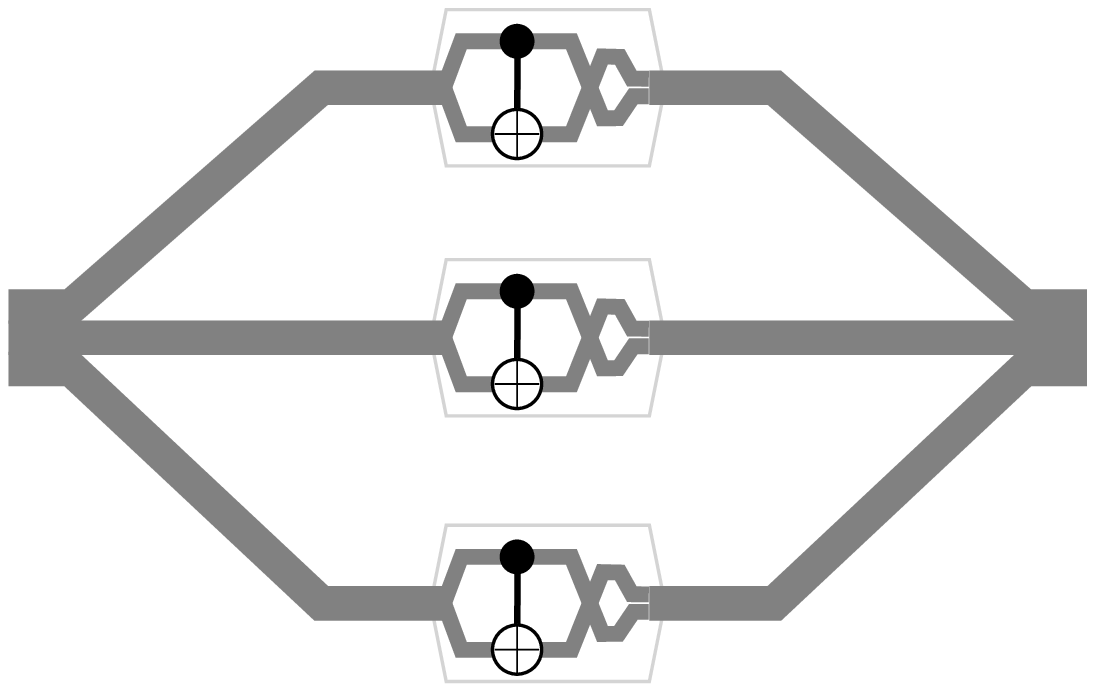}
\end{picture}
\nopagebreak
\herefigcap{Implementation of $*u$ and $*u^2$ on blocks of $C_4$ (top two)
and $C_6$ (bottom two).
}
\end{herefig}
\pagebreak

\begin{herefig}
\label{fig:C_4/C_6_reductions:_tele}
\begin{picture}(0,7.2)(0,-7)
  \nputgr{-1.8,-2.0}{r}{scale=.7}{block4tel}
  \nputbox{-1.4,-2.0}{c}{\scalebox{2.5}{$\Leftrightarrow$}}
  \nputgr{-1,-1.6}{l}{scale=.5}{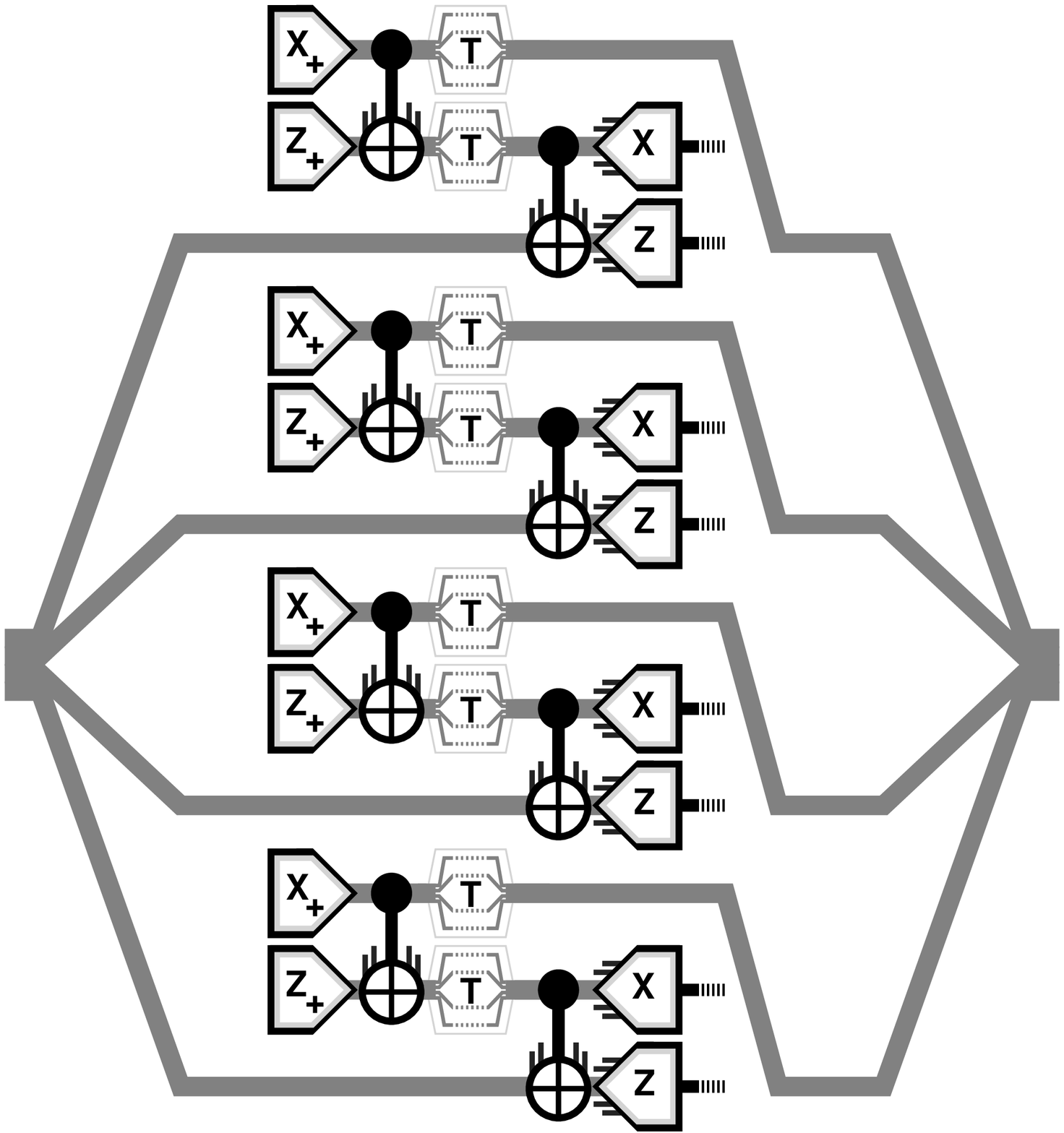}
  \nputgr{-1.8,-6.15}{r}{scale=.7}{block3tel}
  \nputbox{-1.4,-6.15}{c}{\scalebox{2.5}{$\Leftrightarrow$}}
  \nputgr{-1,-5.3}{l}{scale=.5}{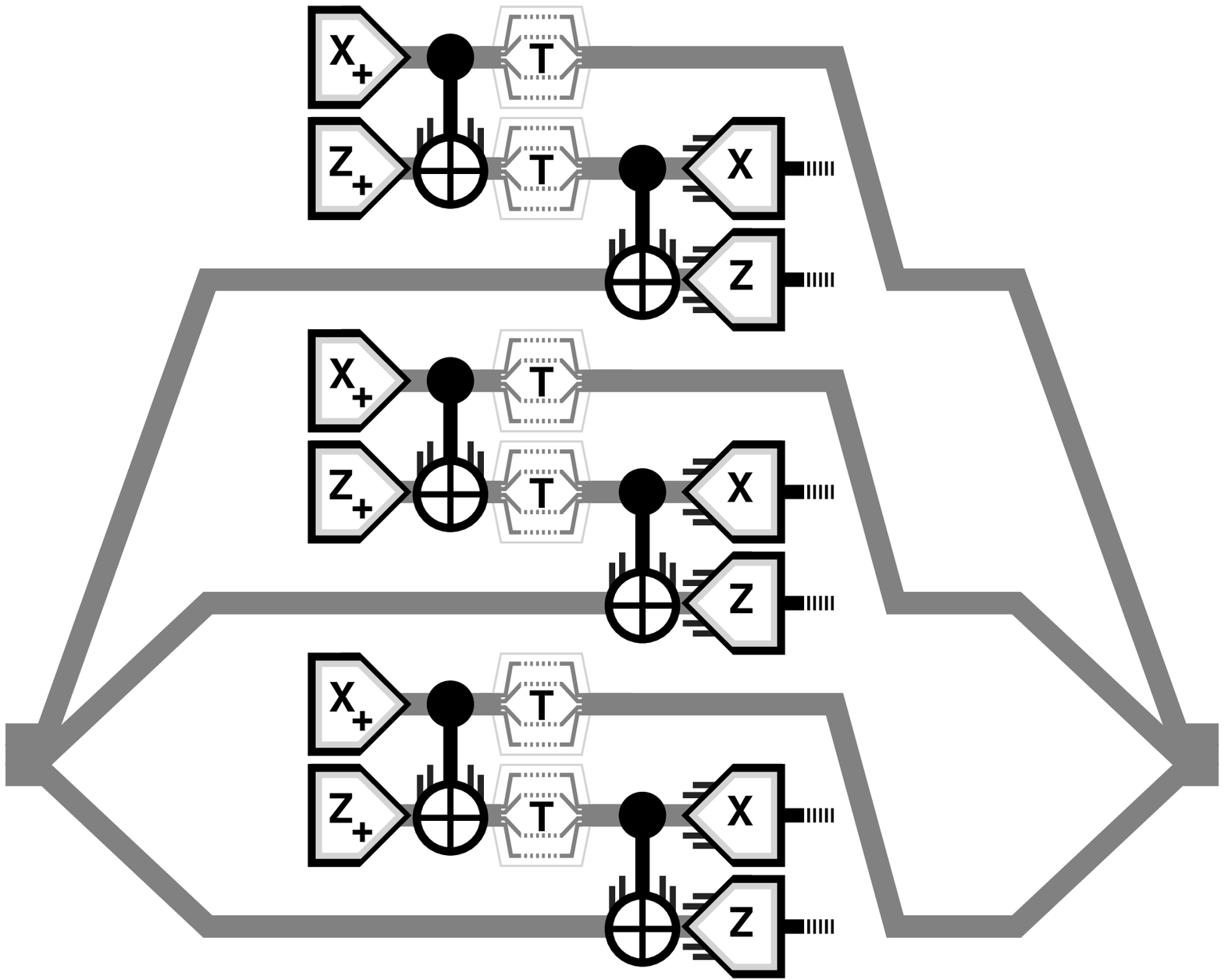}
\end{picture}
\nopagebreak
\herefigcap{Implementation of subblock teleportation.  If the blocks
are physical qubits or pairs of physical qubits, the teleportation
elements on the left are not implemented.  Otherwise the networks
shown on the right are used on the lower-level blocks.  The top
network is for $C_4$ and the bottom for $C_6$. Note that the networks
look like traditional teleportation of each subblock.  However, if the
subblocks are not physical, the encoded Bell states used imply that
the teleportations are error-detecting/correcting for the subblocks.}
\end{herefig}
\pagebreak

\begin{herefig}
\label{fig:C_4/C_6_reductions:_had}
\begin{picture}(0,4.2)(0,-4)
  \nputgr{-1,-1.0}{r}{scale=.7}{blockhad}
  \nputbox{0,-1.0}{c}{\scalebox{2.5}{$\Leftrightarrow$}}
  \nputgr{1,-1}{l}{scale=.5}{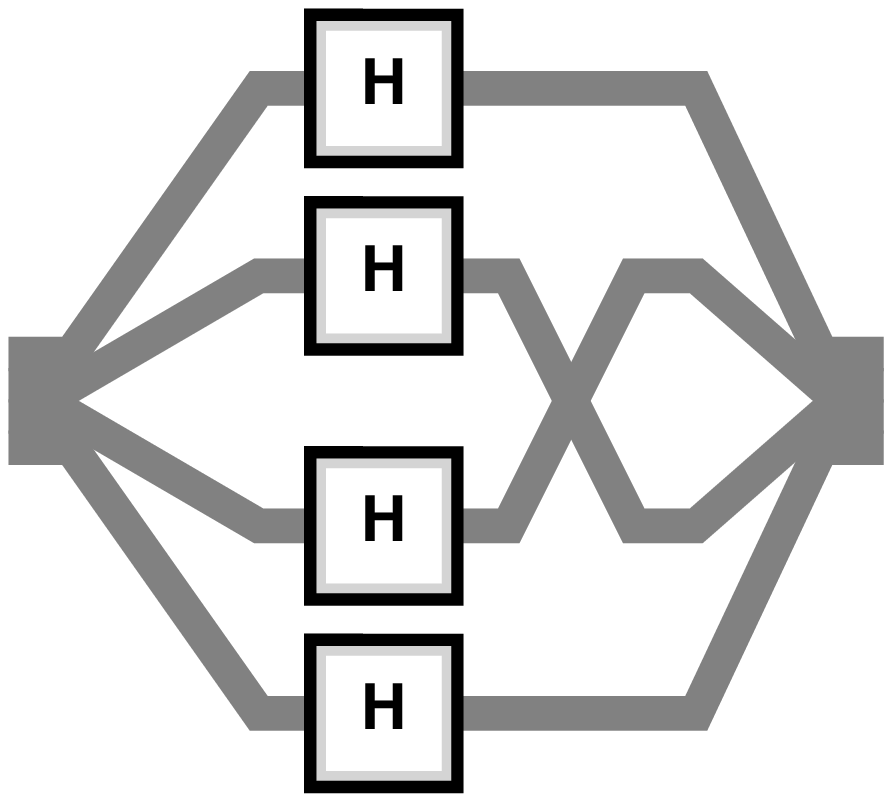}
  \nputgr{-1,-3}{r}{scale=.7}{blockhad}
  \nputbox{0,-3}{c}{\scalebox{2.5}{$\Leftrightarrow$}}
  \nputgr{1,-3}{l}{scale=.5}{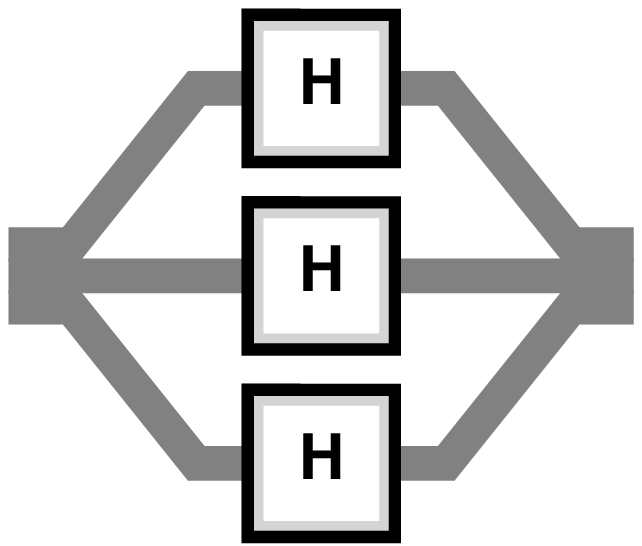}
\end{picture}
\nopagebreak
\herefigcap{Implementation of encoded HADs for $C_4$ and $C_6$.
The top network is for $C_4$ and is transversal except
for an interchange of the middle two qubits. The bottom is for
$C_6$ and is transversal. Using the HAD and
CNOT implementations, it is also possible to implement
the encoded conditional sign flip transversally up to a physical qubit
permutation implementable by relabeling.
}
\end{herefig}
\pagebreak

As can be seen, all preparation networks are based ultimately on Bell
state preparation followed by full or half Bell measurements. As
shown, the networks use teleportation fastidiously.  It may be
possible to delay teleportation in some cases, but this was not
confirmed by simulation.  For postselected computing, there is
no need to wait for measurement outcomes before proceeding to the next
steps.  However, this delays the rejection of states found later to be
faulty, which incurs a large resource cost if the probability of
detecting an error is high.  For standard quantum computing, this
resource cost can be avoided by delaying further processing and
incurring some memory error instead. If error correction is used, at
higher levels the probability of unrecoverable error decreases rapidly so
one can again proceed optimistically, before measurement answers are
known.

\section{Decoding $C_4$ and $C_6$.}
\label{sect:decoding}

There are two reasons to explicitly decode logical states encoded by
concatenating $C_4$ and $C_6$. First, at the highest EPGs, to
implement a standard quantum computation with the postselected
$C_4/C_6$ architecture requires preparing $C_4/C_6$ encoded states
that are themselves states encoded in a code $C_e$ with very good
error-correction capabilities.  Once such a state is prepared, the
$C_4/C_6$ concatenation hierarchy is decoded to obtain a physical
block encoding a state in $C_e$. Second, to implement arbitrary
quantum computations requires preparing special encoded states that
are not reachable using $\minCSS$ and HAD gates alone.  These encoded states
need not be error-free initially, since they can be purified using
low-error logical $\minCSS$ and HAD gates.  A way to prepare these
states with error that is bounded independently of the
number of levels is to prepare a logical Bell state in two blocks,
decode the first block into two physical qubits, and make a
measurement of the physical qubits to project them into the desired
state. (Alternatively, but with more error, the measurement can be
replaced by a teleportation of the desired state prepared in another
pair of physical qubits.) The entanglement between the physical qubits
and the logical ones in the second block ensures that the state is
injected into the logical qubits.

A good method for decoding the $C_4$/$C_6$ concatenation hierarchy is
to decode ``bottom up''.  That is, in the first step, the blocks of
four physical qubits encoding qubit pairs in $C_4$ at the lowest level
of the hierarchy are decoded. Syndrome information becomes available
in a pair of ancillas for each block of $C_4$ and can be used for
error detection. In subsequent steps, six physical qubits encoding
qubit pairs in $C_6$ are similarly decoded. Error information obtained
in previous decoding steps can be combined with new syndrome
information for error detection or correction.  The $C_4$ and $C_6$
decoding networks are shown in Fig.~\ref{fig:c4,c6_decoding}.

\pagebreak

\begin{herefig}
\label{fig:c4,c6_decoding}
\begin{picture}(0,3.6)(0,-3.2)
\nputbox{-2.1,.3}{t}{ Decoding $C_4$:}
\nputgr{-2.1,-1.3}{c}{scale=.65}{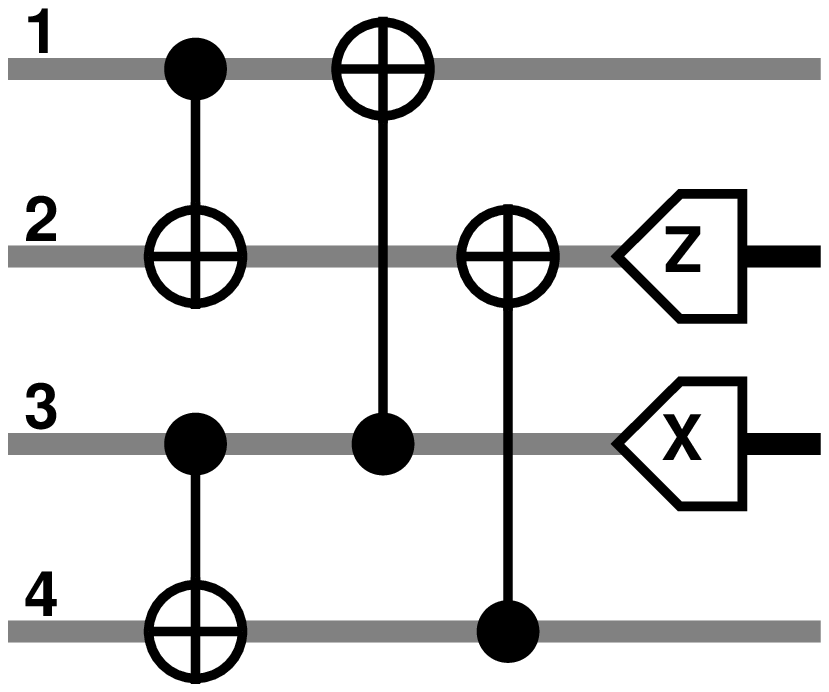}
\nputbox{1.7,.3}{t}{ Decoding $C_6$:}
\nputgr{1.7,-1.3}{c}{scale=.65}{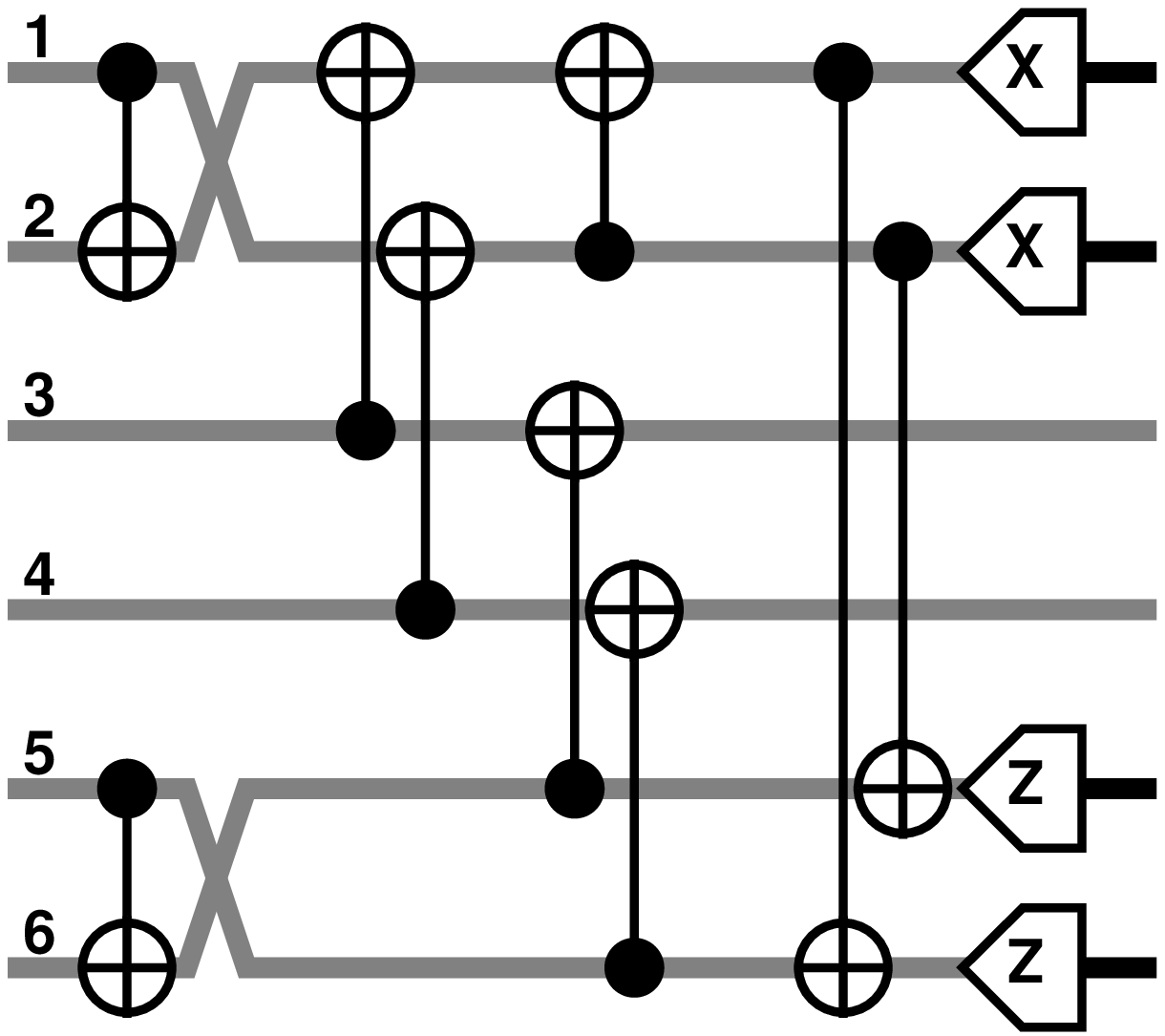}
\end{picture}
\nopagebreak
\herefigcap{Decoding $C_4$ (left) and $C_6$ (right).  The gates are
shown in plain form to indicate that they are physical, not encoded.
The measurements reveal the syndrome and can be used for error
detection.  Error correction can be used if (a) the incoming qubits
were decoded in an earlier step and (b) exactly one of them (for
$C_4$) or one pair of them ($(1,2)$, $(3,4)$ or ($5,6$) for $C_6$) was
detected to have an error.  The $C_6$ decoding can be simplified if
the first level of the full concatenation hierarchy uses $C_4$: The
first step is a $*u^2$ operation on the first and last pair and can be
implemented by relabeling before the level $1$ blocks of $C_4$ are
decoded.  Even if the $C_6$ decoding is implemented with maximum
parallelism and without waiting for measurement outcomes, it has an
initial memory delay on qubits $5$ and $6$ that was not taken into
consideration in the simulations. }
\end{herefig}
\pagebreak

\section{Universal Computing}
\label{sect:universal}

Universal computing with logical qubits encoded with $C_4/C_6$ can be
accomplished by use of the logical $\minCSS$ gates, HADs and
$\ket{\pi/8}$-state preparation.  However, since these operations do
not distinguish between the two logical qubits encoded in one block,
computations are implemented on only one of the two logical qubits in
each block. Because the other one experiences the same evolution, the
computation's output is obtained twice each time it is run. It is
desirable to be able to address the two logical qubits in a block
separately and have the ability to apply a CNOT from one to the
other. One operation that is already available is the $*u$ gate and
its inverse, which acts on a logical qubit pair as a swap followed by
a CNOT.  As with all stabilizer codes, it is also possible to apply
arbitrary combinations of logical Pauli matrices by applying suitable
products of physical Pauli matrices or by making a Pauli frame
change. Fig.~\ref{fig:*u_univ} shows how to use the gates mentioned to
implement a set of gates that is sufficiently rich to address individual
qubits in a pair.

\begin{herefig}
\label{fig:*u_univ}
\begin{picture}(-.2,4.2)(0,-4)
  \nputgr{-1.5,-1.0}{r}{scale=.7}{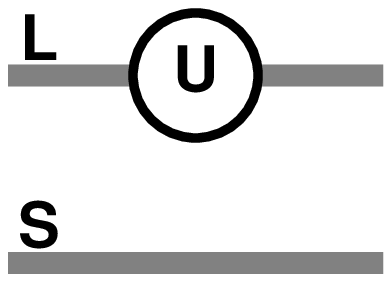}
  \nputbox{-1,-1.0}{c}{\scalebox{2.5}{$\Leftrightarrow$}}
  \nputgr{-.3,-1}{l}{scale=.7}{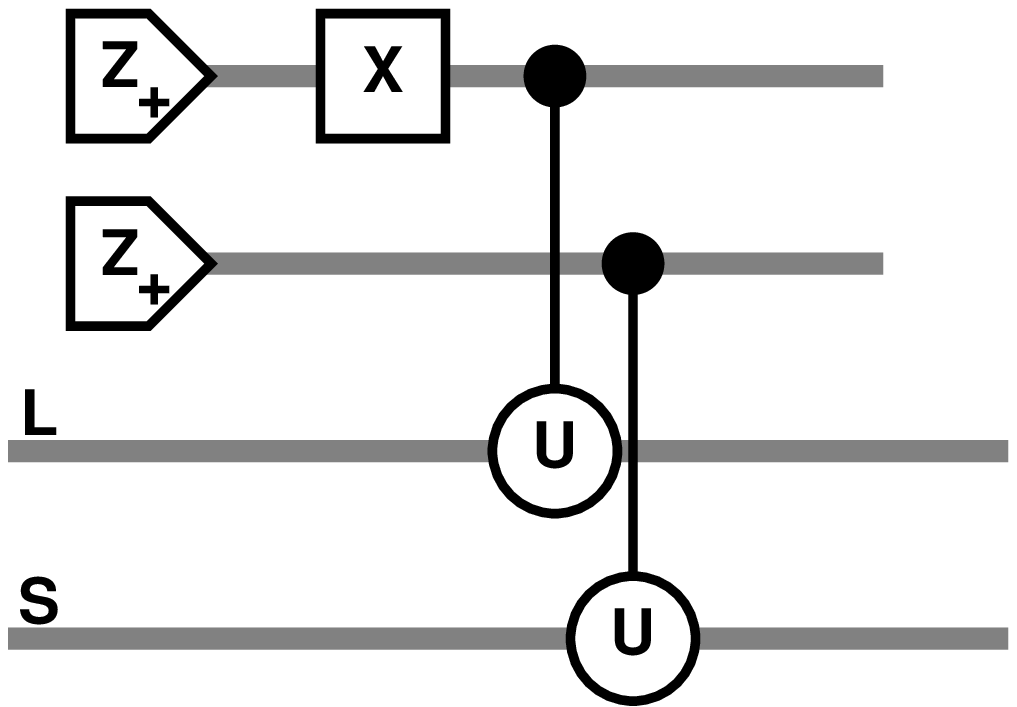}
  \nputgr{-1.5,-3.0}{r}{scale=.7}{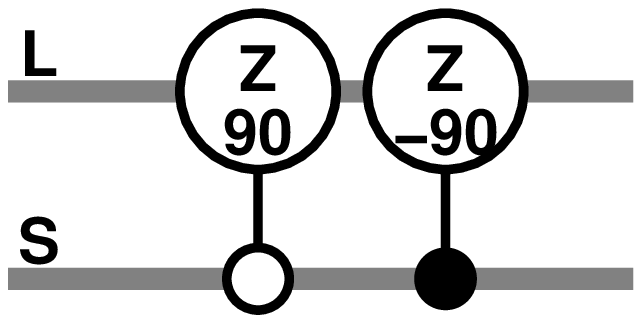}
  \nputbox{-1,-3.0}{c}{\scalebox{2.5}{$\Leftrightarrow$}}
  \nputgr{-.3,-3}{l}{scale=.7}{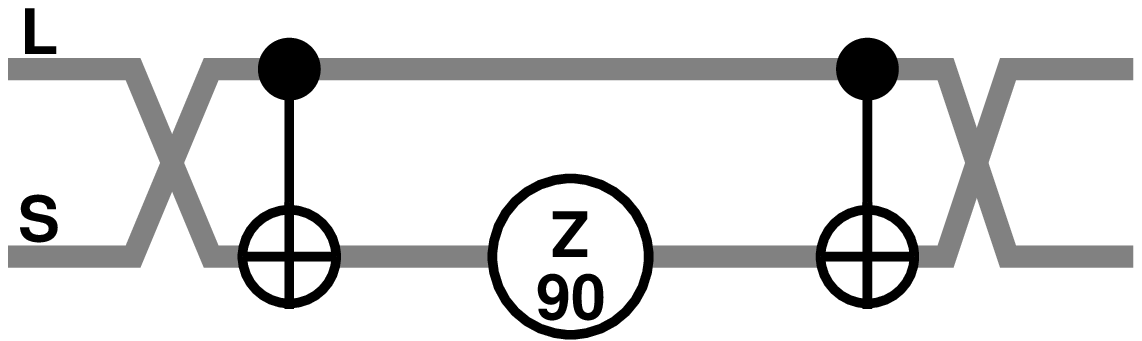}
\end{picture}
\nopagebreak
\herefigcap{Implementing selective gates on qubit pairs.  The networks
are shown for physical qubits, but can be used with logical qubits by
making the appropriate substitutions.  The implementations on the
right use only gates that do not distinguish the qubits in a pair,
$*u$, $*u^2$ and Pauli products.  The top network shows how to
implement any gate $U$ selectively on one qubit in a pair. The
implementation uses a selective Pauli operator and non-selective
controlled-$U$ gates.  The bottom network shows how to implement a type
of controlled phase gate between the two logical qubits in a pair.  It
uses a $*u$ operation, a selective $90\dg$ $z$-rotation (which can be
implemented using the top network) and a $*u^2$ operation. The CNOT
(without swap) between the qubits in a pair can be implemented in
terms of the controlled phase gate shown and selective one-qubit
gates.  }
\end{herefig}

The networks shown in Fig.~\ref{fig:*u_univ} do not result in
particularly efficient ways of implementing gates on individual
logical qubits. An alternative is to inject and purify states needed
for one-qubit teleportation of the desired gates using the techniques
given in~[\citeonline{zhou:qc2000b}]. An example of such a state is
$\ket{0}\ket{+}$. Note that $\ket{0}\ket{+}$ is much more readily
purified than $\ket{\pi/8}$. For example, to purify $\ket{0}\ket{+}$
one can apply the method suggested in the main text to reduce
the preparation error. In the encoded setting, this requires a
measurement of $Z$ and $X$ of the qubits in a pair, which cannot be
done by a transversal encoded measurement.  Instead, a third instance
of $\ket{0}\ket{+}$ is introduced and involved in a transversal Bell
measurement with the block of the qubits to be measured.  As in
error-correcting teleportation, the desired information can be
extracted from parities of the Bell measurements.  At the same time,
syndrome information that can be used for error detection and
correction is obtained.

\section{Simulation of Error Behavior}
\label{sect:simulation}

To simulate the error behavior of fault-tolerant methods based on
stabilizer codes, we use the result that computation with Clifford
gates and feed-forward from $Z$- and $X$-measurements can be
efficiently simulated~\cite{gottesman:qc1997b}.  The Clifford gates
include $Z$- and $X$-state preparations and measurements, HAD, CNOT and
$90\dg$ $Z$-rotations.  Networks using these gates always result in
stabilizer states, which are eigenstates of maximal sets of commuting
check operators (the check matrix).  Simulation requires tracking a 
complete independent set of such check operators and the syndrome
(which gives the state's eigenvalues with respect to the check
operators).  Check operators can be represented by binary vectors (see
Sect.~\ref{sect:tec}).  To simplify the computations required for
updating the check matrix and syndrome after applying gates, we
maintain it in ``graph-state normal
form''~\cite{schlingemann:qc2001b,grassl:qc2002a}. In this form, each
qubit has an associated ``commuting'' operator, which is either $X$ or
$Z$, and there is exactly one check operator acting on the qubit with
an operator different from $I$ or the commuting operator. In addition
to the check matrix and the ideal syndrome, we maintain the Pauli
products representing the current effect of errors (the ``error
vector'') and the Pauli frame. The error vector is known only to the
simulation, not to the user implementing a computation.  The error
vector and Pauli frame are updated with each operation. For
efficiency, blocks that have not yet interacted are associated with
separate check matrices and ``merged'' when needed. Also, since it is
necessary to accumulate as much statistics as possible, an array of
error vectors and corresponding Pauli frames is used to represent
multiple simultaneous preparation attempts without duplicating check
matrices. For rapid prototyping purposes and fast array processing, we
used Octave to implement the simulator.  For simulating measurements
and errors, a random-number generator is needed.  We used the standard
random-number generator provided with Octave.  Because this implies
that there are implicit correlations in the errors for the large-scale
simulations undertaken here, the results obtained do not constitute
full statistical proof.  However, no artifacts not explainable by
statistics were observed.  In particular, in the few cases where an
analytic expression for the data were available, the simulated data
was as expected. This was checked for conditional error probabilities
in postselected computing using concatenation with $C_4$ as
discussed in~[\citeonline{knill:qc2004b}] for up to two levels (data
not shown).

The simulations are used to determine the error behavior of various
logical gates. For the data shown in
Fig.~\ref{fig:condEPG},~\ref{fig:ecEPG} and~\ref{fig:ecChain}, we used
the reference entanglement method~\cite{schumacher:qc1996a} for
determining logical CNOT error probabilities.  This involves applying
the logical CNOT and error-detecting or -correcting teleportations to
the first members of two error-free logical Bell pairs and then
comparing the logical state to what would have been obtained if the
logical CNOT had no error.  The comparison is implemented by applying
error-free CNOTs to disentangle the Bell pairs and making error-free
logical $X$- or $Z$-measurements with error detection or correction
depending on the context.  The procedure was modified by (1) applying
only the CNOT's physical error model associated with the transversal
implementation and (2) applying the error model and error-detecting
teleportation twice and determining the incremental error introduced
the second time. (1) simplifies the verification without affecting the
error probabilities. (2) is required so as to determine the effective
error introduced in the middle of a computation, because the
error-free Bell pairs have no initial error, contrary to what would be
expected later. Using the second of two steps suffices because of the
isolating properties of teleportation, which was verified by taking some
data for more steps as shown in Fig.~\ref{fig:ecChain}.  For detected
error probabilities, the incremental error is determined as the
fraction of trials in which an uncorrectable error was detected during
the teleportations or in the verifying measurements. For conditional
error probabilities, the incremental error is the fraction of trials
with no detected uncorrectable error for which the logical measurement
outcomes are incorrect but there was no undetected logical error in
the preceding steps.

\section{Scalable Quantum Computing via Bootstrapping with Postselection}
\label{sect:scalable}

The fault-tolerant architecture based on a good quantum
error-correcting code $C_e$ using the $C_4/C_6$ architecture with
postselection for state preparation is described in the main text.  We
claimed that if $3.53\gamma \lesssim 0.19$ and $\gamma$ is below the
threshold for fault-tolerant postselected computing with the
$C_4/C_6$ architecture, then the logical errors for the $C_e$
architecture can be made to be below $10^{-3}$, which is below the
threshold for known fault-tolerant architectures.  The estimate
assumes that the decoding error per decoded qubit is $\approx\gamma$,
in which case $\approx 3.53\gamma$ is the effective error per qubit
that determines whether the error-correcting teleportation
successfully corrects. With this assumption, the claim is proven as
follows.  Choose $\epsilon$ such that $3.53\gamma \lesssim
0.19-\epsilon$.  Choose $C_e$ such that if a logical qubit is encoded
in $C_e$ without error and each physical qubit is independently
subjected to an error with probability $0.19-\epsilon$, then the
logical state can be recovered with error at most $10^{-3}/4$.  Such
codes exist~\cite{divincenzo:qc1998c} although their length $n$ grows
as $\epsilon$ goes to zero. Algorithms for encoding needed states in
one to four blocks of $C_e$ require at most $c_1 n^2$ gates for some
constant $c_1$ [\citeonline{cleve:qc1996b}].  Choose the level of the
postselected $C_4/C_6$ architecture so that the logical gate error is
well below $10^{-3}/(2 c_1 n^2)$.  Then, before they are decoded, the
postselected prepared states have logical error at most $10^{-3}/2$,
since they required fewer than $c_1 n^2$ logical gates of the
$C_4/C_6$ architecture. This error persists as a $C_e$-logical error
after the $C_e$-error-correcting teleportation that uses this state
after it is decoded. It adds to the logical error introduced by
failure to error-correct in teleportation. However, because at most
two error-correcting teleportations are involved, the total logical
error is below $10^{-3}$. Note that if $\gamma$ is given and strictly
below the threshold, then the resources required to achieve
$C_e$-logical EPGs below $10^{-3}$ are determined. Because the
fault-tolerant architectures that can be used with EPGs of $10^{-3}$
are known to be theoretically efficient, the combined architecture
starting with $C_e$ is also theoretically efficient. The problem is
that as $\gamma$ approaches the upper limit, the minimum length of the
code $C_e$ grows, and as a result the probability of successfully
preparing the required states by postselection goes down dramatically,
making the combined architecture highly impractical.

\section{Resource Usage}
\label{sect:resources}

The simulations keep track of the number of operations of different
types that are applied in the course of implementing a quantum
network. The resources required depend on whether the networks are
implemented with maximum parallelism or sequentially: If they are
implemented sequentially, one can take advantage of the ability to
abort some computations early, but such implementations require
quantum memory of sufficiently low error.  Here we consider only the
case of maximum parallelism.  At the core of the fault-tolerant
architecture is Bell pair preparation. One can analyze the resources
required to construct a level $l+1$ Bell pair in terms of the number
of level $l$ Bell pairs consumed. As a first step, consider the case
of zero EPG. In this case no error is ever detected and all networks
succeed on the first try. We count only the number of physical qubit
state preparations, $p(l,\gamma)$, and the number of physical CNOTs,
$c(l,\gamma)$. The number of physical qubit measurements is less than
the number of qubit state preparations.  A level $0$ (physical) Bell
pair requires $p(0,0)=2$ qubit state preparations and $c(0,0)=1$
CNOT. A level $1$ Bell pair requires four level $0$ Bell pairs and
four CNOTs for preparing and verifying the initial state of each of
the two blocks to be used.  Combining the blocks requires another four
CNOTs.  Thus $p(1,0) = 8p(0,0)$ and $c(1,0) = 8c(0,0)+12$.  For $l\geq
1$, a level $l+1$ Bell pair requires three level $l$ Bell pairs and
three level $l$ encoded CNOTs for preparing the initial state in each
block.  Combining the blocks to form the level $l+1$ Bell pair
requires $3^{l-1}4\times 3$ physical CNOTs.  To remove lower level
errors, each of the six level $l$ subblocks is teleported.  Each
teleportation uses one level $l$ Bell pair and CNOT.  Thus $p(l+1,0) =
12 p(l,0)$ and $c(l+1,0) = 12 c(l,0)+3^l 20$.

\ignore{
% Octave
function p = preps(l);
  if (l == 0);
    p = 2;
  elseif (l == 1);
    p = preps(l-1)*8;
  else;
    p = preps(l-1)*12;
  endif;
endfunction;
function c = CNOTs(l);
  if (l == 0);
    c = 1;
  elseif (l == 1);
    c = CNOTs(0)*8+12;
  else;
    c = CNOTs(l-1)*12 + 3^(l-1) *20;
  endif;
endfunction;

% Make the table (see below):
function  [f, m]  = fleads(x);
  m = floor(log10(x));
  f = x*10^(-m);
endfunction;
for l = (0:6);
  p = preps(l); c = CNOTs(l);
  [fp, ep] = fleads(p);
  [fc, ec] = fleads(c);
  if (p > 10^4);
    printf('$%d$   &  $%.3f{\\times} 10^{%d}$  & $%.3f{\\times} 10^{%d}$   \\\\\n', \
       l, 
       fp, ep,
       fc, ec);
  else;
    printf('$%d$   &  $%d$  & $%d$   \\\\\n', \
       l, p, c);
  endif;
endfor;
}

\begin{heretab}
\label{tab:zero_error_resources}
\heretabcap{ Table of resources used for logical Bell pair preparation at EPG $\gamma=0$. \\[6pt]}
\begin{tabular}{c|cc}
\hline
Level & Preparations & CNOTs \\
\hline
\expdata{}{Table inserted from octave output above}
$0$   &  $2$  & $1$   \\
$1$   &  $16$  & $20$   \\
$2$   &  $192$  & $300$   \\
$3$   &  $2304$  & $3780$   \\
$4$   &  $2.765{\times} 10^{4}$  & $4.590{\times} 10^{4}$   \\
$5$   &  $3.318{\times} 10^{5}$  & $5.524{\times} 10^{5}$   \\
$6$   &  $3.981{\times} 10^{6}$  & $6.634{\times} 10^{6}$   \\
\hline
\end{tabular}
\end{heretab}

With maximum parallelism, the average resource requirements increase
by factors inversely related to the probability of success at various
points in the preparation process.  Preparing an encoded Bell state
involves two sequential steps that may fail. The first verifies the
initial states of each block before they are combined with CNOTs.  Let
the probability of successful verification of a block at level $l$ be
given by $v(l,\gamma)$.  The second involves teleportation of each
subblock after the two blocks are combined. Let the overall
probability of success of the teleportations be $t(l,\gamma)$.  Note
that both of these probabilities of success are with respect to the
combination of error correction and detection used in state
preparation, which differs from the full error correction used in
logical computation. The above resource formulas are modified as
follows: $p(0,\gamma) = 2$, $c(0,\gamma) = 1$, $p(1,\gamma) =
8p(0,\gamma)/v(1,\gamma)$, $c(1,\gamma) =
8c(0,\gamma)/v(1,\gamma)+12$, $p(l+1,\gamma) =
(6p(l,\gamma)/v(l+1,\gamma) + 6p(l,\gamma))/t(l+1,\gamma)$,
$c(l+1,\gamma) = ((6c(l,\gamma)+3^l 12)/v(l+1,\gamma)+ 6c(l,\gamma)+
3^l 8)/t(l+1,\gamma)$.  These formulas were obtained under the
assumption that the verification of the two blocks proceeds
independently with many simultaneous attempts, where the successful
ones are then combined. This requires waiting for measurement outcomes
and any associated memory error must be accounted for in $\gamma$.
The subblock teleportations are not independent because of the
immediately preceding transversal CNOT, which introduces correlated
errors.
Tables~\ref{tab:gamma_resources1},~\ref{tab:gamma_resources2},~\ref{tab:gamma_resources3}
show the success probabilities up to level 5 for $\gamma=0.01, 0.001,
0.0001$ together with the resources estimated according to these
recursive formulas and the resources determined by the simulation
after averaging over the number of attempts made. The simulation is
expected to show higher resource requirements because it
involves some loss when combining unequal numbers of independently
prepared blocks, as would be expected to occur in a real
implementation.  This was not taken into account in deriving the
formulas.

\ignore{
function p = prepsf(l,v,t);
  if (l == 0);
    p = 2;
  elseif (l == 1);
    p = prepsf(l-1,v,t)*8/v(1);
  else;
    p = (6*prepsf(l-1,v,t)/v(l)+6*prepsf(l-1,v,t))/t(l);
  endif;
endfunction;
function c = CNOTsf(l,v,t);
  if (l == 0);
    c = 1;
  elseif (l == 1);
    c = CNOTsf(0,v,t)*8/v(1)+12;
  else;
    c = ((CNOTsf(l-1,v,t)*6+3^(l-1)*12)/v(l)+6*CNOTsf(l-1,v,t)+3^(l-1)*8)/t(l);
  endif;
endfunction;

function  [f, m]  = fleads(x);
  m = floor(log10(x));
  f = x*10^(-m);
endfunction;
v2 = t2 = v3 = t3 = v4 = t4 = ones(5,1);
for k = (1:5);
  s2{k} = load('-ascii', \
            sprintf('cliff/data/ana6cnts_s1.000_l%d_aa.dat', k));
  v2(k) = (s2{k}.vx(2)+s2{k}.vz(2))/2;
  if (k == 1);
    t2(k) = 1;
  else;
    if (struct_contains(s2{k},'pr'));
      t2(k) = s2{k}.pr;
    else;
      t2(k) = s2{k}.p(2);
    endif;
  endif;
  s3{k} = load('-ascii', \
            sprintf('cliff/data/ana6cnts_s0.100_l%d_aa.dat', k));
  v3(k) = (s3{k}.vx(2)+s3{k}.vz(2))/2;
  if (k == 1);
    t3(k) = 1;
  else;
    if (struct_contains(s3{k},'pr'));
      t3(k) = s3{k}.pr;
    else;
      t3(k) = s3{k}.p(2);
    endif;
  endif;
  s4{k} = load('-ascii', \
            sprintf('cliff/data/ana6cnts_s0.010_l%d_aa.dat', k));
  v4(k) = (s4{k}.vx(2)+s4{k}.vz(2))/2;
  if (k == 1);
    t4(k) = 1;
  else;
    if (struct_contains(s4{k},'pr'));
      t4(k) = s4{k}.pr;
    else;
      t4(k) = s4{k}.p(2);
    endif;
  endif;
endfor;

function tab = restab(v,t,s);
  p = prepsf(1,v,t); c = CNOTsf(1,v,t);
  tab = '';
  tab = sprintf('%s $1$ & $%.3f$ & NA & $%.2f$ & $%.2f$ & $%.2f$ & $%.2f$ & $%d$\\\\\n', \
          tab, v(1), p, s{1}.preps(2), c, s{1}.CNOTs(2), s{1}.tries);
  for l = (2:5);
    p = prepsf(l,v,t); c = CNOTsf(l,v,t);
    if (p > 10^4);
      [fp, mp] = fleads(p); [fc, mc] = fleads(c);
      [tfp, tmp] = fleads(s{l}.preps(2));
      [tfc, tmc] = fleads(s{l}.CNOTs(2));
      tab = sprintf('%s $%d$ & $%.3f$ & $%.3f$ & $%.2f{\\times} 10^{%d}$ & $%.2f{\\times} 10^{%d}$ & $%.2f{\\times} 10^{%d}$ & $%.2f{\\times} 10^{%d}$ & $%d$\\\\\n', \
          tab, l, v(l), t(l), fp, mp, tfp, tmp, fc, mc, tfc, tmc, s{l}.tries);
    else;
      tab = sprintf('%s $%d$ & $%.3f$ & $%.3f$ & $%.1f$ & $%.1f$ & $%.1f$ & $%.1f$ & $%d$\\\\\n', \
          tab, l, v(l), t(l), p, s{l}.preps(2), c, s{l}.CNOTs(2), s{l}.tries);
    endif;
  endfor;
endfunction;

tab2 = restab(v2,t2,s2)
tab3 = restab(v3,t3,s3)
tab4 = restab(v4,t4,s4)
}

\begin{heretab}
\label{tab:gamma_resources1}
\heretabcap{ Table of success probabilities and resources used for 
             Bell state preparation at EPG
             $\gamma = 0.01$. \\[6pt]
             \begin{minipage}{\textwidth}The values $v(l,0.01)$, $t(l,0.01)$
             and the numbers in the ``preparations'' and ``CNOTs''
             columns are obtained by simulation using the number of
             successful Bell pair preparations shown in the
             ``\# Bell pairs'' column. Because only two successful preparations
             were used at level 5, the level 5 data have significant noise.
             \end{minipage}\\[6pt]}
\begin{tabular}{@{}c|ccccccc@{}}
\hline
Level & $v(l,0.01)$ & $t(l,0.01)$ & $p(l,0.01)$ & Preparations & $c(l,0.01)$ & CNOTs & \# Bell pairs\\\hline
$0$ & NA & NA & $2$ & $2$ & $1$ & $1$ & NA
\\
\expdata{}{Table inserted from octave output tab2 above}
$1$ & $0.940$ & NA & $17.01$ & $17.01$ & $20.51$ & $21.01$ & $10345$\\
 $2$ & $0.722$ & $0.247$ & $984.6$ & $1022.2$ & $1485.5$ & $1542.7$ & $2279$\\
 $3$ & $0.602$ & $0.100$ & $1.58{\times} 10^{5}$ & $1.60{\times} 10^{5}$ & $2.40{\times} 10^{5}$ & $2.45{\times} 10^{5}$ & $409$\\
 $4$ & $0.885$ & $0.205$ & $9.84{\times} 10^{6}$ & $9.63{\times} 10^{6}$ & $1.50{\times} 10^{7}$ & $1.48{\times} 10^{7}$ & $70$\\
 $5$ & $0.900$ & $0.500$ & $2.49{\times} 10^{8}$ & $3.08{\times} 10^{8}$ & $3.80{\times} 10^{8}$ & $4.72{\times} 10^{8}$ & $2$\\
\hline
\end{tabular}
\end{heretab}

\begin{heretab}
\label{tab:gamma_resources2}
\heretabcap{ Table of success probabilities and resources used for Bell
             state preparation at EPG
             $\gamma = 0.001$. \\[6pt]
}
\begin{tabular}{@{}c|ccccccc@{}}
\hline
Level & $v(l,0.001)$ & $t(l,0.001)$ & $p(l,0.001)$ & Preparations & $c(l,0.001)$ & CNOTs & \# Bell pairs\\\hline
$0$ & NA & NA & $2$ & $2$ & $1$ & $1$ & NA
\\
\expdata{}{Table inserted from octave output tab3 above}
$1$ & $0.994$ & NA & $16.10$ & $16.11$ & $20.05$ & $20.11$ & $10927$\\
 $2$ & $0.970$ & $0.870$ & $225.6$ & $226.6$ & $351.2$ & $352.8$ & $2014$\\
 $3$ & $0.957$ & $0.815$ & $3395.6$ & $3434.9$ & $5513.4$ & $5576.1$ & $401$\\
 $4$ & $1.000$ & $0.970$ & $4.20{\times} 10^{4}$ & $4.67{\times} 10^{4}$ & $6.88{\times} 10^{4}$ & $7.61{\times} 10^{4}$ & $64$\\
 $5$ & $1.000$ & $1.000$ & $5.04{\times} 10^{5}$ & $5.61{\times} 10^{5}$ & $8.27{\times} 10^{5}$ & $9.17{\times} 10^{5}$ & $2$\\
\hline
\end{tabular}
\end{heretab}

\begin{heretab}
\label{tab:gamma_resources3}
\heretabcap{ Table of success probabilities and resources used for 
             Bell state preparation at EPG
             $\gamma = 0.0001$. \\[6pt]
}
\begin{tabular}{@{}c|c@{\hspace*{3pt}}c@{\hspace*{3pt}}c@{\hspace*{3pt}}c@{\hspace*{3pt}}c@{\hspace*{3pt}}c@{\hspace*{3pt}}c@{}}
\hline
Level & $v(l,0.0001)$ & $t(l,0.0001)$ & $p(l,0.0001)$ & Preparations & $c(l,0.0001)$ & CNOTs & \# Bell pairs\\\hline
$0$ & NA & NA & $2$ & $2$ & $1$ & $1$ & NA
\\
\expdata{}{Table inserted from octave output tab4 above}
$1$ & $0.999$ & NA & $16.01$ & $16.02$ & $20.01$ & $20.02$ & $10987$\\
 $2$ & $0.995$ & $0.984$ & $195.7$ & $204.9$ & $305.6$ & $317.2$ & $2155$\\
 $3$ & $0.994$ & $0.982$ & $2398.6$ & $2556.0$ & $3929.9$ & $4158.1$ & $429$\\
 $4$ & $1.000$ & $1.000$ & $2.88{\times} 10^{4}$ & $3.13{\times} 10^{4}$ & $4.77{\times} 10^{4}$ & $5.16{\times} 10^{4}$ & $66$\\
 $5$ & $1.000$ & $1.000$ & $3.45{\times} 10^{5}$ & $3.92{\times} 10^{5}$ & $5.74{\times} 10^{5}$ & $6.50{\times} 10^{5}$ & $2$\\
\hline
\end{tabular}
\end{heretab}

Resources for implementing logical gates transversally are dominated
by those required for logical Bell state preparation.  For example,
the logical CNOT includes error-correcting teleportation and therefore
requires two logical Bell states and three transversal CNOTs.  The
number of physical CNOTs in a transversal CNOT grows by a factor of
$3$ for each level after the first, whereas the number of physical
CNOTs required for logical Bell state preparation grows by a factor
greater than $12$.  This justifies focusing attention on the resources
required for logical Bell state preparation.  The biggest resource
overhead is incurred when implementing non-Clifford gates such as
$\ket{\pi/8}$-preparation (see below) or Toffoli gates. Note that two
$\ket{\pi/8}$ states are needed to implement a Toffoli gate up to a
reversible phase in the logical basis, which is all that is required
for most uses of Toffoli gates.  We have not attempted to optimize
$\ket{\pi/8}$-preparation.  Furthermore, it is possible that gates
such as the Toffoli gate can be implemented more efficiently using
other states, for example, using Steane's
adaptation~\cite{steane:qc1999a} of Shor's method~\cite{shor:qc1996a}.

For completeness and to obtain an upper bound on the requirements for
a minimal non-trivial quantum algorithm at $\gamma=0.01$, we outline
one method for preparing good logical $\ket{\pi/8}$ states, discuss
why the error when using these states is expected to be similar to
that of one logical CNOT and estimate the average number of logical
CNOTs required.  A straightforward method for preparing a noisy
logical $\ket{\pi/8}$-state is to prepare a logical Bell state, decode
the first block and make a measurement in the basis
$\ket{\pi/8},\ket{5\pi/8} = -\sin(\pi/8)\ket{0}+\cos(\pi/8)\ket{1}$ of
each of the two decoded, now physical qubits. Note that $\ket{5\pi/8}$
differs from $\ket{\pi/8}$ by a $Y$ operator, so any measurement
outcome is acceptable and can be accounted for by a change in Pauli
frame if necessary.  The simulations indicate that if the measurement
has the same error probability as a $Z$- or $X$-measurement, then the
error $\epsilon_{\pi/8}$ in the logical prepared state is near the EPG
parameter $\gamma$.  To reduce the noise in the logical $\ket{\pi/8}$
states, one can purify them.  The simplest purification method known
so far involves using $15$ prepared $\ket{\pi/8}$ states. One is
encoded into the $[[7,1,3]]$ code~\cite{steane:qc1995a} (a code that
encodes $1$ qubit in $7$ qubits with minimum distance $3$, which
implies that it can correct any $(3-1)/2=1$ qubit error or detect any
$(3-1)=2$ qubit errors).  The other $2\times 7$ $\ket{\pi/8}=14$
states are used to implement a conditional logical HAD from an ancilla
to realize an encoded HAD measurement. Note that $\ket{\pi/8}$ is the
$+1$ eigenstate of HAD.  In the last step, the $[[7,1,3]]$ code is
decoded.  If the measurement outcomes are as would be expected if no
error had occurred, the state is accepted and has much reduced
conditional error.  The method is equivalent to Bravyi and Kitaev's
scheme~\cite{bravyi:qc2004a} (Reichardt, private communication) and
can be analyzed using their formulas.  With no error in the $\minCSS$
and HADs used to implement the procedure, the probability of error in
successfully purified gates is $\epsilon^{(1)}_{\pi/8} =
35\epsilon_{\pi/8}^3+O(\epsilon_{\pi/8}^4)$.  The probability of
success is $p^{(1)}_{\pi/8} =
1-35\epsilon_{\pi/8}-O(\epsilon_{\pi/8}^2)$.  Using the exact
formulas, for $\epsilon_{\pi/8} = 0.01, 0.001, 0.0001$ we obtain
$\epsilon^{(1)}_{\pi/8} = 3.6\times 10^{-5}, 3.5\times 10^{-8}, 3.5\times
10^{-11}$ and $p^{(1)}_{\pi/8} = 0.860, 0.985, 0.999$.  The purification
can be iterated, but for the example considered below, the dominant
errors are from the logical $\minCSS$ and HAD gates, so we use only one
purification stage.

\ignore{
% octave
function e1 = eps8(e);
  e1 = (1-15*(1-2*e)^7+15*(1-2*e)^8-(1-2*e)^15)/ \
       (2*(1+15*(1-2*e)^8));
endfunction;
function p1 = p8(e);
  p1 = (1+15*(1-2*e)^8)/16;
endfunction;

}

Consider the effect of logical gate error on the error in the purified
$\ket{\pi/8}$.  We conjecture that by using state injection with
Steane's fault-tolerant methods for preparing states, the additional
error on the purified $\ket{\pi/8}$ state is dominated by a decoding
error of the order of the logical CNOT error.  Specifically, one can
encode one noisy logical $\ket{\pi/8}$ by teleportation into the
$[[7,1,3]]$ code using a Bell state correlating a logical qubit and a
$[[7,1,3]]$-encoded qubit. (Strictly speaking, our architecture
requires the use of logical qubit pairs associated with blocks of the
$C_4/C_6$ codes, but we treat each qubit in a pair identically.) This
Bell state has minimum distance $4$, so that any combination of Pauli
errors on up to three qubits results in an orthogonal state. It can
therefore be well verified using Steane's methods. The error in the
state teleported into the $[[7,1,3]]$ code is due to the initially
prepared $\ket{\pi/8}$ state, initial error in the logical qubit of
the Bell state used, and the CNOT and measurements needed for the
teleportation Bell measurement.  Because we are operating with logical
qubits of the $C_4/C_6$ architecture, all but the first of these
errors are comparatively small, assuming that the $C_4/C_6$ encoding
level is chosen so as to significantly decrease CNOT errors.  The
errors have two effects. One is to modify the encoded state, which can
be subsumed by considering this as additional error in the initial
$\ket{\pi/8}$ state. The other is to perturb the syndrome of the
encoded state. If two or fewer errors occurred, this can be detected
in the decoding stage.  The encoded state is verified using the
controlled-HADs implemented with the other $14$ noisy logical
$\ket{\pi/8}$ states. Each of these controlled-HADs involves at most
five CNOTs~\cite{knill:qc2004a}.  The error is dominated by that in
the $\ket{\pi/8}$ states used.  Additional error due to the logical
CNOT either has a smaller effect, to be detected in decoding, or
results in the wrong outcome in the encoded HAD measurement.  The
latter event could cause unintentional acceptance of the final state,
but only if additional error occurred elsewhere.  At the end of the
procedure, the $[[7,1,3]]$-encoded qubit is decoded and the syndrome
verified.  One can decode directly or by reverse teleportation through
the same type of Bell state used for the initial teleportation,
verifying the syndrome in the teleportation process. The latter method
may be more robust.  In all cases, the effect of additional errors are
either suppressed by the fault-tolerant methods used to encode and
decode the $[[7,1,3]]$ code, or can be subsumed as a relatively small
amount of additional error in the initial $\ket{\pi/8}$ states due to
at most five logical CNOTs.  Based on experience with $C_4/C_6$ codes,
it is likely that the additional error from encoding and decoding is
of the order of that of a logical CNOT, whereas the effective
additional $\ket{\pi/8}$ error should be sufficiently small (because
of significant decrease in CNOT errors at the level chosen) to have
little effect on the error in the purified $\ket{\pi/8}$.

We estimate the number of logical CNOTs needed for the $\ket{\pi/8}$
purification process.  The Bell state needed for injection into the
$[[7,1,3]]$ code can be prepared from a logical Bell state by encoding
one of the two blocks into the $[[7,1,3]]$ code. $11$ CNOTs suffice
for encoding. The resulting state can be verified using Steane's
methods.  There are eight syndromes each of weight $4$ to check, 
each requires an ancilla preparation with five CNOTs and four CNOTs
for the syndrome check.  If memory is an issue, it may be necessary to
add error-correcting teleportations not associated with a gate. We do
not consider this here but note that this may add another four logical
Bell states per syndrome check to the resources required.  If the
robust decoding scheme is used, two of the injection Bell states are required
overall.  The verification process using controlled-HADs requires
about $5\times 7$ CNOTs.  This gives a total of $201$ CNOTs, but does
not take into account the probability of failure in the various
checks. We can estimate this probability as $1-(1-p)^{201}$,
where $p$ is the probability of detected error in a logical CNOT.  For
the relevant parameters, $p$ is below $\expdata{0.003}{detected error
probability at $\gamma=0.01$ and level $4$}$.  Taking the average
number of trials required due to logical gate failure as
$1/(1-p)^{201}$, we can upper bound the average number of CNOTs
required as $\expdata{370}{computed here}$.

An obvious optimization of the $\ket{\pi/8}$-purification method in
the context of the fault tolerant $C_4/C_6$ architecture is to
concatenate with the $[[7,1,3]]$ code as a last level, lifting all
logical states accordingly, but injecting $\ket{\pi/8}$ states to the
last $C_4/C_6$ level as before for $\ket{\pi/8}$ purification
purposes.  This avoids having to decode the purified $\ket{\pi/8}$
states while achieving significantly lower error probabilities.
Within the $C_4/C_6$ scheme, if more than one purification stage is
required, it may be worthwhile injecting and purifying states at
intermediate levels before injecting and purifying at the top.

As an example, consider $\gamma=0.01$, aiming for implementing a
non-trivial quantum computation. The smallest non-trivial quantum
computation must be one involving more qubits than can be directly
simulated on existing classical computers.  $100$ qubits is a safe
number for this property. Such a quantum computation should also apply
sufficiently many gates for a classical simulation with current
computers not to be able to predict the output of the quantum
computation by taking advantage of restrictions on the reachable
states.  If the number of gates applied involves sufficiently many
parallel steps of non-Clifford gates involving all qubits, this is
expected to be the case. Short of having an explicit example of a
computation whose output is unknown and not believed to be accessible
to classical computers, we assume that $10$ steps involving parallel
CNOTs, HADs and $\ket{\pi/8}$-preparations suffice\footnote{Finding a
computation with as few as $100$ qubits and fewer than $10^4$ gates
with a definite and convincing answer of interest independent of
quantum information theory would be very helpful and could be a boon
for quantum information processing. For comparison, all fully worked
out computations of this sort seem to require that the number of gates
greatly exceeds $10^9$.}. We therefore take $1000$ as a minimal number
of gates in a non-trivial quantum algorithm. Note that with EPGs of
$0.01$ it is not possible to combine this many physical gates and
still expect that a computation's output can be discerned.  If more
than $68$ physical gates at this EPG are applied, the probability that
the output is correct cannot be guaranteed to be strictly greater than
$0.5$. Although the output of a computation with such few gates may
already be difficult to simulate with current classical computers, it
is conceivably possible to do so.

Consider level $4$ of our scheme at $\gamma=0.01$.  The detected error
probability of a logical CNOT is
$p_d=\expdata{2.4\aerrb{1.0}{0.7}\times
10^{-3}}{cliff/finaldata/2teleec_l4_de.tex[0.01,:]}$.  The conditional
probability of a logical error is much lower and estimated 
as $p_c=\expdata{2.3\times 10^{-5}}{see below}$ (see below).  The logical
$\ket{\pi/8}$-purification method ensures that similar error
probabilities apply to uses of these states in the algorithm.  The
probability that there is a detected failure in $1000$ gates is 
$1-(1-p_d)^{1000} \approx \expdata{0.91}{computed here}$.  Thus, the algorithm
needs to be applied $\expdata{11.1}{computed here}$ times 
on average before a successful answer
is obtained. Once an answer is obtained, its error probability is
$1-(1-p_c)^{1000} \approx \expdata{0.02}{computed here}$.  The average
number of physical CNOTs required for 
obtaining an answer can now be estimated as
\begin{equation}
\expdata{1.23\times 10^{14}}{computed here} = 
   \begin{array}[t]{lclcl}
       \underbrace{1000}_{
         \textrm{\begin{tabular}{c}logical gates\\ in computation\end{tabular}}
         } &\times& 
       \underbrace{370}_{
         \textrm{\begin{tabular}{c} logical CNOTs per\\ purified $\ket{\pi/8}$-prep\end{tabular}}
         } &\times& 
       \underbrace{1/0.09}_{
         \textrm{\begin{tabular}{c}prob.${}^{-1}$~of \\ overall success\end{tabular}}
         } \\[.7in]  &\times & 
       \underbrace{2\times 1.5\times 10^7}_{
         \textrm{\begin{tabular}{c}physical CNOTs \\ per logical CNOT\end{tabular}}
         }
    \end{array}
\end{equation}
Although this number of physical CNOTs vastly exceeds current
capabilities, it may be compared to typical resources available today
for classical computation.  For example, today's central processing
units in desktop computers have more than
$\expdata{10^8}{http://computer.howstuffworks.com/microprocessor2.htm}$
transistors and operate at rates above $10^9$ bit operations per
second~\cite{intel.com:qc2004a}.

Resource requirements for implementing a given computation decrease
significantly with $\gamma$. Simulation is too inefficient for
resolving the dependence of resource requirements on $\gamma$,
particularly when error probabilities are extremely small.  We
therefore obtain and verify simple models for resources and errors as
a function of $\gamma$ and level of concatenation.  Ideally, we would
like to obtain analytic expressions, however this is difficult to do,
particularly since our schemes are not strictly concatenated, and the
combination of error-detection and correction behaves differently
depending on the level.  Nevertheless, it is possible to derive 
functional forms for Bell state preparation resources and logical CNOT
error behavior that are asymptotically valid as $\gamma$ goes to $0$.

We model the number of physical CNOTs required for preparing logical
Bell states at level $l$ as $\textrm{rbell}(l,\gamma) =
P(l)/(1-\gamma)^{k(l)}$.  This is a naive model based on assuming that
the resources are determined by applying a network with $P(l)$
physical CNOT gates, $k(l)$ of which fail independently
with probability $\gamma$ each, and the network is repeatedly
applied until no failure is detected.
Perhaps surprisingly, this model matches the
simulations well in the range shown in Fig.~\ref{fig:res_naive}.

\begin{herefig}
\label{fig:res_naive}
\begin{picture}(0,3.7)(0,-3.5)

\nputgr{0,0}{t}{height=3in}{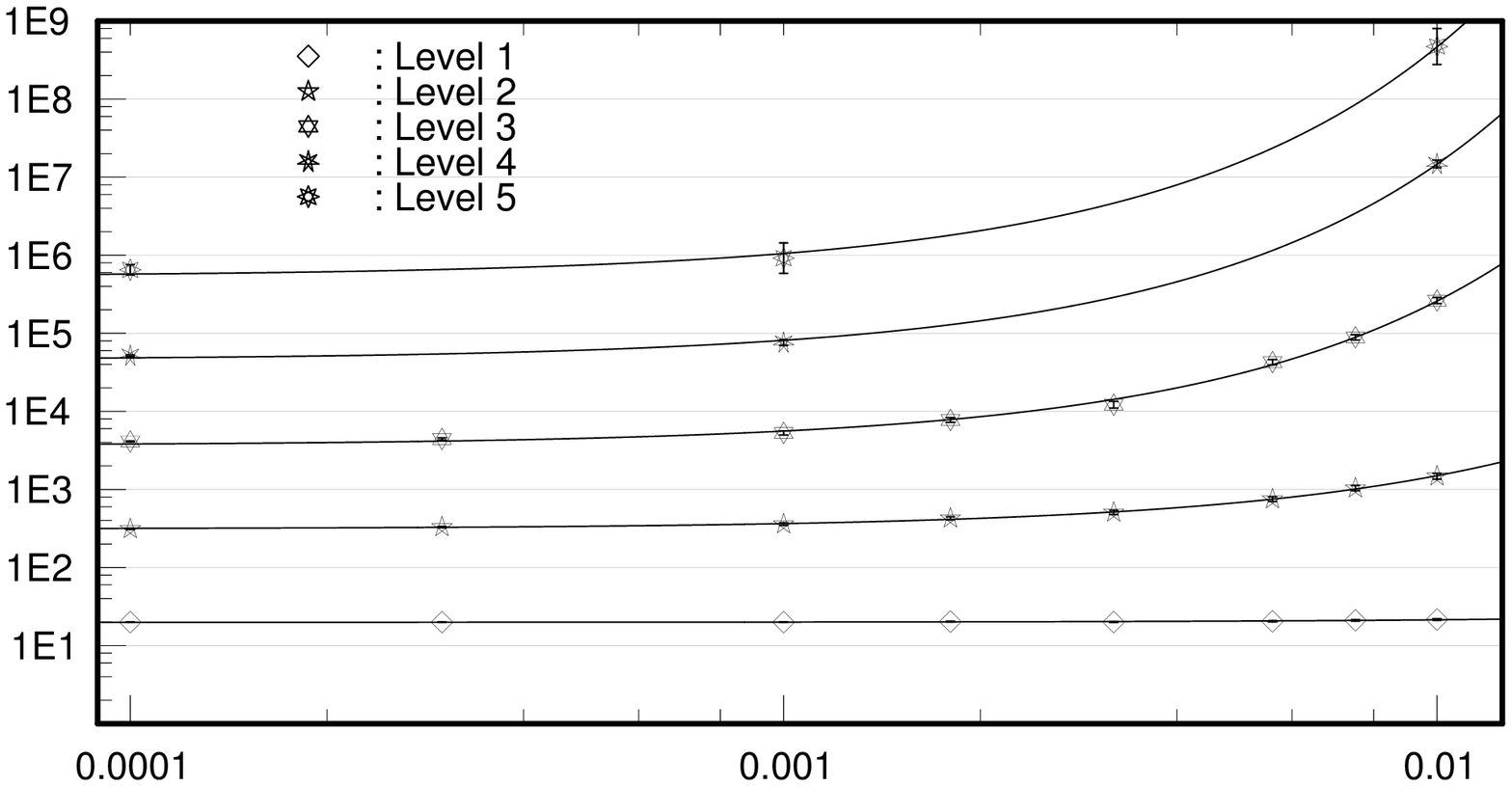}
\nputbox{-3,0}{tr}{\rotatebox{90}{\textsf{\small Number of physical CNOTs per logical Bell pair}}}
\nputbox{3,-3.0}{tr}{\textsf{\small Physical CNOT error probability $\gamma$.}}

\end{picture}
\herefigcap{Graphs of resources for logical Bell pair preparation.
The points are obtained by simulation and counting the number of physical
CNOTs used in preparing a number of logical Bell pairs at different
values of $\gamma$. The error bars are not statistically
rigorous. They are standard deviations computed from the number of
prepared Bell pairs and assuming the naive model described in the
text. The curves are least-squares fits to the data with
functions of the form $P(l)/(1-\gamma)^{k(l)}$.
}
\end{herefig}

To understand the error behavior of the $C_4/C_6$ architecture,
suppose more generally that we have a fault-tolerant scheme $\cA$ for
implementing an encoded gate, which results in a detected,
uncorrectable error with probability $p_d$, or an undetected logical
error with probability $p_c$, conditional on not having detected an
error. Suppose that this is concatenated with a one-error detecting
(minimum distance $2$) code $C$ and used in a scheme similar to the
ones used here. $C$ can correct any error at a known location.  If the
implementation of $C$-encoded gates is fault tolerant and includes
error-correcting teleportation or another method for determing the
$C$-syndrome, any one error detected by $\cA$ can be corrected with no
resulting encoded error. The event that an error is detected but not
correctable during implementation of a $C$-encoded gate therefore
requires at least one undetected error or at least two detected
errors. The conditional event that an undetected error occurs requires
that the $\cA$ gates used have one detected and one undetected error
or two or more undetected errors.  To lowest order, the detected and
conditional error probabilities for $C$-encoded gates are therefore of
the form $p'_d = D_c p_c + D_d p_d^2$ and $p'_c = L_c p_c^2 + C_d p_d
C_c $ In our case, $p_d$ and $p_c$ depend on one parameter $\gamma$.
After level $1$, the order in $\gamma$ of $p_c$ is always between
$p_d$ and $p_d^2$, so the expressions can be simplified to $p'_d = D
p_c$, $p'_c = C p_d p_c$, to lowest order in $\gamma$.  Let $p_d(l)$
and $p_c(l)$ be the detected and logical error probabilities at level
$l$ in the $C_4/C_6$ architecture. Using the above, we can
write $p_d(l+1) = D(l)p_c(l)$ and $p_c(l+1)=C(l)p_d(l)p_c(l)$.  At
level $1$, $p_d(1) = d(1)\gamma$ and $p_c(1) = c(1)\gamma^2$, to
lowest order in $\gamma$.  Expanding the recursion at higher levels,
we obtain $p_d(2) = d(2)\gamma^2, p_c(2) = c(2)\gamma^3$, $p_d(3) =
d(3)\gamma^3, p_c(3) = c(3)\gamma^5$. The Fibonacci sequence $f(0) = 0,
f(1) = 1, f(n+1) = f(n)+f(n-1)$ emerges as the relevant exponent so
that $p_d(l) = d(l)\gamma^{f(l+1)}, p_c(l)= c(l)\gamma^{f(l+2)}$.  As is
typical of concatenation schemes, the exponent grows exponentially.

In view of the previous paragraph, we examine the data shown in
Fig.~\ref{fig:ecEPG} to determine $C(l), D(l)$ for $l=1,2$ and $c(l),
d(l)$ for $l=1,2,3$. The results are shown in
Table~\ref{tab:errcnsts}.  We computed the values of $c(l)$ and $d(l)$
by fitting the model curves to the error probabilities obtained by
simulation.  The points at $\gamma = 0.01$ were omitted for levels
$1$, $2$, and $3$ to reduce the chance of introducing optimistic
biases by the curves' leveling off at higher $\gamma$, although this
effect has not been observed.  We obtained the fits by starting with a
least-squares fit of the log-log plots and then using a
fastest-descent method to optimize the likelihood.  We computed
standard deviations by resampling the data according to the fitted
curve and repeating the fitting process.  The fitted curves are shown
with the data in Fig.~\ref{fig:ecfibfits}.  $C(l)$ and $D(l)$ were
computed from $c(l)$, $c(l+1)$, $d(l)$ and $d(l+1)$ by solving the
equations. Their uncertainty intervals are found by linear error
analysis.

\begin{heretab}
\label{tab:errcnsts}
\heretabcap{ Table of $d(l)$, $c(l)$, $D(l)$ and $C(l)$ with uncertainty
intevals based on standard deviations.
  \\[6pt] }
$\begin{array}{@{}c|cccc@{}}
\hline
  \textrm{Level} & d(l) & c(l) & D(l) & C(l) \\\hline
\expdata{}{data produced by octave code following this table.}
1 & 37.0\serrb{0.1} & 35.2\serrb{1.5} & 29.94\serrb{1.13} &  3.43\serrb{0.01} \\
 2 & 1.06\serrb{0.01}{\times}10^{3} & 4.47\serrb{0.18}{\times}10^{3} &  4.87\serrb{0.14} &  1.69\serrb{0.14} \\
 3 & 2.18\serrb{0.02}{\times}10^{4} & 7.95\serrb{1.01}{\times}10^{6} & 3.01\serrb{0.70} & \textrm{NA} \\
 4 & 2.39\serrb{0.86}{\times}10^{7} & \textrm{NA} & \textrm{NA} & \textrm{NA} \\
\hline
\end{array}$
\end{heretab}
\ignore{
% Octave.
% Computing Table~\ref{tab:errcnsts}
% Load files:
stat1d = load('cliff/finaldata/2teleec_to0.9pwr_l1_de.ipol');
stat1c = load('cliff/finaldata/2teleec_to0.9pwr_l1_be.ipol');
stat2d = load('cliff/finaldata/2teleec_to0.9pwr_l2_de.ipol');
stat2c = load('cliff/finaldata/2teleec_to0.9pwr_l2_be.ipol');
stat3d = load('cliff/finaldata/2teleec_to0.9pwr_l3_de.ipol');
stat3c = load('cliff/finaldata/2teleec_to0.9pwr_l3_be.ipol');
stat4d = load('cliff/finaldata/2teleec_topwr_l4_de.ipol');

function  [f,df, m]  = flead(x,dx);
  m = floor(log10(x));
  f = x*10^(-m);
  df = dx*10^(-m);
endfunction;

function [D,C] = concrate(d0, c0, d1, c1);
  % arguments are given as pairs of  [value, stddv].

  vD = d1(1)/c0(1);
  vC = c1(1)/(d0(1)*c0(1));

  % D = x3/x2; C = x4/(x1*x2);
  % Jac([D; C]) = 
  %   [0,-x3/x2^2,1/x2,0;
  %    -x4/(x1^2*x2), -x4/(x1*x2^2), 0, 1/(x1*x2)]

  dD = abs((-d1(1)/(c0(1)^2))*c0(2) + \
       (1/c0(1))*d1(2));
  dC = abs((-c1(1)/(d0(1)^2*c0(1)))*d0(2) + \
       (-c1(1)/(d0(1)*c0(1)^2))*c0(2) + \
       (1/(d0(1)*c0(1)))*c1(2));

  D = [vD, dD]; C = [vC, dC];

endfunction;

tab = '';
sd1 = stat1d(3); dsd1 = sqrt(stat1d(6)); md1 = 0;
sc1 = stat1c(3); dsc1 = sqrt(stat1c(6)); mc1 = 0;
[sd2, dsd2, md2] = flead(stat2d(3), sqrt(stat2d(6)));
[sc2, dsc2, mc2] = flead(stat2c(3), sqrt(stat2c(6)));
[D1, C1] = concrate([sd1,dsd1]*10^md1, [sc1, dsc1]*10^mc1, \
                    [sd2, dsd2]*10^md2, [sc2, dsc2]*10^mc2);
[sd3, dsd3, md3] = flead(stat3d(3), sqrt(stat3d(6)));
[sc3, dsc3, mc3] = flead(stat3c(3), sqrt(stat3c(6)));
[D2, C2] = concrate([sd2,dsd2]*10^md2, [sc2, dsc2]*10^mc2, \
                    [sd3, dsd3]*10^md3, [sc3, dsc3]*10^mc3);
[sd4, dsd4, md4] = flead(stat4d(3), sqrt(stat4d(6)));
[D3, C3] = concrate([sd3,dsd3]*10^md3, [sc3, dsc3]*10^mc3, \
                    [sd4, dsd4]*10^md4, [1, 0]);

tab = sprintf('%s %d & %.1f\\serrb{%.1f} & %.1f\\serrb{%.1f} & %.2f\\serrb{%.2f} &  %.2f\\serrb{%.2f} \\\\\n', \
         tab, 1, sd1, dsd1, sc1, dsc1, \
         D1(1), D1(2), C1(1), C1(2) \
       );
tab = sprintf('%s %d & %.2f\\serrb{%.2f}{\\times}10^{%d} & %.2f\\serrb{%.2f}{\\times}10^{%d} &  %.2f\\serrb{%.2f} &  %.2f\\serrb{%.2f} \\\\\n', \
         tab, 2, sd2, dsd2, md2, sc2, dsc2, mc2, \
         D2(1), D2(2), C2(1), C2(2) \
       );

tab = sprintf('%s %d & %.2f\\serrb{%.2f}{\\times}10^{%d} & %.2f\\serrb{%.2f}{\\times}10^{%d} & %.2f\\serrb{%.2f} & \\textrm{NA} \\\\\n', \
         tab, 3, sd3, dsd3, md3, sc3, dsc3, mc3, \
         D3(1), D3(2) \
       );

tab = sprintf('%s %d & %.2f\\serrb{%.2f}{\\times}10^{%d} & \\textrm{NA} & \\textrm{NA} & \\textrm{NA} \\\\\n', \
         tab, 4, sd4, dsd4, md4 \
       );

% Print value to insert into table:
tab

% Transfer to threshold guess.
global D; 
D = [sd3*10^md3, D2(1)*ones(1,1000)];
global C; 
C = [sc3*10^mc3, C2(1)*ones(1,1000)];
% Recursion:
function [pd, pc]  = pes(l, g);
  global D;
  global C;
  if (l < 3); return; endif;
  if (l == 3);
    pd = D(1)*g^3;
    pc = C(1)*g^5;
  else;
    [pdp, pcp] = pes(l-1, g);
    pd = D(l-2)*pcp;
    pc = C(l-2)*pdp*pcp;
  endif;
endfunction;

% Try it.
[pd, pc] = pes(30,0.028066); % low
[pd, pc] = pes(30,0.028067); % high

% Output for use in optimum resource graph.
probfid = fopen('cliff/finaldata/probs_pl.ipol','w');
  fprintf(probfid, '%d %.16g %.16g\n', 0, 0, 1);
  fprintf(probfid, '%d %.16g %.16g\n', 1, sd1*10^md1, sc1*10^mc1);
  fprintf(probfid, '%d %.16g %.16g\n', 2, sd2*10^md2, sc2*10^mc2);
  trnsD = D2(1);
  trnsC = C2(1);
  dl = sd2*10^md2;
  cl = sc2*10^mc2;
  for l = (3:10);
    cln = dl*cl*trnsC;
    dln = cl*trnsD;
    dl = dln; cl = cln;
    fprintf(probfid, '%d %.16e %.16e\n', l, dl, cl);
  endfor;
fclose(probfid);

}

\pagebreak
\begin{herefig}
\label{fig:ecfibfits}
\begin{picture}(0,7.2)(0,-7)
\nputgr{0,0}{t}{height=3in}{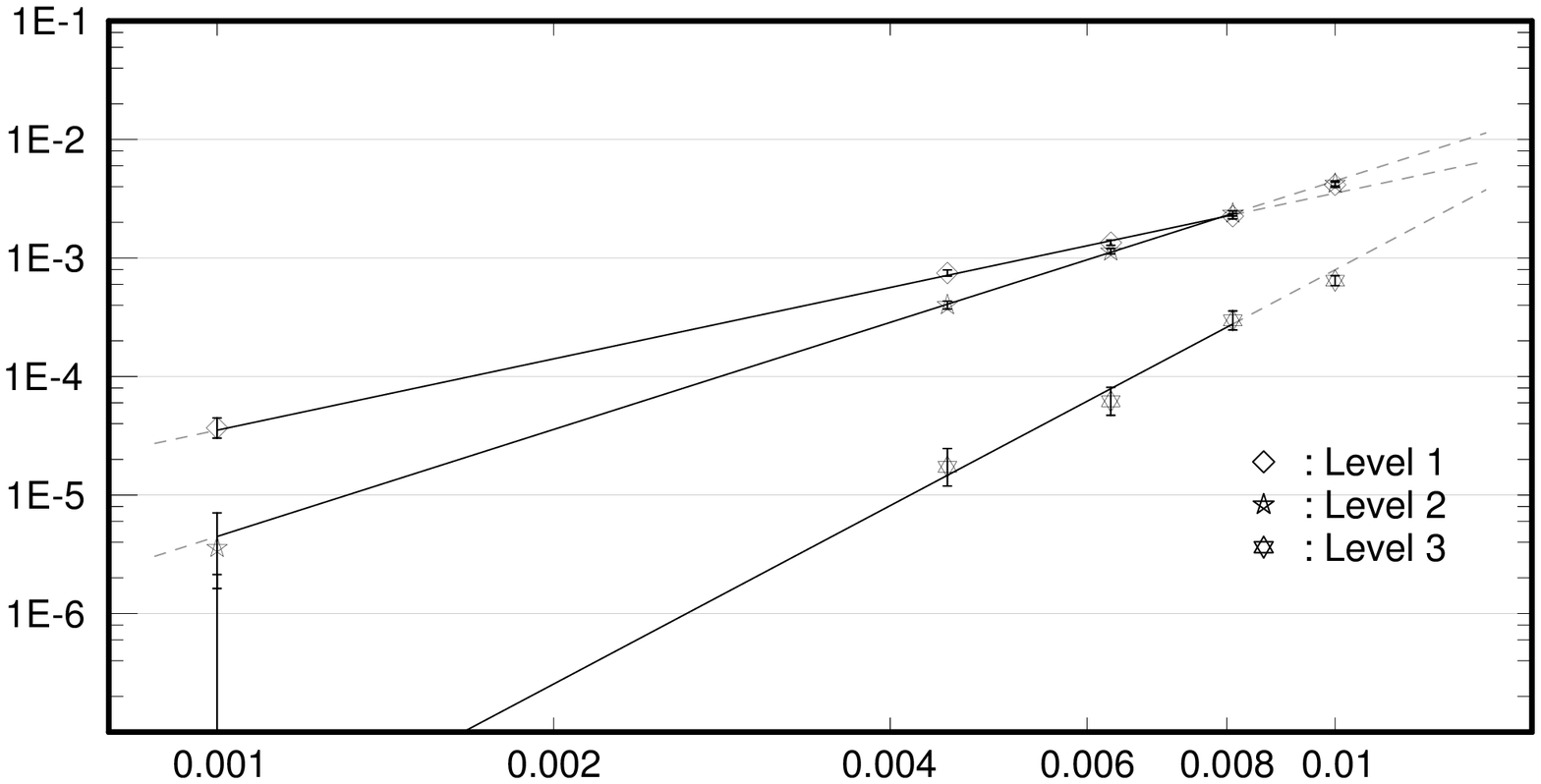}
\nputbox{-3.0,0}{tr}{\rotatebox{90}{\textsf{\small Logical CNOT conditional error probability}}}

\nputbox{-3.0,-3.3}{tr}{\rotatebox{90}{\textsf{\small Logical CNOT detected error probability}}}
\nputgr{0,-3.3}{t}{height=3in}{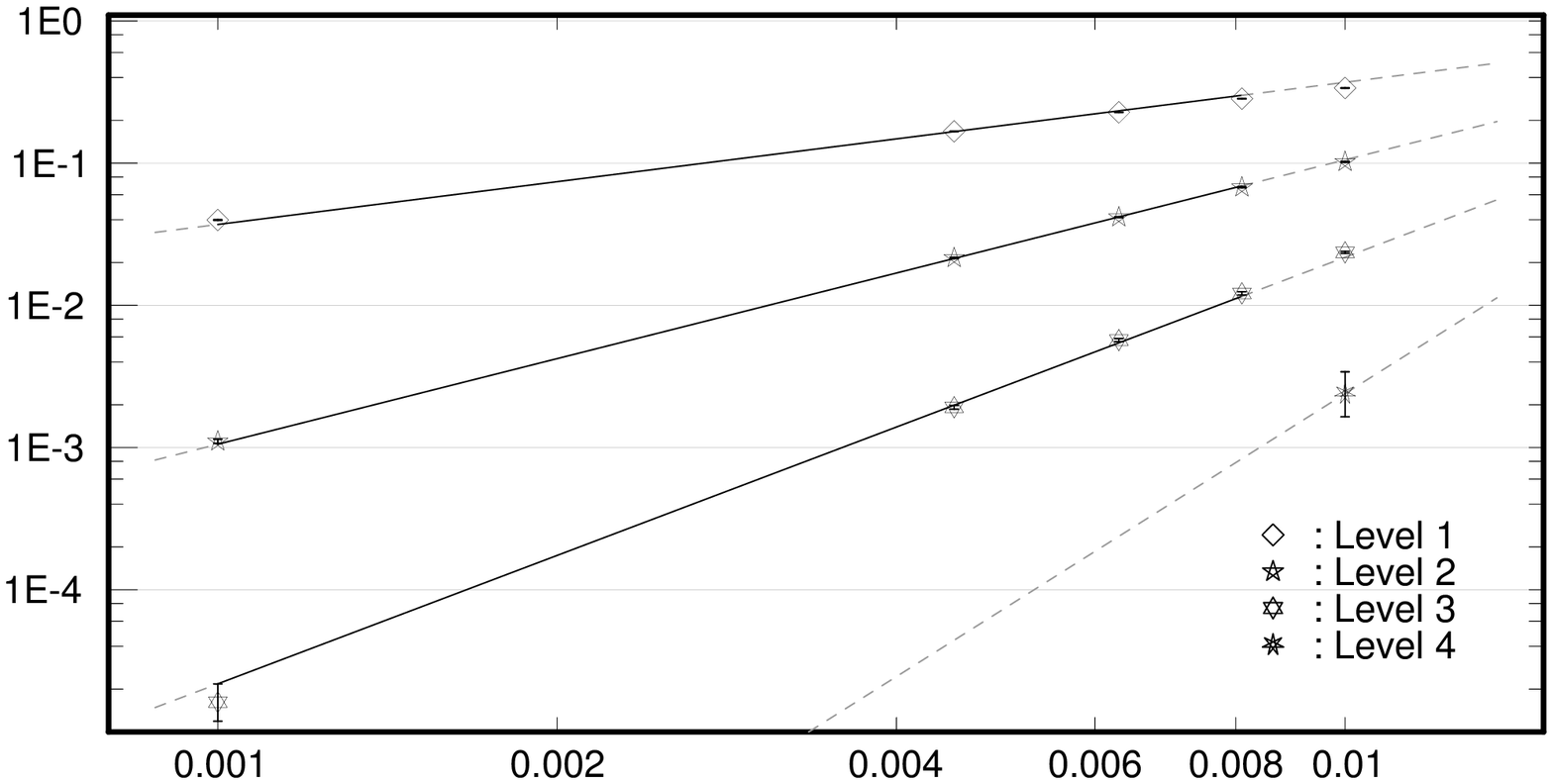}
\nputbox{3.0,-6.4}{tr}{\textsf{\small Physical CNOT error probability $\gamma$.}}
\end{picture}
\herefigcap{Fits to the
error data for the logical CNOT. The model
assumed is $p_d(l) = d(l)\gamma^{f(l+1)}, p_c(l)= c(l)\gamma^{f(l+2)}$,
where $f$ is the Fibonacci sequence. The constants
$d(l)$ and $c(l)$ are obtained by a maximum-likelihood
method from the data points in the range of the solid
lines. The gray dashed lines are extrapolations.}
\end{herefig}
\pagebreak

The constants $D(l), C(l)$ are significantly reduced for going from
level $2$ to level $3$ compared to going from level $1$ to $2$.  Level
$2$ is the first stage of using $C_6$ and the first where error
correction can be used. One may conjecture that the level $2$ to level
$3$ behavior persists or improves at higher levels, as is the case for
$D(3)$ compared to $D(2)$. For the purposes of modeling errors we use
this conjecture to recursively obtain $d(l+1)$ and $c(l+1)$ with
$D(2)$ and $C(2)$ in place of $D(l)$ and $C(l)$ for $l>2$. It is an
interesting exercise to use the recursion implied by the $D(l)$ and
$C(l)$ to obtain a threshold. The threshold thus obtained is
conjectural, because the approximations made are not strictly valid,
particularly at high $\gamma$, and because of the extrapolation of
$D(l)$ and $C(l)$.  By implementing the recursion numerically, we
obtained a threshold of ${\approx}\expdata{0.028}{octave above}$ for
this architecture, which does not seem unreasonable in view of the
data shown in Fig.~\ref{fig:ecEPG}. Of course, the resource overheads
diverge as any such threshold is approached from below.

We return to the question of resource requirements for implementing
gates at $\gamma < 0.01$.  As $\gamma$ decreases, the physical
resources required per logical CNOT are reduced in two ways. First,
the state preparation success probabilities at a given level of
concatenation increase, see
Tables~\ref{tab:gamma_resources1},~\ref{tab:gamma_resources2}
and~\ref{tab:gamma_resources3} and Fig.~\ref{fig:res_naive}.  This
increase is particularly notable near the upper limit for
$\gamma$. Second, fewer levels of concatenation suffice for achieving
sufficiently low logical errors.  Consider implementing a computation
$\cC$ with the product of the number of logical gates and average
number of qubits per gate given by $KQ$. For computations that are not
maximally parallel, this quantity should include memory delays in the
gate count. To simplify the resource estimates, logical errors and
physical gate counts are given in terms of ``effective'' error and
physical gate counts per (logical) qubit and gate.  For example,
consider the logical cnot in the $C_4/C_6$ architecture.  It acts on
two logical qubit pairs, so its effective error per qubit is $1/4$ of
its total error. Similarly, its effective physical gate count per
qubit is $1/4$ of the total gate count.  With this simplification, we
can estimate the total error and number of physical gates for
implementing the computation $\cC$ by multiplying $KQ$ by the the
appropriate effective quantity and a nontransversal-gate state
preparation overhead. In making these estimates, we assume that (1)
each of the logical gates needed by $\cC$ can be implemented with
effective error similar to that of the logical CNOT, (2) the
implementation can take advantage of both logical qubits in the
logical qubit pairs and (3) overhead for addressing individual logical
qubits in the pairs is accounted for in the nontransversal-gate state
preparation overhead.  The assumptions require that the
nontransversal-gate state preparations have the property that logical
gates used in the preparations do not contribute additional error, as
is the case for the $\ket{\pi/8}$ state preparation described above.
The reason for not including the nontransversal-gate state preparation
overhead in the effective quantities per qubit and gate is that this
overhead can be optimized independent of the architecture and depends
on the choice of elementary nontransversal gates.  It is expected to
add one to two orders of magnitude to the total implementation
resources.

We estimate the optimal effective number $\textrm{pcnot}(KQ,\gamma)$
of physical CNOTs per qubit and gate as a function of the size $KQ$ of
$\cC$ and the EPG parameter $\gamma$.  As noted above, other physical
resources such as state preparation and measurement are comparable.
We optimize $\textrm{pcnot}(KQ,\gamma)$ by choosing the level $l$ of
the $C_4/C_6$ architecture and use it to repeatedly implement $\cC$
until no uncorrectable error is detected in the logical gates.  At
this point the output of $\cC$ must be correct with probability at
least $2/3$. The value of $2/3$ is chosen to be strictly between $1/2$
and $1$ but otherwise not crucial.  At the minimizing level,
$\textrm{pcnot}(KQ,\gamma)$ is computed as the product of the average
number of times $\cC$ must be implemented until no error is detected
and $1/2$ of the number of physical CNOTs, $\textrm{rbell}(l,\gamma)$,
needed to prepare a logical Bell state (neglecting the relatively
small additional number of physical cnots needed for transversal gates
and for using the Bell state in an error-correcting teleportation).
The factor of $1/2$ accounts for having two qubits in each block of
the $C_4/C_6$ concatenated codes.  The probability of success of a
single instance of $\cC$ can be estimated as
$(1-p_d(l,\gamma)/4)^{KQ}$, which is approximately correct for our
accounting using effective errors per qubit and gate, provided that
$p_d(l,\gamma)$ is small.  On average, $\cC$ must be tried
$1/(1-p_d(l,\gamma)/4)^{KQ}$ times to successfully obtain the output.
The conditional probability of a successful output's being correct is
$(1-p_c(l,\gamma)/4)^{KQ}$.  Thus, given $KQ$, the optimal
$\textrm{pcnot}(KQ,\gamma)$ is obtained as the minimum over $l$ of ${1
\over 2}\textrm{rbell}(l,\gamma)/(1-p_d(l,\gamma)/4)^{KQ}$ subject to
$(1-p_c(l,\gamma)/4)^{KQ}\geq 2/3$.  Curves for
$\textrm{pcnot}(KQ,\gamma)$ for various $KQ$ as a function of $\gamma$
are plotted in Fig.~\ref{fig:resource_graph}.

The quantity $\textrm{pcnot}(KQ,\gamma)$ gives the overall ``work''
overhead for implementing a computation using the $C_4/C_6$
architecture, but does not differentiate between parallel and
sequential resources or indicate the number of physical qubits needed
per logical qubit (``scale-up'').  The $C_4/C_6$ architecture does not
determine these resources uniquely, as they depend on how the
trade-off between parallelism and requirements for memory is resolved.
In the case of maximum parallelism, the scale-up is close to
$\textrm{pcnot}(KQ,\gamma)$.  If minimum parallelism is used, this can
be reduced to a small multiple of the minimum scale-up associated with
the $C_4/C_6$ concatenated code at the level $l$ that is used. This
minimum scale-up is given by $3^{l-1}2$ (taking into account that
there are two qubits per block of $3^{l-1}4$ qubits).  If there is no
memory error at all, the additional overhead per block can be
minimized by operating on only one block at a time.  Otherwise, for
each block, two additional blocks are needed in error-correcting
teleportation. Logical Bell state preparation requires an additional
overhead depending on the degree of parallelism required.  If the
subblock teleportations in the preparation are done in parallel, and
taking into accounting lower level Bell state preparations, two more
blocks or equivalent are needed for each level other than the first.
This means that $1+2(l-1)$ blocks are needed per computational block.
The $\ket{\pi/8}$-state preparation has additional overhead. Depending
on how it is implemented it may require up to $14$ blocks with their
own overhead of $1+2(l-1)$ or more blocks each. The contribution of
$\ket{\pi/8}$-state preparation can be minimized by implementing the
logical part of the computation sequentially but using memory steps to
remove the effects of memory error as needed.  Based on these
estimates, the scale-up for low but not minimum parallelism is
$\approx 3^{l-1}2(1+2(l-1))$.  At levels $2$, $3$, $4$, this evaluates
to $18$, $90$, $378$, respectively.

The error-correcting $C_4/C_6$ architecture is relatively simple and
designed to work well at high EPGs. However, there is a minimum
resource cost (of order $10^3$ per gate and qubit) to use it since
error-correction kicks in only at level $2$. As a result, at low EPGs,
architectures such as Steane's~\cite{steane:qc2002a} based on more
efficient codes with little or no concatenation are more efficient and
have more flexibility in achieving the desired logical error
probabilities. This effect can be quantified by comparing the
$C_4/C_6$ architecture to that of Steane using the illustrative
example at $\gamma\approx 10^{-4}$ worked out
in~[\citeonline{steane:qc2002a}]. Steane's error model differs from
ours in that preparation, measurement and one-qubit gates all have
error probability $\gamma$. In our analysis, preparation and
measurement errors are $4\gamma/15$, which we justified with a
purification scheme. This scheme could also be used in the context of
Steane's error model. We compare the two architectures based on the
resources per logical qubit of one logical step such as a CNOT, for
which the $C_4/C_6$ architecture does not require one-qubit gates
other than preparation and measurement.  Steane's error model also
includes memory error ($\gamma/100$ per step) and accounts for
measurement times in excess of gate times ($25$ times the gate time).
In our model and in the maximally parallel setting, this would require
an additional error of $\gamma/4$ per qubit at the end of state
preparation to delay for measurement outcomes that determine whether
the state is good or not.  The comparison is also complicated by
Steane's method deferring some error correction to later steps (we do
not account for the implicit overhead in this) and by our method
having both detected and conditional logical error, with the latter
typically being much lower (we use only the detected error for
comparison).

Steane's example is based on a $[[127,43,13]]$ code, which encodes
$43$ logical qubits.  Full error correction of a block requires about
$1.8\times 10^4$ physical CNOTs on average and has a probability of
logical error (called ``crash probability''
in~[\citeonline{steane:qc2002a}]) of $\approx 3\times 10^{-10}$.  This
translates to $\approx 420$ physical CNOTs per qubit and gate and an
effective error of $\approx 7\times 10^{-12}$ per logical qubit.  The
$C_4/C_6$ architecture at level 3 uses $\expdata{4158.1}{from
table~\ref{tab:gamma_resources3}}$ physical CNOTs for an
error-correcting teleportation. Including $36$ physical gates for a
transversal operation, this gives $\approx \expdata{2100}{a little
more than the previous number /2}$ physical CNOTs per qubit and gate.
The detected error probability for a logical CNOT was estimated above
as $\expdata{2.2\times 10^{-8}}{C*.0001^3 with C from
cliff/finaldata/2teleec_to0.9pwr_l3.ipol[3]}$, which translates to
$\approx \expdata{5.5 \times 10^{-9}}{previous number /4}$ effective
error per qubit and gate.  To meet the effective error probability
achieved by Steane requires another level of encoding. At level $4$,
the $C_4/C_6$ architecture uses $\expdata{2.6{\times}10^4}{from
table~\ref{tab:gamma_resources3}, /2}$ physical CNOTs per qubit and
gate and with a detected error probability of $\expdata{6\times
10^{-14}}{C*.0001^5/4 with C from
cliff/finaldata/2teleec_to0.9pwr_l4_de.ipol[3]}$ per qubit and gate.  One
can also compare the scale-up for the two architectures at
$\gamma=10^{-4}$: Steane's example has a scale-up of between $10$ and
$20$ compared to from $378$ to over $2000$ for the $C_4/C_6$
architecture at level $3$, depending on parallelism.  As expected,
Steane's architecture requires fewer resources at low EPGs.  It is
however notable that the $C_4/C_6$ architecture requires only two
orders of magnitude more resources at EPGs as low as $\gamma=10^{-4}$.
The $C_4/C_6$ architecture has the advantage of simplicity and of
yielding more reliable answers, conditional on having no detected
errors.

\ignore{
Steane's ``illustrative example'' parameters from \cite{steane:qc2002a}.
43 logical qubits.
n_{rep} = 2.5: number of ancilla blocks maintained per block per X or Z.
w = 47: Maximum row/column weight of latin square.
N_A = 1802: Total weight. Number of CNOT/csgn in verification.
N_{GV} = 3689: Total number of gates in the G and V networks.
N_{h} = 8893: Number of memory steps in the G and V networks.
r = 5:
r' = 4:
r'' = 3:
t_R = 143: recovery time steps.
N_A/w = 38: gates per time step.

Consider one X or Z recovery step (need one of each per block):
alpha = 0.74: Fraction of prepared ancillas passing verification on first try.
beta = 0.8: Fraction of syndromes that are zero on first extraction try.
(1-beta)*(r-1): Additional syndrome extractions.

Recovery gate count: 2*(N_{GV}/0.74)(0.8+0.2*5) = 17946

\bar p = 3 10^{-10} : Apparently the total probability of error for both
recovery steps for one block, but not clear in the example.

This is ignoring the problem of trying to catch errors at later
recoveries (delayed error recovery).

Regarding Steane's resources in \cite{steane:qc1999a}:
Focuses on ``scale-up'' rather than overall work requirement.
Scale-up of 22 for KQ \approx 10^16.

From \cite{steane:qc1997a}, apparently improving \cite{preskill:qc1998a}:
KQ = 2*10^10 (*2 for this being two qubit gates, includes a factor
  of 10 for Toffoli gate state preps) 
T(total number of CNOTs or equivalent, but includes
the Toffoli overhead) = 4*10^16 at \gamma=3.2*10^{-5}.
Hence, for comparison, the overhead is about 2*10^5.
Generally, \cite{steane:qc1997a} comparable
overheads are of the order of 10^5.

}

\end{document}